\documentclass[12pt]{article}
\usepackage{jheppub}

\usepackage{amsmath,bbm,array,cleveref,amsfonts,subcaption,graphicx,wrapfig,lscape,float,slashbox,multirow,longtable,rotating,epstopdf}
\usepackage[export]{adjustbox}
\newcommand{\be}{\begin{equation}}
\newcommand{\ee}{\end{equation}}
\newcommand{\beq}{\begin{equation}}
\newcommand{\beql}[1]{\begin{equation}\label{#1}}
\newcommand{\eeq}{\end{equation}}
\newcommand{\ba}{\begin{array}}
\newcommand{\ea}{\end{array}}
\newcommand{\bea}{\begin{eqnarray}}
\newcommand{\beal}[1]{\begin{eqnarray}\label{#1}}
\newcommand{\eea}{\end{eqnarray}}
\newcommand{\ben}{\begin{enumerate}}
\newcommand{\een}{\end{enumerate}}
\newcommand{\bean}{\begin{eqnarray*}}
\newcommand{\eean}{\end{eqnarray*}}
\newcommand{\eref}[1]{(\ref{#1})}
\newcommand{\sref}[1]{\S\ref{#1}}

\newcommand{\fref}[1]{Figure \ref{#1}}
\newcommand{\btab}[1]{\begin{tabular}{#1}}
\newcommand{\etab}{\end{tabular}}

\newcommand{\comment}[1]{}

\newcommand{\IC}{\mathbb{C}}

\newcommand{\qed}{\nobreak \ifvmode \relax \else
      \ifdim\lastskip<1.5em \hskip-\lastskip
      \hskip1.5em plus0em minus0.5em \fi \nobreak
      \vrule height0.75em width0.5em depth0.25em\fi}

\def\beqa{\begin{eqnarray}}
\def\eeqa{\end{eqnarray}}	

\newcolumntype{C}[1]{>{\centering\arraybackslash}m{#1}}

\def\makeatletter{\catcode`\@=11}
\makeatletter
\def\mathbox#1{\hbox{$\m@th#1$}}%
\def\math@ccstyles#1#2#3#4#5#6#7{{\leavevmode
     \setbox0\mathbox{#6#7}%
     \setbox2\mathbox{#4#5}%
     \dimen@ #3%
     \baselineskip\z@\lineskiplimit#1\lineskip\z@
     \vbox{\ialign{##\crcr
            \hfil \kern #2\box2 \hfil\crcr
            \noalign{\kern\dimen@}%
            \hfil\box0\hfil\crcr}}}}
\def\mathaccstyles{\math@ccstyles\maxdimen}
\def\maththroughstyles{\math@ccstyles{-\maxdimen}}
\def\unity%
{\maththroughstyles{.45\ht0}\z@\displaystyle {\mathchar"006C}\displaystyle 1}

\title{Bipartite field theories and D-brane instantons}

\author[a,b]{Sebasti\'an Franco}
\author[c,d]{Eduardo Garc\'{\i}a-Valdecasas}
\author[d]{Angel M. Uranga}

\affiliation[a]{
Physics Department, The City College of the CUNY \\
160 Convent Avenue, New York, NY 10031, USA}

\affiliation[b]{The Graduate School and University Center, The City University of New York  \\
365 Fifth Avenue, New York NY 10016, USA}

\affiliation[c]{Departamento de F\'isica Te\'orica, Facultad de Ciencias, Universidad Aut\'onoma de Madrid \\ 28049 Madrid, Spain}

\affiliation[d]{Instituto de F\'isica Te\'orica IFT-UAM/CSIC \\
C/ Nicol\'as Cabrera 13-15, Universidad Aut\'onoma de Madrid, 28049 Madrid, Spain
}

\emailAdd{sfranco@ccny.cuny.edu, eduardo.garcia.valdecasas@gmail.com, angel.uranga@uam.es}

\abstract{
We study D-brane instantons in systems of D3-branes at toric CY 3-fold singularities. The instanton effect can be described as a backreaction modifying the geometry of the mirror configuration, in which the breaking of $U(1)$ symmetries by the instanton translates into the recombination of gauge D-branes, which also directly generates the instanton-induced charged field theory operator. In this paper we describe the D-brane instanton backreaction in terms of a combinatorial operation in the bipartite dimer diagram of the original theory. Interestingly, the resulting theory is a general Bipartite Field Theory (BFT), defined by a bipartite graph tiling a general (possibly higher-genus) Riemann surface. This provides the first string theory realization of such general BFTs. We study the general properties of the resulting theories, including the construction of the higher-dimensional toric diagrams and the interplay between backreaction and Seiberg duality. In cases where the non-perturbative effects relate to complex deformations, we show that the procedure reproduces and explains earlier existing combinatorial recipes. The combinatorial operation and its properties generalize to an operation on the class of general BFTs, even including boundaries, relating BFTs defined on Riemann surfaces of different genus.}

\preprint{
\begin{flushright}
CCNY-HEP-18-04 \\
IFT-UAM/CSIC-18-44
\end{flushright} 
}

\begin{document}

\maketitle


\section{Introduction}

The present paper combines new results in two interesting areas of D-brane physics: non-perturbative D-brane instantons and the realization of Bipartite Field Theories (BFTs) using D-branes.

D-brane instantons have become a centerpiece in the understanding of string theory beyond perturbation theory (see e.g. \cite{Becker:1995kb,Witten:1996bn,Harvey:1999as,Witten:1999eg}) and in model building applications to moduli stabilization (see e.g. \cite{Kachru:2003aw,Balasubramanian:2005zx}) or the generation of charged field theory operators (see e.g. \cite{Blumenhagen:2006xt,Ibanez:2006da,Florea:2006si} and \cite{Blumenhagen:2009qh,Ibanez:2012zz} for reviews). These field theory operators arise as 't Hooft couplings required by the saturation of fermion zero modes charged under the gauge groups carried by the D-branes \cite{Ganor:1996pe}, coming from the open sector between gauge D-branes and D-brane instantons. 

A particularly interesting setup in which they can be studied is systems of D3-branes at toric Calabi-Yau (CY) 3-fold singularities, which are described in terms of {\it dimer diagrams}, also known as {\it brane tilings} \cite{Hanany:2005ve,Franco:2005rj,Feng:2005gw} (see \cite{Kennaway:2007tq} for a review). These are bipartite tilings of a 2-torus, namely graphs whose nodes can be colored black and white, such that white nodes are connected only to black nodes and vice-versa. These diagrams also allow to describe D-brane instantons and easily read out the charged fermion zero modes and their couplings, either directly on the dimer diagram or in the mirror picture. In the latter, the configuration corresponds to a set of intersecting D6-branes and D2-brane instantons, eventually encoded in a set of 1-cycles in the mirror punctured Riemann surface. In a recent development in this framework, \cite{Tenreiro:2017fon} showed that in this mirror picture the generation of non-perturbative charged field theory operators can be obtained as a perturbative coupling in a modified geometry, triggered by the backreaction of the D-brane instanton on the mirror CY. In the resulting configuration, the D2-brane instanton is geometrized, along the lines of \cite{Koerber:2007xk,Koerber:2008sx,Garcia-Valdecasas:2016voz}, and we are left with a set of recombined D6-branes in the modified geometry. 

In this paper we show that the resulting gauge theory (and thus the charged field theory operators) can be encoded in a Bipartite Field Theory, albeit in general not defined on a 2-torus (as the original dimer diagram) but on a general (possibly higher-genus) Riemann surface, thus of the kind introduced in \cite{Franco:2012mm}. This resulting BFT is related to the original one by a simple operation, which can be regarded as the direct backreaction of the D-brane instanton on the gauge theory. For the simplest case of a D-brane instanton located on a face of the original dimer diagram, it essentially corresponds to the removal of the face and its edges, and the recombination of nodes of the same color. In general, avoiding crossing of edges requires the introduction of handles, so that the new BFT is in general defined in higher genus. 

Considered abstractly, this operation can be carried out also by taking a general BFT as starting point. From this perspective, a main result of the present paper is the definition of a new operation on BFTs, relating theories defined on Riemann surfaces of different genus. This is thus a particularly interesting new insight in the field of BFTs.

BFTs are $4d$ $\mathcal{N}=1$ supersymmetric gauge theories whose Lagrangians are defined by bipartite graphs embedded into a Riemann surface, possibly with boundaries \cite{Franco:2012mm}.\footnote{Closely related theories were introduced in \cite{Xie:2012mr} and studied further in \cite{Heckman:2012jh}.} The special subclass of BFTs defined on a torus without boundaries are the brane tilings or dimer diagrams mentioned before, which describe the worldvolume theory of D3-branes probing toric CY$_3$ singularities \cite{Hanany:2005ve,Franco:2005rj,Franco:2005sm,Kennaway:2007tq}. Extensive catalogues of explicit BFT examples have been provided in e.g. \cite{Franco:2012mm,Franco:2012wv} for general BFTs and \cite{Hanany:2012vc,Cremonesi:2013aba,He:2014jva} for higher genus examples without boundaries.

General BFTs and their associated graphs have received several physical interpretations, in particular in connection with the reformulation of $4d$ $\mathcal{N}=4$ SYM in terms of {\it on-shell diagrams} \cite{ArkaniHamed:2012nw}. The new approach makes all symmetries of the theory manifest and sheds new light on previous results \cite{ArkaniHamed:2012nw,ArkaniHamed:2010kv,Britto:2004ap,Britto:2005fq}. The connection of on-shell diagrams with bipartite graphs and BFTs has been extensively studied in  \cite{Franco:2012mm,Franco:2012wv,Franco:2013ana,Franco:2013pg,Franco:2013nwa,Franco:2014nca}.

On the other hand, there has been no direct realization of BFTs in string theory beyond the very restricted subclass of theories associated to graphs with vanishing curvature \cite{Heckman:2012jh,Franco:2013ana}. Our work provides precisely that link, regarding higher genus BFTs as the result of D-brane instanton backreactions in lower genus theories. We expect that the new operation we have obtained linking BFTs in different genus has interesting implications both for the further study of general BFTs in string theory, and for the complementary physical realizations of the corresponding graphs.

Our work takes first steps in this direction, for instance by computing the toric CYs associated to the new BFTs and establishing how they are connected to the original ones. We expect this to be a very useful tool towards a general dictionary, and the study of dual theories, the inverse problem, etc. 

\medskip

The paper is organized as follows: In \sref{sec:review} we review the description of systems of D3-branes at singularities and their D-brane instantons in terms of dimers (\sref{sec:dimer-instanton-review}), and the backreaction description of the latter in the mirror geometry (\sref{sec:backreaction-mirror}). In \sref{sec:dimer-backreaction} we derive the description of the D-brane instanton backreaction in the dimer, and its properties. In \sref{sec:dimer-backreaction-general} we prove that in general it leads to a higher genus BFT, and provide illustrative examples in \sref{sec:example-pdp2} and \sref{sec:example-pdp4}. In \sref{sec:dimer-backreaction-special} we discuss instances in which the backreaction does not result in an increase of the genus, which correspond to dimers where the global $\mathbb{T}^2$ topology implies certain identifications among faces. In \sref{section_extension} we apply the combinatorial recipe of dimers to general BFTs, thus defining a new operation relating BFTs in different genus Riemann surfaces. In \sref{section_BFT_genus_backreaction} we introduce a useful graphical depiction of the handle attachment surgery which simplifies the discussion of the genus increase.
In \sref{section_toric_geometry_backreacted_dimers} we describe the computation of the new toric data corresponding to the backreacted BFTs. The change in perfect matchings between the original and final theories is discussed in \sref{section_removed_perfect_matchings}, yielding the construction of the new toric diagram in \sref{section_new-toric-diagram}. A direct construction based on relating new toric coordinates with the bridges identifying formerly different nodes of the original theory is provided in \sref{section_coordinates-bridges}. These concepts are illustrated in a detailed example in \sref{section_dp3}. In \sref{section_seiberg-duality} we describe instances of the interplay of instanton backreaction and Seiberg duality: in \sref{section_seiberg_same}, when they are applied to the same dimer face,  and in  \sref{section_seiberg_neighbor} when applied to neighboring faces. In \sref{section_multi-complex} we consider the generalization to backreaction of multi-instantons, focusing on cases corresponding to complex deformations of the original CY 3-fold, and recover earlier results in the literature. In \sref{section_inverse} we show the non-uniqueness of the inverse problem of reconstructing initial theories for a given final one. We present our concluding remarks in \sref{section_conclusions}.

\section{Review of D-brane Instanton Backreaction on D-Branes at Singularities}
\label{sec:review}

In this section we review some background material on dimer diagrams as tools to describe systems of D3-branes at toric singularities. Subsequently we review the backreaction of instantons in the mirror of these systems, to lay the ground for their novel discussion in the dimer diagram, and the connection with general BFTs.

\subsection{Dimers and Instantons}
\label{sec:dimer-instanton-review}

\subsubsection{Overview of Dimers}
\label{sec:dimer-review}

The gauge theory on Type IIB D3-branes probing toric CY  3-fold singularities is given by a set of unitary gauge factors, bifundamental or adjoint chiral multiplets, and a superpotential. Much information on these gauge theories, and properties of the underlying D-branes, can be encoded in a brane tiling or dimer diagram,\footnote{The description of $4d$ $\mathcal{N}=1$ gauge theories in terms of tilings is complementary to that of quiver diagrams, in which gauge groups are represented by nodes, and chiral multiplets by arrows. However, brane tilings also encode the superpotentials, and thus facilitate a deeper understanding of these theories.} see e.g. \cite{Hanany:2005ve,Franco:2005rj}, and \cite{Kennaway:2007tq} for a review. A dimer diagram is a tiling of $\mathbb{T}^2$ defined by a bipartite graph. Faces in the dimer correspond to gauge factors in the field theory, edges describe bifundamental fields, and nodes represent superpotential terms. The bipartite character of the graph underlies the assignment of chirality for the bifundamental matter in terms of the edge orientation, e.g. clockwise and counterclockwise around black and white nodes, respectively. The node colors also determine the signs of the corresponding superpotential terms. Several well-known theories are described in the examples later on.

For future convenience we emphasize that these theories are easily generalized to Bipartite Field Theories (BFTs). These are $4d$ $\mathcal{N}=1$ supersymmetric gauge theories whose Lagrangians are defined by bipartite graphs
embedded into a Riemann surface, possibly with boundaries \cite{Franco:2012mm}.
For the purposes of this section, however, we stick to BFTs defined on $\mathbb{T}^2$. Throughout the paper, we restrict the meaning of the term ``dimer diagram" to such theories. Additional ingredients and extensions for general BFTs will arise at different points of the paper.

\begin{figure}[!ht]
\begin{center}
\includegraphics[width=9.5cm]{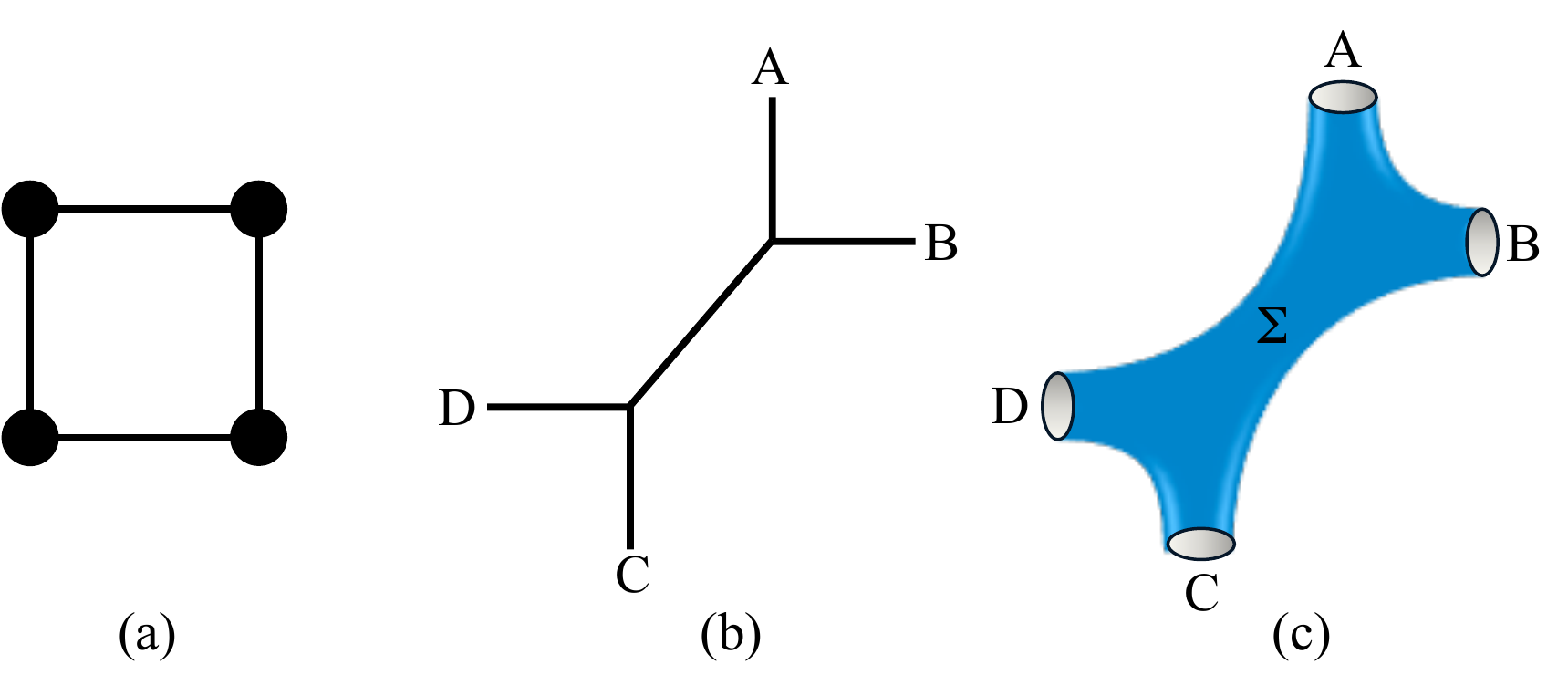}
\caption{a) Toric diagram, b) $(p,q)$-web and c) mirror Riemann surface $\Sigma$ for the conifold. External legs of the web map to punctures in $\Sigma$.}
\label{toric_web_Riemann}
\end{center}
\end{figure}

For D3-brane systems, the type IIA mirror configuration can be constructed in terms of combinatorics of the dimer diagram, as described in \cite{Feng:2005gw}. The mirror corresponds to a double fibration over the complex plane, with fibers given by a $\IC^*$ and a genus-$g$ Riemann surface $\Sigma$. This is a smooth punctured Riemann surface which can be thought of as a thickening of the $(p,q)$-web diagram \cite{Aharony:1997ju,Aharony:1997bh,Leung:1997tw} dual to the toric  diagram, see \fref{toric_web_Riemann}. 

The information about the gauge theory is encoded in a set of 1-cycles on $\Sigma$ (which are part of the 3-cycles wrapped by the D6-branes in the mirror picture). Each 1-cycle corresponds to a gauge factor, and their intersections support bifundamental chiral multiplets associated to the edges in the dimer. Oriented disks suspended among intersections provide worldsheet instantons producing the superpotential terms of the dimer nodes. 

\begin{figure}[!ht]
\begin{center}
\includegraphics[width=13cm]{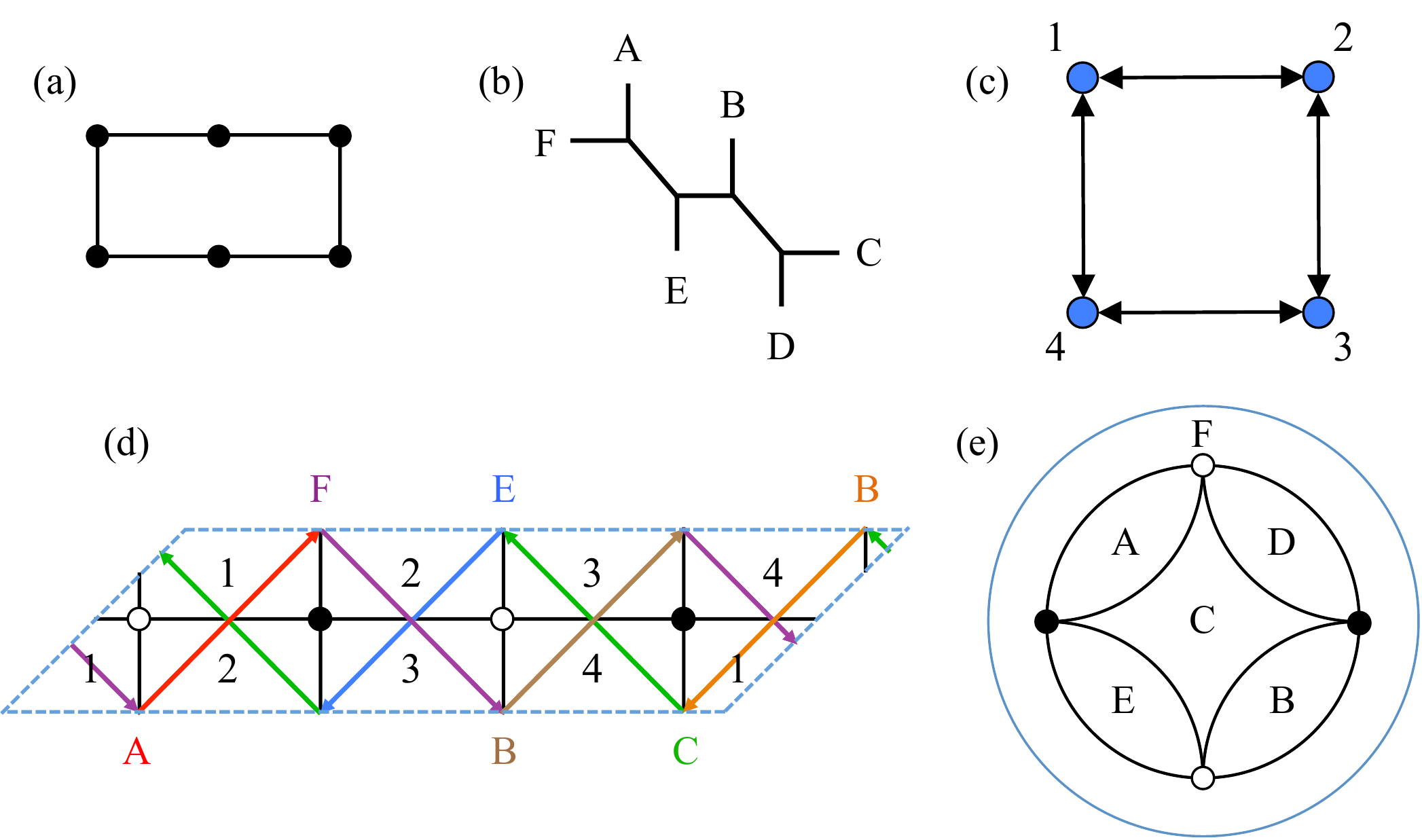}
\caption{Various diagrams for the double conifold. a) Toric diagram. b) $(p,q)$-web. c) Quiver. d) Dimer with zig-zag paths. The blue dashed parallelogram indicates the unit cell. Its opposite sides are identified to form a $\mathbb{T}^2$. e) The mirror Riemann surface $\Sigma$ is a sphere with 6 punctures. Here we represent it as the complex plane with the point at infinity, indicated by the blue circle, added.}
\label{double_conifold_diagrams}
\end{center}
\end{figure}

The Riemann surface $\Sigma$ and these 1-cycles can be systematically obtained from the dimer that defines the gauge theory as follows. Given a dimer diagram, we introduce the so-called zig-zag paths \cite{Hanany:2005ss}, as paths composed of edges that turn maximally to the right at e.g. black nodes and maximally to the left at white nodes. They can be conveniently depicted as oriented lines that cross once at each edge and turn at each vertex. Notice that the two zig-zag paths that intersect every edge must have opposite orientations.  As shown in \cite{Feng:2005gw}, the zig-zag paths of the dimer diagram associated to D3-branes at a singularity lead, by an untwisting procedure, to a tiling of the Riemann surface $\Sigma$ in the mirror geometry. Specifically, each zig-zag path encloses a face of the tiling of $\Sigma$ which includes a puncture, and the $(p,q)$ charge of the associated leg in the web diagram is the $(p,q)$  homology charge of the zig-zag path in $\mathbb{T}^2$. The Riemann surface $\Sigma$ can be regarded as a thickening of this web diagram into a genus $g$ surface. \fref{double_conifold_diagrams} illustrates all these objects in an explicit example, a non-chiral $\mathbb{Z}_2$ orbifold of the conifold, also known as the double conifold.

By construction, the 1-cycles in the mirror Riemann surface associated to the different gauge factors are given by zig-zag paths of the tiling of $\Sigma$. This description allows to easily classify supersymmetric wrapped branes in toric singularities and their mirrors. In fact, they can be used to describe gauge D-branes (i.e. D-branes spanning the $4d$ Minkowski directions) or D-brane instantons (i.e. Euclidean D-branes localized in the $4d$ dimensions), as extensively exploited in the next sections.

\subsubsection{D-brane Instantons on Dimers}

As just described, D-brane instantons in systems of D3-branes at toric CY$_3$ singularities can be described as D-branes wrapped on faces of dimer diagrams. As explained later, the dimer combinatorics allows an easy description of these instantons and some features of the non-perturbative field theory operators they produce. We should note that in general, such D-brane instantons do not generate $4d$ superpotential terms, due to the existence of additional neutral fermion zero modes. Although these can be subsequently removed by further ingredients (orientifolds, fluxes, etc), our main goal in the present paper is to understand the breaking of $U(1)$ symmetries by the appearance of charged $4d$ fields in the non-perturbative instanton operator, even if the latter is not a superpotential term. Hence, we focus on the pattern of instanton fermion zero modes charged under the $4d$ gauge group, independently of any additional bosonic or neutral fermionic zero modes.

The realization of instantons in the dimer makes it easy to read out the content of fermion zero modes charged under the $4d$ gauge group, from edges between the instanton faces and other gauge faces. Although the general discussion is straightforward, for concreteness we focus on the case of a single instanton on a single dimer face. The pattern of breaking of $U(1)$ symmetries by the instanton follows from the pattern of bifundamentals defined by the edges around the instanton face. Also, the couplings of these fermion zero modes to $4d$ bifundamental fields is given by the superpotential couplings (see e.g. \cite{Franco:2007ii}, as a particular case of \cite{Kachru:2008wt}).

In the mirror, the instanton face maps to a 1-cycle on $\Sigma$, which corresponds to a zig-zag path of its tiling. The charged fermion zero modes are supported at the intersections between this 1-cycle and those of the $4d$ gauge D-branes. Thus, the breaking of $U(1)$ symmetries is determined by the intersection numbers of the instanton cycle, as shown in \cite{Blumenhagen:2006xt,Ibanez:2006da,Florea:2006si,Blumenhagen:2009qh}. Also, their couplings with $4d$ bifundamental matter are determined by the corresponding worldsheet instanton disks.

The computation of the $4d$ charged field theory operator by saturation of charged fermion zero modes is also simplified by the bipartite character of the dimer. For simplicity, we focus on the abelian case, in which all relevant gauge factors correspond just to $U(1)$'s. Most of the discussion extends to the non-abelian case, with the proviso of taking determinants of certain combination of fields. It is easy to realize that there are two instanton-induced $4d$ field theory operators, arising from exactly two ways of saturating fermion zero modes, obtained by taking the couplings corresponding to either all the white or all the black nodes, and ``forgetting'' the fermion zero modes. This procedure has a more direct physical interpretation in terms of the instanton backreaction on the mirror geometry, as we review in the next section. In the mirror, black and white nodes fall on different sides of the instanton 1-cycle, so the two $4d$ charged operators induced by the instanton are read out from the disks at either side of the instanton 1-cycle.

\subsection{D-Brane Instanton Backreaction in the Mirror}
\label{sec:backreaction-mirror}

As proposed in \cite{Garcia-Valdecasas:2016voz} building on \cite{Koerber:2007xk,Koerber:2008sx}, D-brane instanton effects can be described in terms of a backreacted geometry. In the context of D-branes at singularities, this description was achieved in terms of the mirror setup with both gauge D6-branes and D2-brane instantons in \cite{Tenreiro:2017fon}. In this section we review the description of the instanton backreaction in terms of simple graph operations in the mirror geometry. The direct description in terms of the original systems of D-branes at singularities was not provided in \cite{Tenreiro:2017fon}, and is in fact one of the main points of the present paper.

In the mirror, backreaction is captured by the recombination of the D6-branes intersecting the instanton and the appearance of the non-perturbative D-brane instanton superpotential in terms of purely worldsheet instanton effects in the backreacted geometry. The recombination of the D6-branes provides a direct physical realization of the breaking of $U(1)$ symmetries by the instanton. We refer the reader to \cite{Tenreiro:2017fon} for further details.

As mentioned above, we restrict the discussion to the case of a single instanton, namely the D2-brane on a 3-cycle $\Pi_3$ associated to a single 1-cycle in the mirror Riemann surface. Also, note that, although the combinatorial recipe does not care about the ranks of the gauge factors, the physical process as described below corresponds to the abelian case. With this condition, the backreacted configuration can be generated with very simple steps on this graph:

\begin{itemize}

\item {\bf Step 1. Cut:} The instanton 3-cycle $\Pi_3$ should disappear from the geometry, so we cut $\Sigma$ by removing a small strip around the instanton 1-cycle, and seal off the two resulting boundaries in $\Sigma$ by identifying each of them to a point. Any 3-cycle formerly intersecting $\Pi_3$ turns into a 3-chain in the backreacted geometry, so correspondingly any 1-cycle intersecting the cut is split, and turns into a chain with boundary points.  These 1-chains will be glued in the next step.

\item {\bf Step 2. Recombine:} 
The bipartite character of the graphs implies that the instanton 1-cycle in $\Sigma$ has an equal number of positive and negative orientation intersections with the other 1-cycles. Thus, on each side of the cut there is an equal number of incoming and outgoing 1-chains, which we must recombine to form 1-cycles. The recombination should be carried out without crossing edges of the underlying tiling of $\Sigma$, which becomes possible due to the bipartite nature of the graph.

\item {\bf Step 3. Field theory operators:} The previous two steps already define the backreacted geometry. This last step merely establishes that the $4d$ non-perturbative field theory operator induced by the original instanton arises as a worldsheet instanton on the backreacted geometry, bounded by the recombined D-branes. Such world sheet instantons are easily identified by considering disks bounded by recombined 1-cycles (and the cut, which recall is regarded as shrunk to a point). There are always two such couplings, which nicely agrees with the fact that the saturation of charged fermion zero modes in the open string picture can always occur in two ways, as a consequence of the dimer being bipartite.

\end{itemize}

\begin{figure}[!ht]
\begin{center}
\includegraphics[width=8cm]{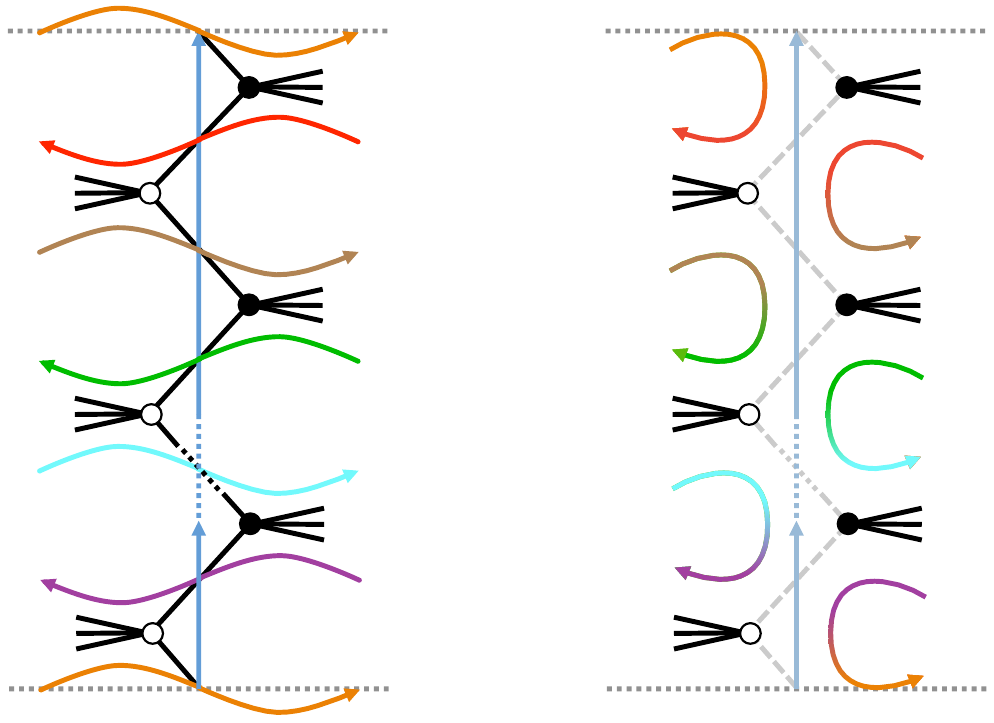}
\caption{{\bf Left:} Local piece of the mirror Riemann surface around the 1-cycle wrapped by a D2-brane instanton (blue), consisting of $2k$ edges. The top and bottom dotted lines are identified. The collections of half edges sticking out of each node represent general superpotential couplings. The orange, brown and cyan, and the red, green and purple arrowed lines are pieces of additional zig-zag paths that correspond to  D6-branes intersecting the D2-brane instanton, with positive or negative intersection numbers, respectively.
{\bf Right:} The D2-brane instanton has backreacted, so the blue line has disappeared. D6-brane paths have been cut at their intersection with the former D2-brane instanton cycle, and recombined. The black nodes and the white nodes should be regarded as recombined into a single black and a single white node, respectively.}
\label{mirror_general}
\end{center}
\end{figure}

The general structure of the process is shown in \fref{mirror_general} for an instanton on a $2k$-sided face of the dimer.

\section{D-brane Instanton Backreaction on the Dimer}
\label{sec:dimer-backreaction}

In this section we present a main result of this paper. We show that the description of the instanton backreaction can be carried out directly on the dimer diagram, and that it generically turns it into a general BFT. This motivates the definition of a general combinatoric operation, which extends automatically to the whole class of BFTs, and which relates BFTs on different Riemann surfaces.

\subsection{General Idea}
\label{sec:dimer-backreaction-general}

The procedure in  \sref{sec:backreaction-mirror} to describe the instanton backreaction in the mirror configuration, turns a system of intersecting D6-branes into a different one, which is nevertheless still described in terms of a bipartite graph tiling of the backreacted Riemann surface. The remarkable fact that we obtain a bipartite structure implies that we can reconstruct faces, edges and nodes of a BFT describing the resulting set of D6-branes, including the instanton superpotential. 

The fact that in general this corresponds not to a 2-torus dimer diagram but rather to a generically different genus BFT is easily derived. Using the recipe for the generic case summarized by \fref{mirror_general}, the change of genus in the corresponding gauge theory is as follows. First, the number of edges is reduced by $\Delta E=-2k$; the number of vertices $V$ changes, since each set of $k$ black/white nodes turns into a single black/white node, hence $\Delta V=-2k+2$; finally, the number of faces in the BFT is determined by the disappearance of the instanton 1-cycle in the mirror, and the recombination of the $2k$ D6-brane 1-cycles into a single one (see \sref{sec:dimer-backreaction-special} for other possibilities in non-generic cases), resulting in $\Delta F=-2k$. Since the Euler formula gives
\beqa
F+V-E=2-2g \, ,
\label{Euler_formula}
\eeqa
we have a change in the BFT genus
\beqa
\Delta g= \frac 12\, (\, \Delta E - \Delta F - \Delta V\,)\, =\, k-1 \, .
\label{Delta_g}
\eeqa

The above analysis exploits the fact that the gauge theory resulting after the transformation in the mirror discussed in \sref{sec:backreaction-mirror} still corresponds to a bipartite graph. In fact, it is easy to check that there is a simple operation that can be carried out in the dimer and which reproduces the different steps in the mirror, and that yields a BFT on a Riemann surface of the appropriate genus. 

These steps are:

\begin{itemize}

\item {\bf Step 1. Remove:} Remove the face corresponding to the D-brane instanton, and its edges, leaving the adjacent faces open. This reproduces the operation of removing the instanton mirror 1-cycle and cutting the 1-cycles intersecting it.

\item {\bf Step 2. Fuse:} 
Declare that all black nodes of the former instanton face are identified into a single black node, and similarly for all white nodes. The edges ending on the initial nodes remain as edges ending on the final node, and their ordering is preserved (this can be done by performing the identification of nodes sequentially according to their ordering as one circles the original face).
In order to avoid edge crossings after the identifications of nodes, it is in general necessary to introduce $k$ handles for an original instanton face of  $2k$ sides. This provides the required increase in the genus of the resulting BFT. This step closes off the former open faces, in general into a single recombined one.

\item {\bf Step 3. Field theory operators:} The above two steps already define the backreacted BFT. This last step merely establishes that the $4d$ non-perturbative field theory operators induced by the original D-brane instanton are simply the superpotential terms corresponding to the combined black and white nodes.

\end{itemize}

\begin{figure}[!ht]
\begin{center}
\includegraphics[width=14cm]{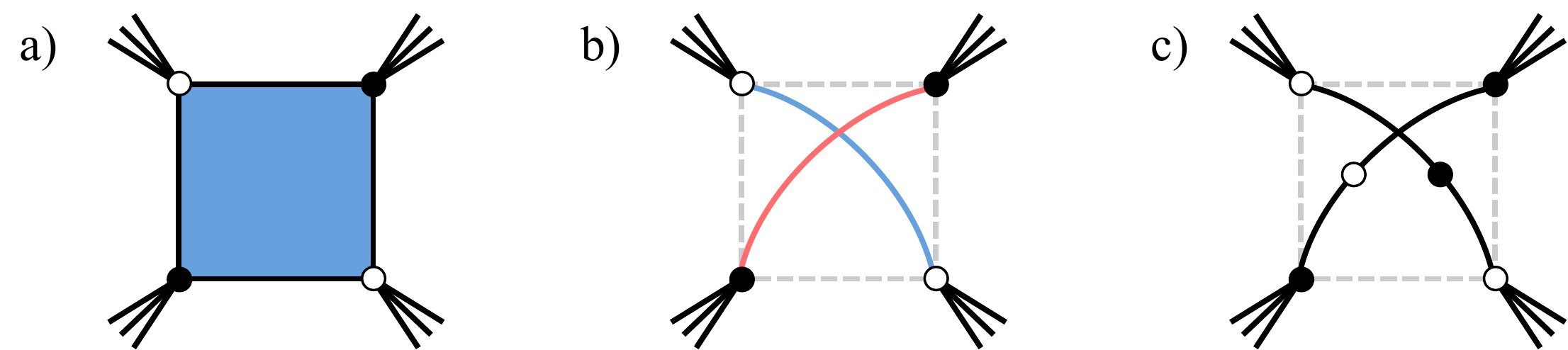}
\caption{a) Local piece of the BFT surface around a face wrapped by the D-brane instanton (blue). The collections of half edges sticking out of each node indicate general superpotential couplings. b) After backreaction, the blue face and its edges disappear. The black and white nodes are recombined into single black and white nodes, respectively; this is indicated by bridges. The crossing of bridges can be avoided by embedding the BFT in a surface with an additional handle, in agreement with $\Delta g=1$. Generically, the faces around the original instanton recombine into a single one. c) An alternative representation of the backreacted BFT, in which bridges are replaced by pairs of actual edges, joined by 2-valent nodes corresponding to superpotential mass terms.}
\label{dimer-generic-4side}
\end{center}
\end{figure}

\begin{figure}[!ht]
\begin{center}
\includegraphics[width=14cm]{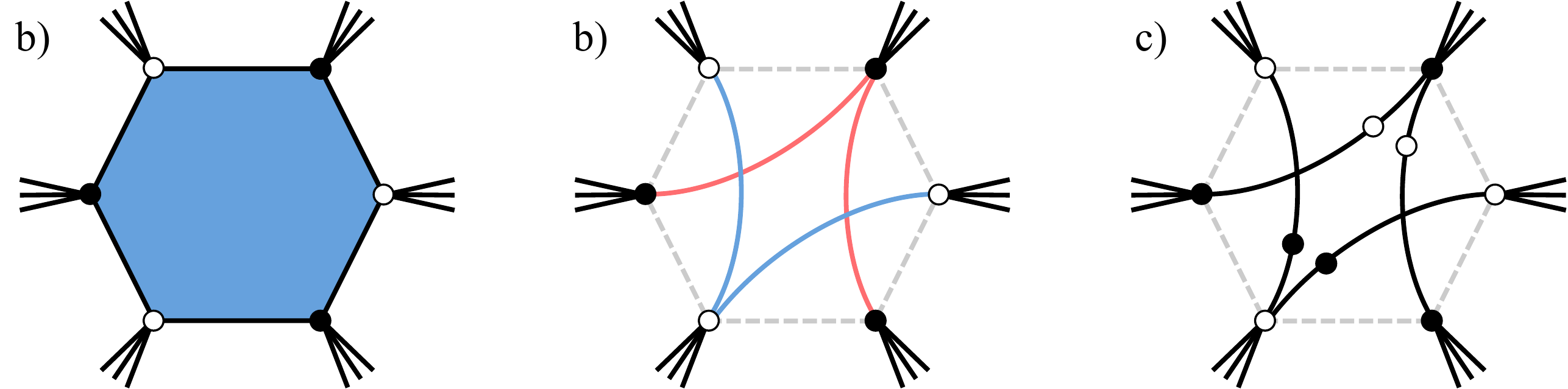}
\caption{Backreaction of a local piece of a BFT surface around a D-brane instanton on a 6-edge face. Similar remarks to \fref{dimer-generic-4side} apply, with the difference that $\Delta g=2$ in this case.
}
\label{dimer-generic-6side}
\end{center}
\end{figure}

Figures \ref{dimer-generic-4side} and \ref{dimer-generic-6side} illustrate this operation in the case of instantons on 4- and 6-sided faces. The recombination of black and white nodes is indicated by blue (for white nodes) and red (for black nodes) {\it bridges}. Whenever a bridge connects two nodes of the same color, we understand that there is a 2-valent node of the opposite color in the middle of it, making the graph bipartite. In other words, bridges correspond to massive pairs of chiral fields, which lead to the desired identifications of nodes if they are integrated out. For clarity, we leave such intermediate nodes implicit in the figures that follow.

Clearly, although we have introduced this operation for bipartite graphs defined on 2-tori, the procedure extends to general BFTs, thus defining an operation relating BFTs on Riemann surfaces generically of different genus. In the generic case (see \sref{sec:dimer-backreaction-special} for non-generic situations) the operation corresponds to a local surgery, whose characterization we study in more detail in \sref{section_BFT_genus_backreaction}. The application of the operation in general BFTs is discussed in \sref{section_extension}.

\begin{figure}[!ht]
\begin{center}
\includegraphics[width=10cm]{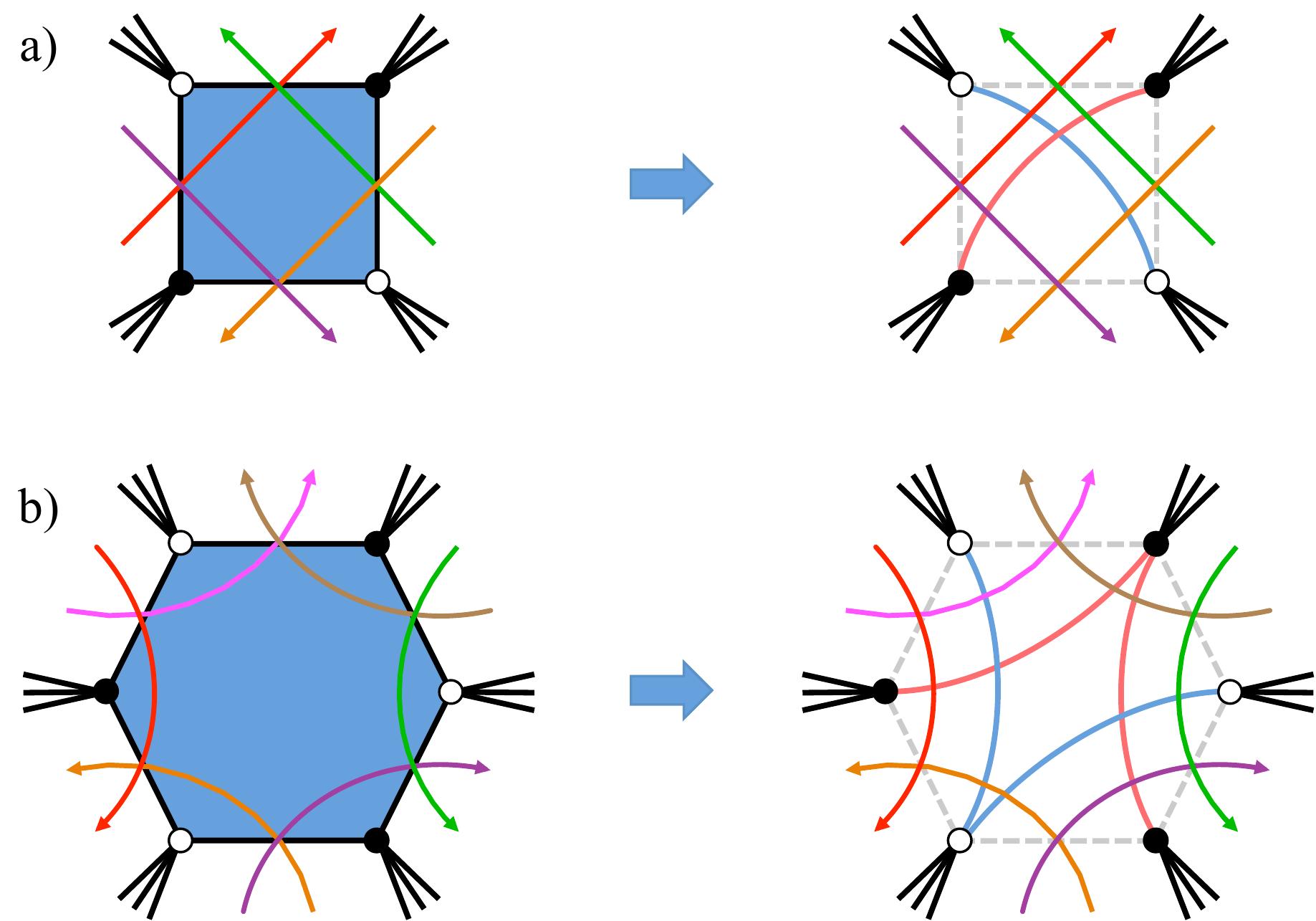}
\caption{Structure of zig-zag paths around faces in a dimer, and in its backreacted version. Zig-zag paths maintain their structure, in agreement with the fact that punctures in the mirror Riemann surface are unchanged by the backreaction process.}
\label{backreacted_zz_square_hexagon}
\end{center}
\end{figure}

It is easy to check that the above transformation rule on the dimer preserves the structure of zig-zag paths, see \fref{backreacted_zz_square_hexagon}. This dovetails the fact that the instanton backreaction in the mirror preserves the punctures of the Riemann surface. 

\subsection{Examples}

Below we present two examples illustrating the ideas introduced in the previous section.

\subsubsection{A $PdP_2$ Example}

\label{sec:example-pdp2}

\begin{figure}[!ht]
\begin{center}
\includegraphics[width=14cm]{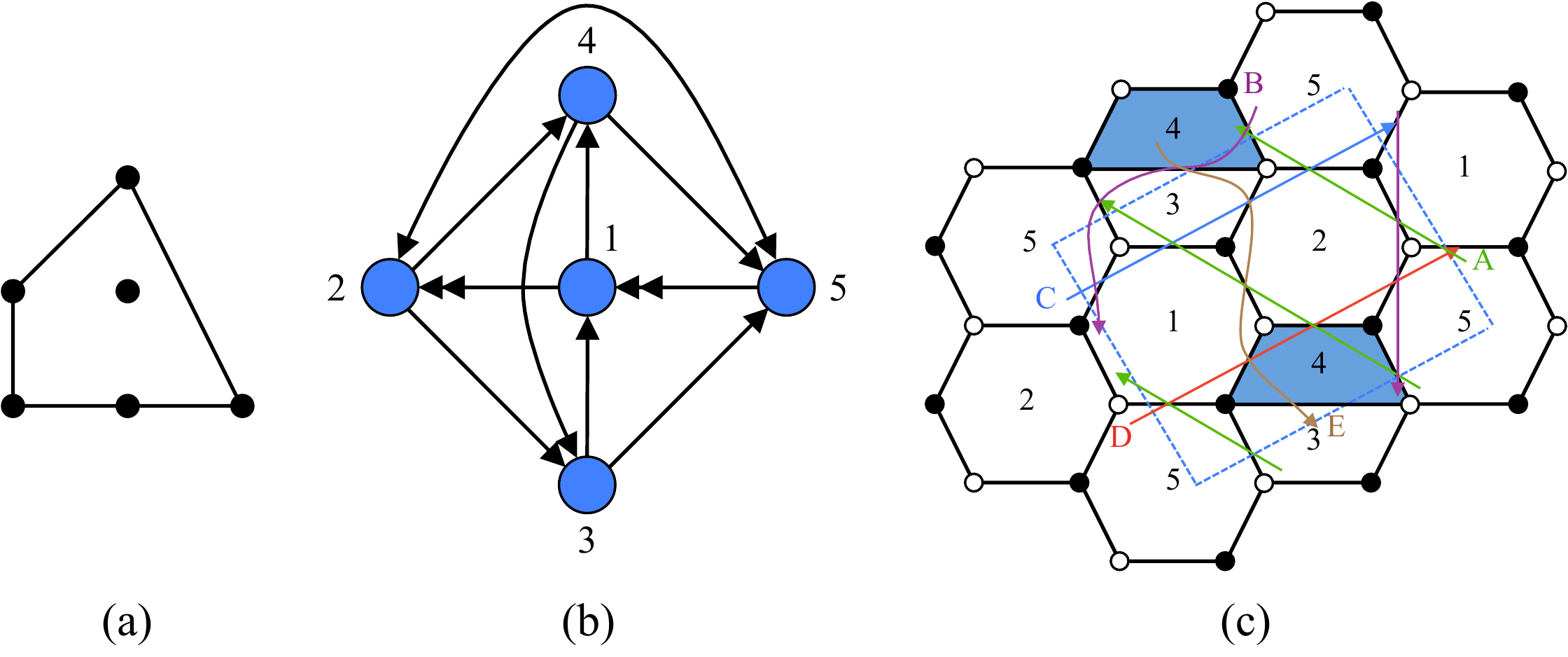}
\caption{Diagrams for $PdP_2$: a) toric diagram, b) quiver and c) dimer with zig-zag paths.}
\label{diagrams_PdP2}
\end{center}
\end{figure}

Let us consider the pseudo del Pezzo 2 ($PdP_2$) theory. By this we mean the theory obtained by placing D3-branes at the tip of a complex cone over the $PdP_2$ surface. In the examples that follow, we will often use this abbreviated way of referring to the full CY$_3$ and the corresponding gauge theory. This geometry, which corresponds to a blowup of $\mathbb{CP}^2$ at two non-generic points, was originally studied in \cite{Feng:2004uq}, where it was determined that it has a single toric phase. \fref{diagrams_PdP2} shows the toric diagram for $PdP_2$, the corresponding quiver and the dimer with zig-zag paths. 

\begin{figure}[!ht]
\begin{center}
\includegraphics[width=10cm]{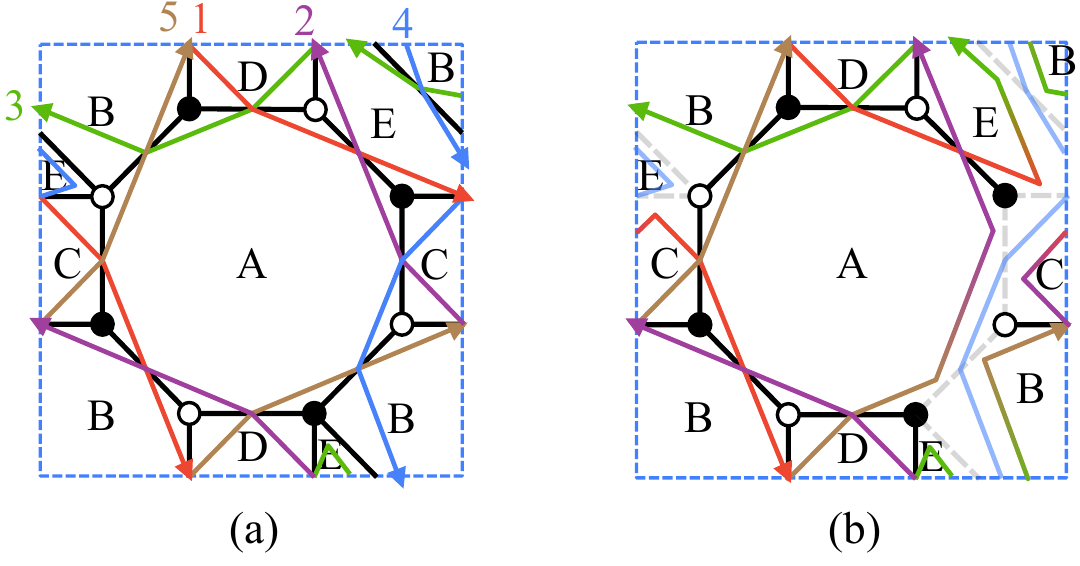}
\caption{a)  Mirror surface for $PdP_2$. b) Backreaction of an instanton on face 4 of the dimer, which maps to the blue zig-zag path. All zig-zags recombine into a single one. \comment{\textcolor{blue}{Redo this figure without the zig-zag labels.}}}
\label{mirror_PdP2}
\end{center}
\end{figure}

\begin{figure}[!ht]
\begin{center}
\includegraphics[width=11cm]{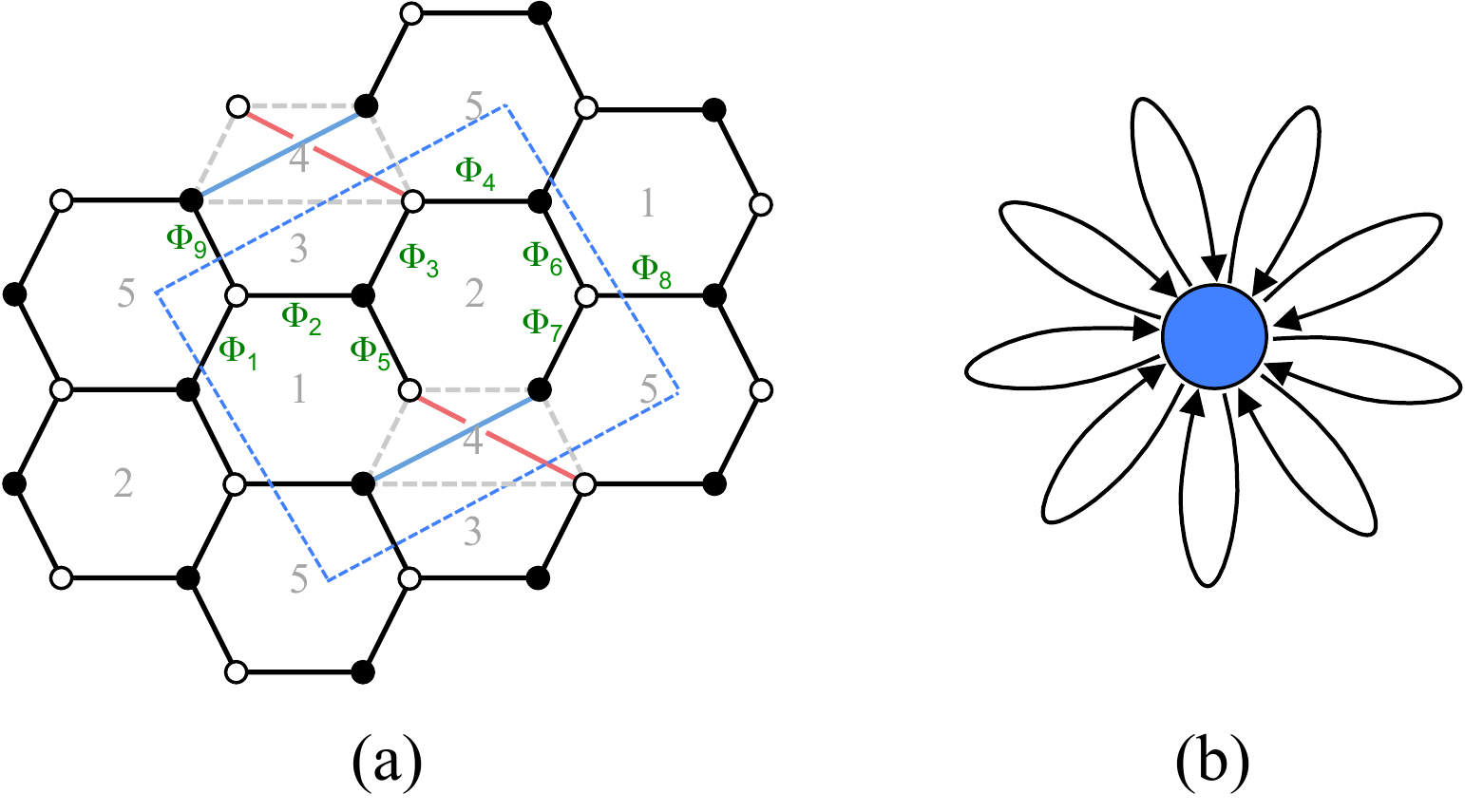}
\caption{a) Backreaction of an instanton on face 4 of the $PdP_2$ dimer and b) quiver for the resulting BFT.}
\label{dimer_quiver_backreaction_PdP2}
\end{center}
\end{figure}

The mirror surface is presented in \fref{mirror_PdP2}.a.  Consider introducing a D-brane instanton on face 4 of the dimer, which corresponds to the zig-zag path 4 on the mirror Riemann surface. Let us first perform the backreaction in the mirror, following the prescription reviewed in \sref{sec:backreaction-mirror}. The result is shown in \fref{mirror_PdP2}.b. The final theory is described by the recombined 1-cycles, their intersections and worldsheet instanton disks. Interestingly, all the zig-zags in the mirror fuse into a single one, i.e. the corresponding BFT has a single face. We obtain a BFT with $F=1$, $V=6$ and $E=9$, i.e. with 1 gauge group 6 superpotential terms and 9 chiral fields. From the Euler formula \eref{Euler_formula}, we conclude that it is a BFT defined on a genus-2 surface. Let us describe the final gauge theory explicitly. The 9 chiral fields transform in the adjoint representation of the single gauge group, as shown in the quiver in \fref{dimer_quiver_backreaction_PdP2}.b. Below we will discuss the superpotential in further detail. 

Let us now see how the same theory is recovered by implementing backreaction directly on the dimer, as described in section \ref{sec:dimer-backreaction-general}. Applying this recipe to an instanton on face 4, the edges of the instanton face disappear and the nodes of the same color are identified. The resulting diagram is shown in \fref{dimer_quiver_backreaction_PdP2}.a, where we see that a new handle needs to be introduced for the identification, nicely reproducing the expected genus-2 result. The original face numbers are shown in grey for reference. Integrating out the massive edges associated to the bridges, we obtain a BFT with $V=6$ and $E=9$, which combined with the Euler formula imply that $F=1$. We thus replicate the result of the mirror. As mentioned earlier, the theory has a single gauge group and 9 adjoint chirals as shown in \fref{dimer_quiver_backreaction_PdP2}.b. The superpotential contains 6 terms, which can be read from \fref{dimer_quiver_backreaction_PdP2}.a, where we have labeled the chiral fields associated with the edges, and is given by
\beq
W =  \Phi_1 \Phi_2 \Phi_9 + \Phi_3 \Phi_4 \Phi_5 + \Phi_6 \Phi_7 \Phi_8 - \Phi_2 \Phi_3 \Phi_5 - \Phi_1 \Phi_6 \Phi_4 - \Phi_7 \Phi_9 \Phi_8 \, .
\label{W_backreacted_PdP2}
\eeq

\subsubsection{A $PdP_4$ Example}

\label{sec:example-pdp4}

\begin{figure}[H]
\begin{center}
\includegraphics[width=12cm]{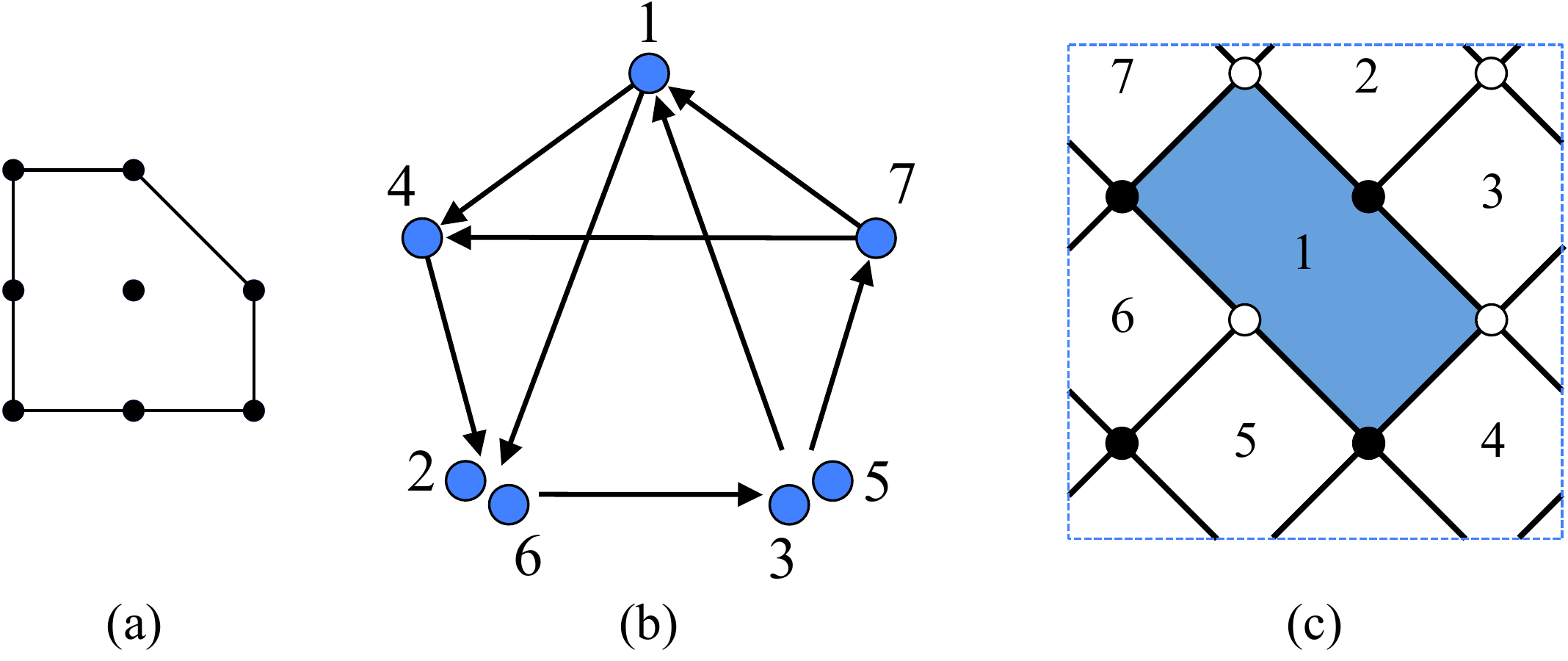}
\caption{a) Toric diagram for $PdP_4$. b) Quiver and c) dimer for its phase 1.}
\label{Fig:PdP4Diagrams}
\end{center}
\end{figure}

We now focus on $PdP_4$, which is a blowup of $\mathbb{CP}^2$ at four non-generic points. This geometry was first considered in \cite{Feng:2002fv}, where it was established that it has three toric phases. \fref{Fig:PdP4Diagrams} shows the toric diagram for $PdP_4$, and the quiver and dimer for its phase 1, in the classification of \cite{Feng:2002fv}.

\begin{figure}[!ht]
\begin{center}
\includegraphics[width=8cm]{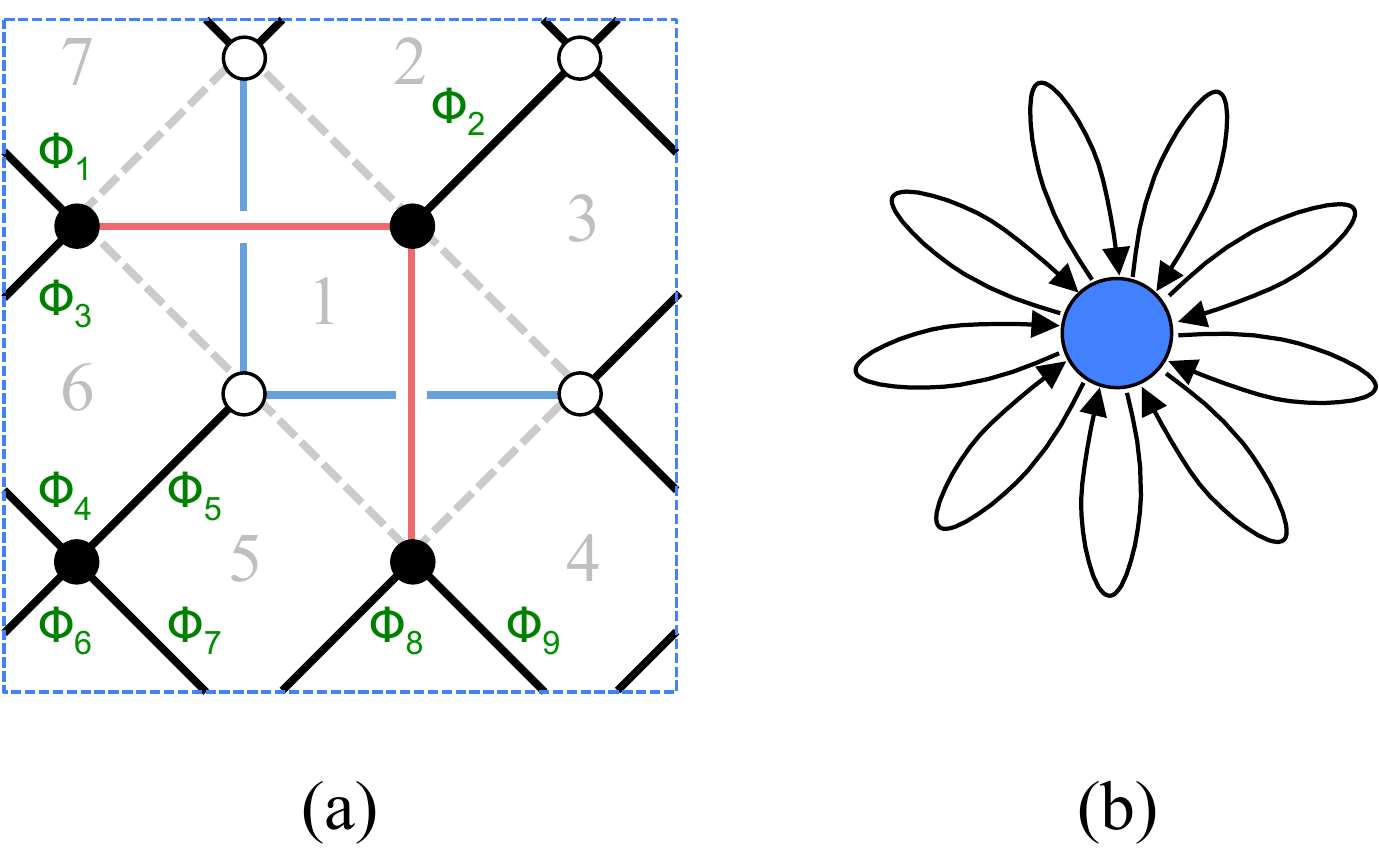}
\caption{a) Backreaction of an instanton on face 1 of the dimer for phase 1 of $PdP_4$ and b) quiver for the resulting BFT.}
\label{PdP4Back}
\end{center}
\end{figure}

Let us consider an instanton on the hexagonal face 1 of the dimer, indicated in blue in \fref{Fig:PdP4Diagrams}. Its backreaction is shown in \fref{PdP4Back}.a. Note that it is necessary to add two handles, so the resulting BFT is in genus 3. This theory has $F=1$, $V=4$ and $E=9$. The gauge theory has 1 gauge group with 9 adjoint chiral fields, as illustrated in the quiver in \fref{PdP4Back}.b. While this is the same quiver that we obtained for the example in the previous section, shown in \fref{dimer_quiver_backreaction_PdP2}.b, we know that the two BFTs are fundamentally different; in particular the first theory has $g=2$ and the second one has $g=3$. The distinction between both theories comes from the superpotential. Instead of the 6 cubic terms of \eref{W_backreacted_PdP2}, the superpotential of the new theory is given by
\beq
W =  \Phi_5 \Phi_4 \Phi_3 \Phi_8 \Phi_7+ \Phi_2 \Phi_1 \Phi_6 \Phi_9 - \Phi_3 \Phi_1 \Phi_2 \Phi_9 \Phi_8 - \Phi_4 \Phi_5 \Phi_7 \Phi_6 \, .
\eeq
In the examples that follow, we will not write the superpotentials explicitly, since it is straightforward to read them from the corresponding bipartite graphs.

\subsection{Non-Generic Situations: Global Identifications}

\label{sec:dimer-backreaction-special}

In the previous discussion we have implicitly assumed that in the dimer diagram all the faces adjacent to the instanton face are different. In terms of the mirror, this implies that any 1-cycle intersects the 1-cycle wrapped by the instanton at most once. Therefore, when the mirror Riemann surface $\Sigma$ is cut along the instanton 1-cycle, the formerly intersecting 1-cycles become {\em connected} 1-chains (namely, they do not split into several disjoint pieces). All these 1-chains combine into a single 1-cycle, as accounted for in the change of the number of faces of the BFT that we used in the Euler formula \eref{Delta_g}. If this condition is not satisfied, the genus of the resulting BFT need not be higher than the original one. More generally, it is possible for the change in genus to be in the range
\beq
0\leq \Delta g \leq k-1 \, .
\eeq
We refer to this situation as cases with {\it global identifications}, in the sense that faces adjacent to the instanton are identified due to the global topology on the dimer 2-torus.

\subsubsection{Examples}

\label{section_examples_global_identifications}

We now present various examples with global identifications and explain how to implement backreaction at the level of the dimer in such cases.

\subsubsection*{From $F_0$ to the Conifold}

\begin{figure}[!ht]
\begin{center}
\includegraphics[width=10.5cm]{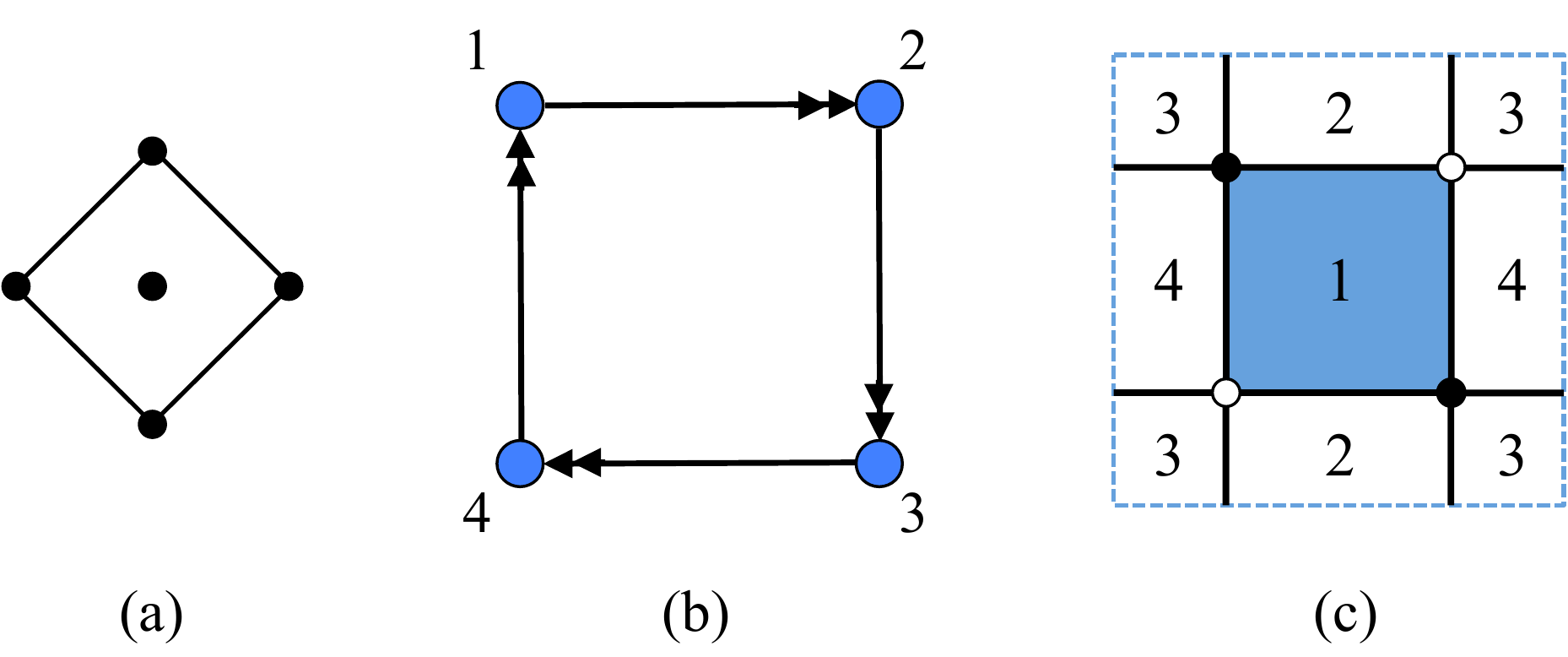}
\caption{a) Toric diagram for $F_0$. b) Quiver and c) dimer for its phase 1.}
\label{F0_1_toric_quiver_dimer}
\end{center}
\end{figure}

Let us consider $F_0$, which admits two toric phases (see e.g. \cite{Feng:2001xr}). \fref{F0_1_toric_quiver_dimer} shows the toric diagram for $F_0$ and the quiver and dimer for its phase 1.

\begin{figure}[!ht]
\begin{center}
\includegraphics[width=13cm]{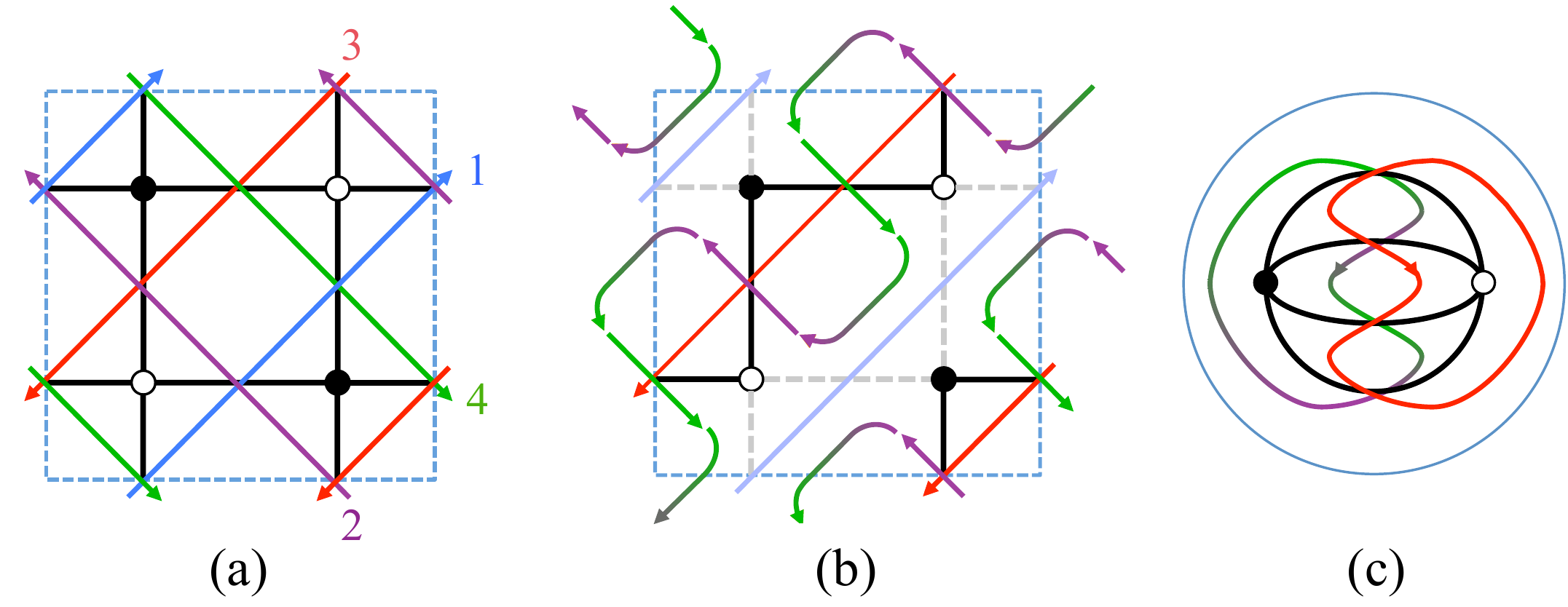}
\caption{Backreaction in the mirror of $F_0$. The instanton wraps the blue 1-cycle, which corresponds to face 1 of the original dimer. The final result is the mirror of the conifold.}
\label{F0_1_mirror}
\end{center}
\end{figure}

Let us consider an instanton on face 1 of the dimer (the symmetry of the theory implies that single instantons on any of the faces are equivalent). We start with the mirror description. The mirror for phase 1 of $F_0$ is again a square lattice and is given in \fref{F0_1_mirror}.a.\footnote{This particular theory is special in that the mirror is identical to the original dimer, see e.g. \cite{Feng:2005gw} While this behavior is not generic, it is certainly common for simple toric phases (namely those with the smallest number of chiral fields) for geometries for which $\Sigma$ is a punctured 2-torus, i.e. when the toric diagram has a single internal point. We expect the original dimer and the tiling of the mirror are not confused and that the distinction between them becomes clear from the context.} Zig-zags have been labeled according to the corresponding faces in the original dimer. An instanton on face 1 of the dimer corresponds to an instanton on the blue 1-cycle in the mirror. The backreaction is shown in \fref{F0_1_mirror}.b. Upon rearranging the diagram and integrating out massive fields associated to some of the 2-valent nodes, we recognize the result (c) as the mirror of the conifold theory. This is a theory that can be defined by a dimer diagram on $\mathbb{T}^2$, so in this case instanton backreaction does not increase the genus of the BFT. 

As anticipated, the reason for this behavior is that the original dimer contains faces that intersect the instanton more than once. This implies that some of the 1-cycle pieces in \fref{mirror_general} are actually not different. In this case, the counting in the Euler formula needs to be modified to take into account that there are two 1-cycles intersecting the instanton, each of them with intersection number $\pm 2$. These 1-cycles recombine into a single one, leading to $\Delta F=-2$ (including the disappearance of the instanton 1-cycle), instead of the generic $\Delta F=-2k=-4$. This implies that $\Delta g=0$ and the resulting BFT remains defined by a standard dimer on $\mathbb{T}^2$.

Let us now explain how this can be understood directly at the level of the dimer.
The procedure introduced in \sref{sec:dimer-backreaction-general} is still valid, with a minor clarification. The underlying feature of these non-generic cases is that some of the faces intersecting the instanton are globally identified. Therefore, in the process of identifying the black/white corners of the instanton face into a single black/white node, we should not insist in doing so in the local patch given by the instanton face (as implicit in Figures \ref{dimer-generic-4side} and \ref{dimer-generic-6side}). Such local procedure would lead to a higher genus BFT. Instead, we should always pick the identifications that minimize $\Delta g$. In other words, we should choose bridges such that the number of crossings is minimal. We refer to the original and the new prescriptions as the {\it local} and {\it global} recipes, respectively. The global prescription is the correct one and must always be used. The local and global prescriptions agree whenever there are no global identifications.

\begin{figure}[!ht]
\begin{center}
\includegraphics[width=\textwidth]{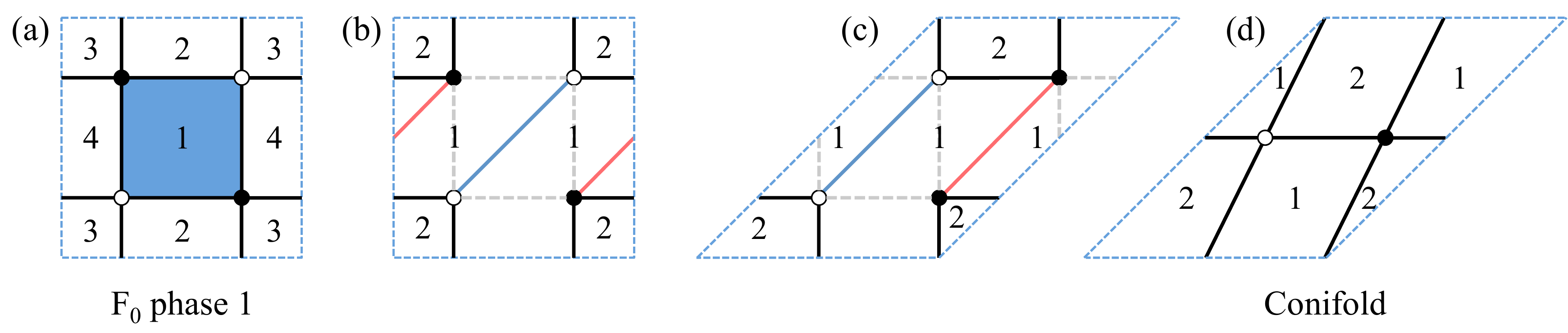}
\caption{Instanton backreaction from phase 1 of $F_0$ to the conifold.}
\label{F0_to_conifold}
\end{center}
\end{figure}

In \fref{F0_to_conifold} we illustrate this phenomenon for the $F_0$ example. (a) shows the instanton. In (b), we removed the corresponding face and added bridges, taking advantage of the periodicity of the $\mathbb{T}^2$ to evade crossings without increasing the genus. We also labeled the new faces in black. In the rest of the paper, we will apply a similar relabeling in those examples that remain on $\mathbb{T}^2$ after backreaction, for which visualizing the recombined faces is trivial. In (c) we switched to a different (but fully equivalent) unit cell, in order to bring the final theory to a more standard form. Finally, in (d) we condensed the bridges, obtaining the dimer for the conifold.  As this example shows, the global properties of the dimer can sometimes lead to $\Delta g < k-1$, which would be the naive result of the local recipe.

\subsubsection*{From $dP_0$ to $\mathbb{C}^3$}

\begin{figure}[!ht]
\begin{center}
\includegraphics[width=11.5cm]{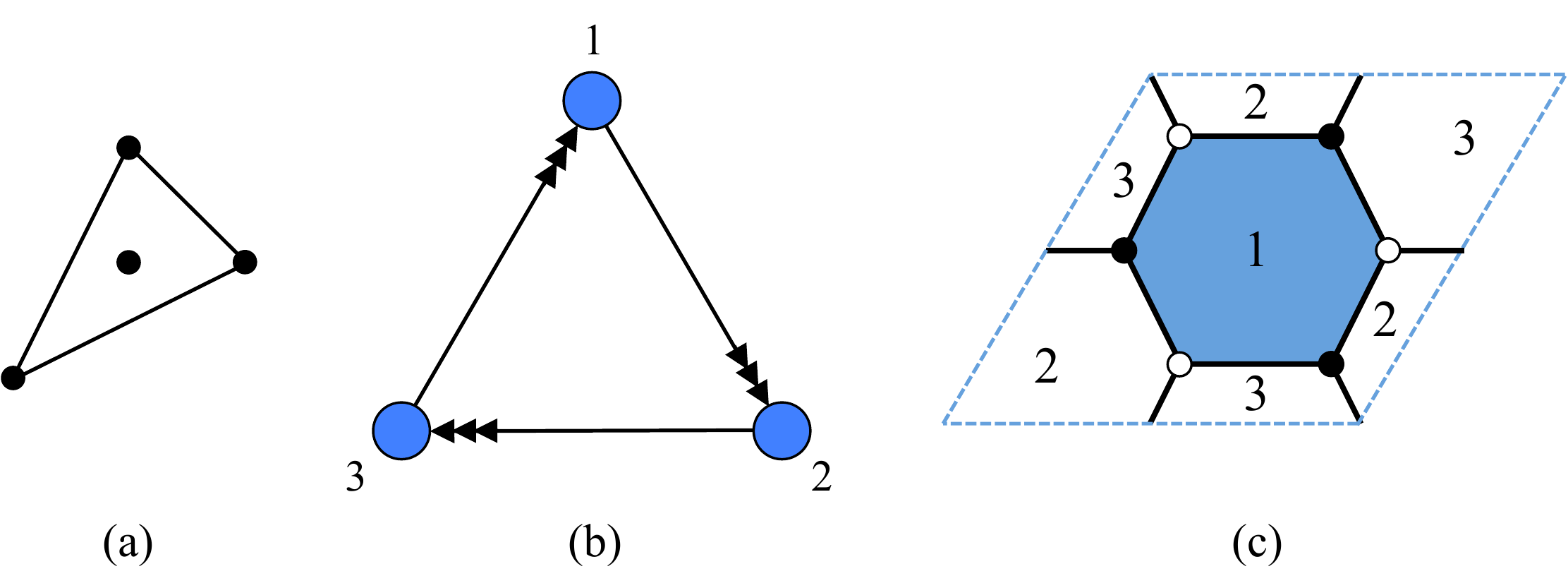}
\caption{Diagrams for $dP_0$: a) toric diagram, b) quiver and c) dimer.}
\label{dP0_toric_quiver_dimer}
\end{center}
\end{figure}

As another example, let us consider $dP_0$. Its toric diagram, and the quiver and dimer for its only toric phase are presented in \fref{dP0_toric_quiver_dimer}. 

Let us consider an instanton on face 1 of the dimer (the two other faces are equivalent by symmetry). The instanton has six edges but they represent intersections with only two gauge factors, since each of them intersects the instanton three times. The number of faces decreases from 3 to 1, i.e. $\Delta F=-2$, instead of the generic $\Delta F=-2k=-6$. This implies that the resulting BFT has $\Delta g=0$, instead of the generic $\Delta g=k-1=3$, and remains on $\mathbb{T}^2$. 

\begin{figure}[!ht]
\begin{center}
\includegraphics[width=\textwidth]{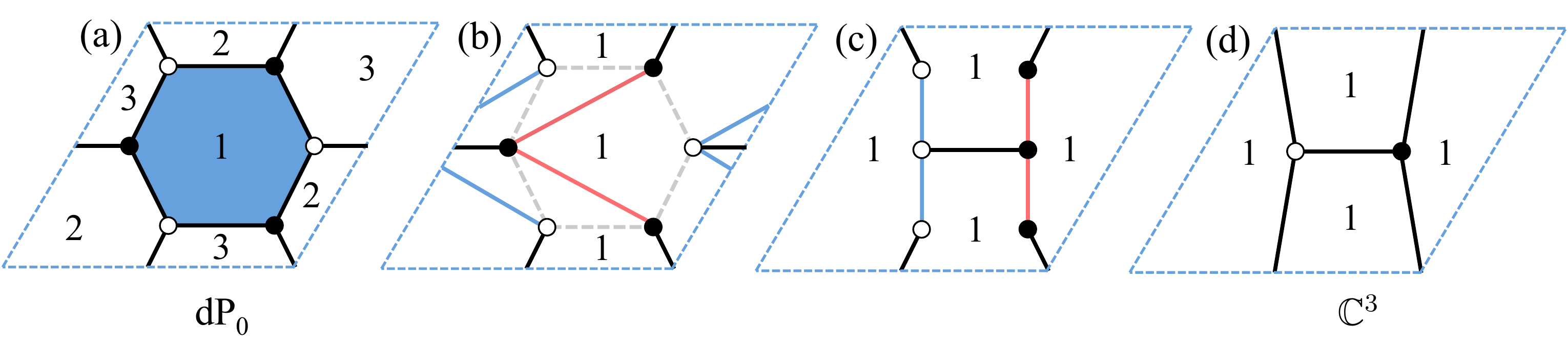}
\caption{Instanton backreaction from $dP_0$ to $\mathbb{C}^3$.}
\label{dP0_to_C3}
\end{center}
\end{figure}

We are now ready to implement the backreaction directly on the dimer, as shown in \fref{dP0_to_C3}. The instanton under consideration is given in (a). (b) shows a choice of bridges that exploits the periodicity of $\mathbb{T}^2$ to avoid crossings. (c) shows a continuous deformation of the diagram, which moves the two middle nodes horizontally. Upon condensation of the bridges we obtain the dimer for the $\mathbb{C}^3$ theory, i.e. for $\mathcal{N}=4$ SYM, as shown in (d). This result is fully reproduced by the mirror, as explicitly worked out in \cite{Tenreiro:2017fon}.

\subsubsection*{From $dP_1$ to $\mathbb{C}^2/\mathbb{Z}_2\times \mathbb{C}$}

\begin{figure}[!ht]
\begin{center}
\includegraphics[width=11.5cm]{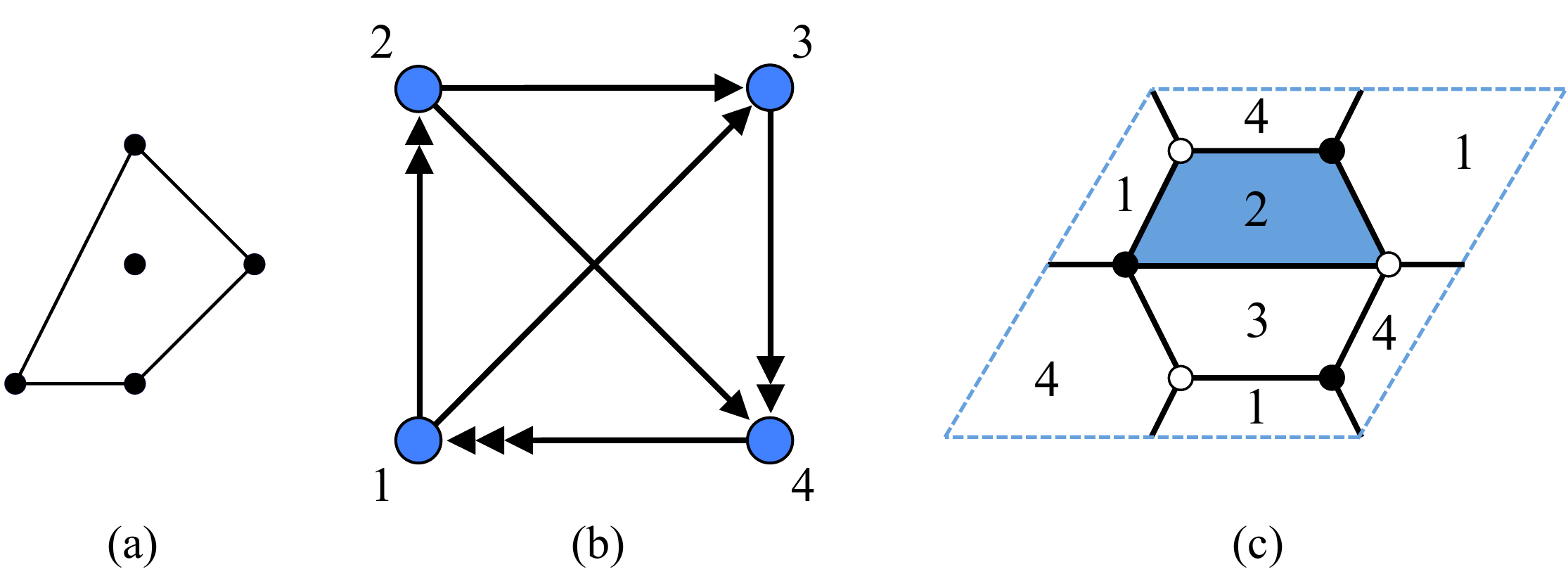}
\caption{Diagrams for $dP_1$: a) toric diagram, b) quiver and c) dimer.}
\label{dP1_toric_quiver_dimer}
\end{center}
\end{figure}

Let us finally consider $dP_1$, whose toric diagram, quiver and dimer are presented in \fref{dP1_toric_quiver_dimer}.

\begin{figure}[!ht]
\begin{center}
\includegraphics[width=\textwidth]{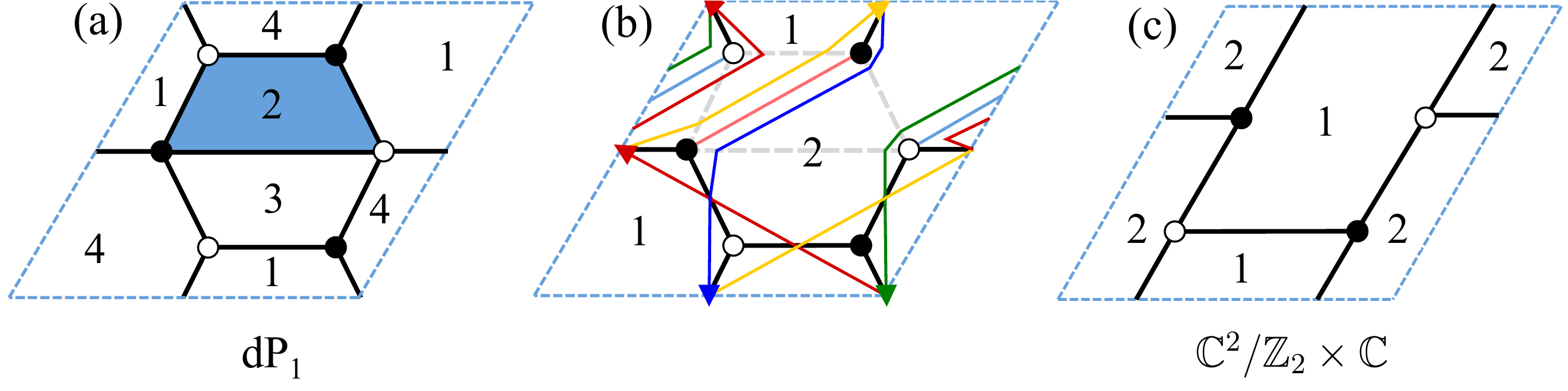}
\caption{Instanton backreaction from $dP_1$ to $\mathbb{C}^2/\mathbb{Z}_2 \times \mathbb{C}$.}
\label{dP1_to_C2Z2xC}
\end{center}
\end{figure}

Let us consider an instanton on face 2. \fref{dP1_to_C2Z2xC} shows the backreaction in the dimer. As shown in (b), it is possible to pick bridges such that there are no crossings. The final result is the dimer for $\mathbb{C}^2/\mathbb{Z}_2 \times \mathbb{C}$.

\begin{figure}[!ht]
\begin{center}
\includegraphics[width=6cm]{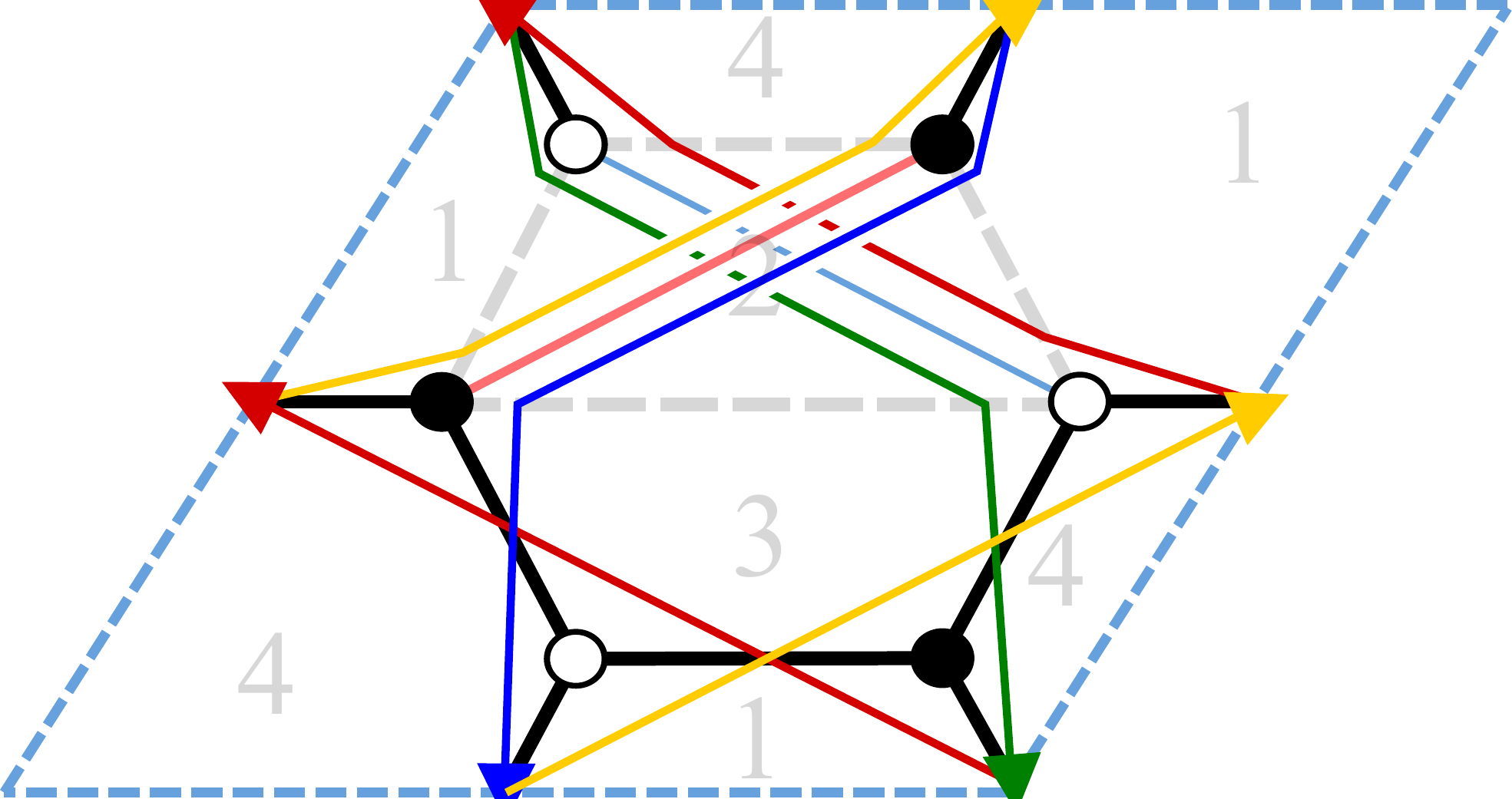}
\caption{Backreaction using the naive local recipe. This figure should be compared with \fref{dP1_to_C2Z2xC}.b, which correctly exploits global identifications.}
\label{DimerdP1UnitBack}
\end{center}
\end{figure}

It is interesting to use this example to discuss in further detail how the local recipe for backreaction in the dimer fails when there are global identifications. \fref{DimerdP1UnitBack} shows the {\it incorrect} backreaction that would be obtained by naively applying the local recipe. The zig-zag paths in this figure should be compared to the correct ones, which appear in \fref{dP1_to_C2Z2xC}.b. Note that the local and global backreactions do not change the intersections between the zig-zag paths. They however differ in the topology of their windings around cycles. As we will see in \sref{Sec:FakeTilings}, the local recipe generates a diagram that is not a consistent tiling.

\subsubsection{Fake Tilings}
 \label{Sec:FakeTilings}

We may provide a deeper insight into why the local recipe fails in cases with global identifications. 
Since the local recipe preserves the correct intersections of zig-zag paths, e.g. Figures \ref{dP1_to_C2Z2xC}.b and \ref{DimerdP1UnitBack}, it would naively appear that both pictures correspond to the same mirror. On the other hand, there is a problem with the counting of faces in the configuration obtained with the local recipe, so they cannot agree. The conundrum is solved by noticing that in cases with global identifications the local recipe gives rise to a graph which is not a consistent tiling of the corresponding Riemann surface, so it actually is not a consistent BFT.

Mathematically, a graph embedded in a Riemann surface provides a tiling of it if the Riemann surface is cut into regions which are, topologically, disks bounded by a concatenation of edges. Namely, faces must necessarily have the topology of a disk.
It is easy to show that in cases with global identifications, the local recipe produces what we call {\it fake tilings} of the resulting higher genus Riemann surfaces, in which some of the ``faces" are not disks but rather correspond to cylinders or other topologies. In particular, they contain non-trivial cycles precisely defined by exploiting the global identifications, as we now show.

\begin{figure}
    \centering
    \begin{subfigure}[t]{0.23\textwidth }
        \begin{center} 
		\includegraphics[width=\textwidth]{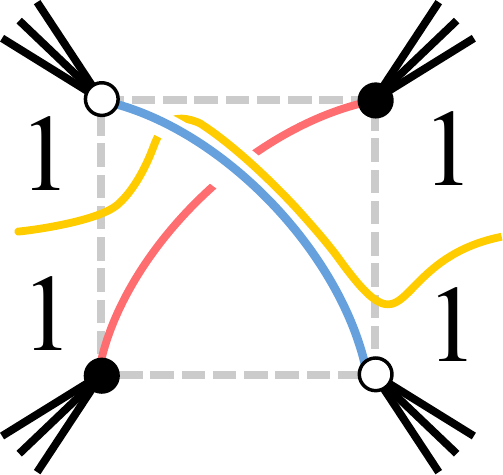}
		\caption{}
		\label{Fig:GlobalIdentification}
		\end{center}
    \end{subfigure} \hspace{10mm}
	\begin{subfigure}[t]{0.23\textwidth }
		\begin{center} 
		\includegraphics[width=\textwidth]{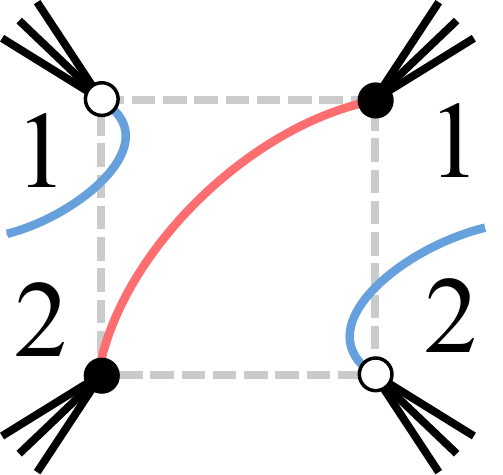}
		\caption{}
		\label{Fig:GlobalIdentification2}
		\end{center}
    \end{subfigure}\hspace{10mm}
    	\begin{subfigure}[t]{0.37\textwidth }
		\begin{center} 
		\includegraphics[width=\textwidth]{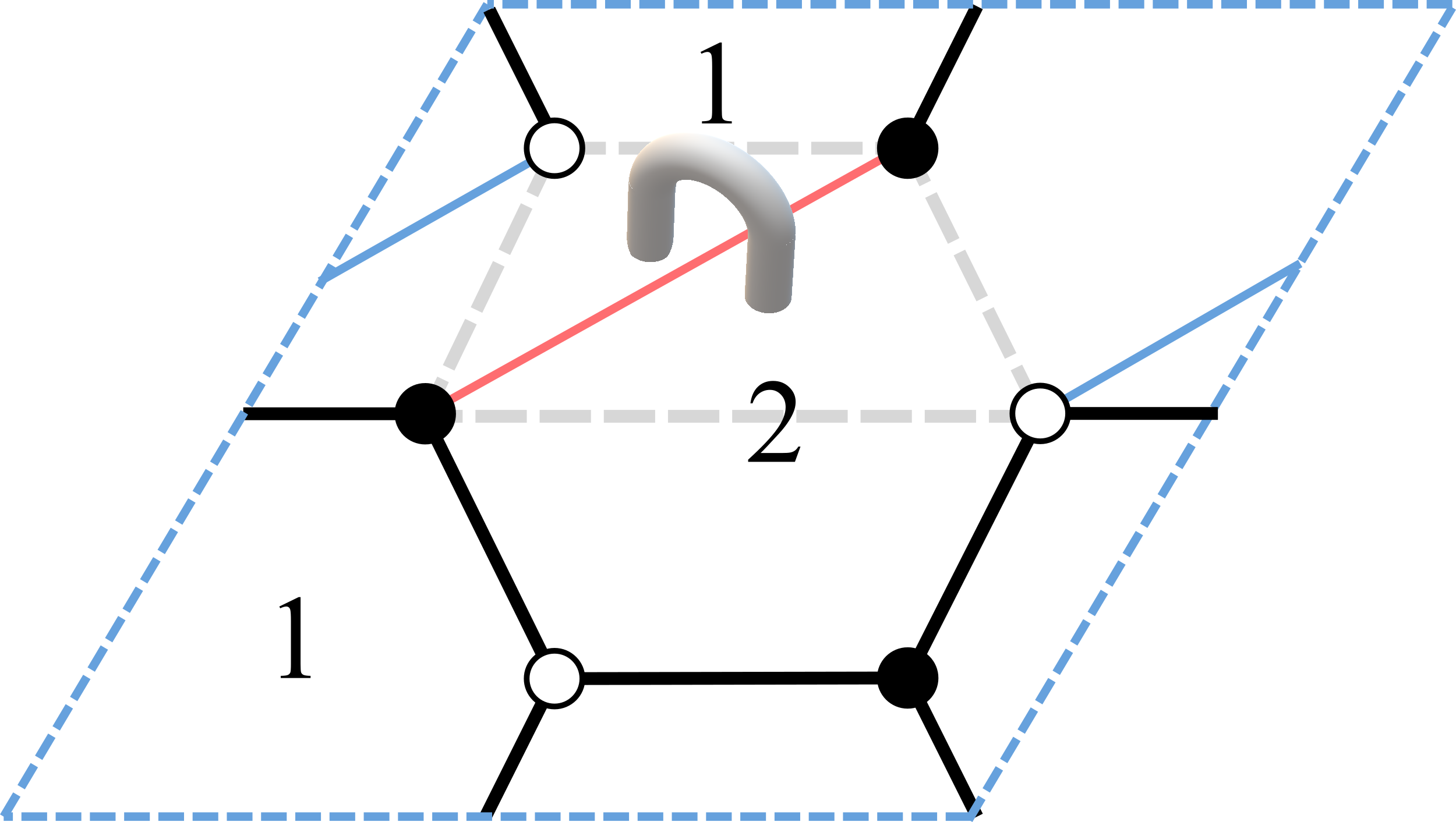}
		\caption{}
		\label{Fig:Bridge}
		\end{center}
    \end{subfigure}
    \caption{a) The local recipe for backreaction leads to a non-contractible cycle, shown in yellow. b) The cycle is not present if the global recipe is used. In both cases, the numbers label the resulting ``faces". c) Spurious handle connecting faces 1 and 2 for the example of the backreacted $dP_1$ theory.}\label{Fig:GlobalIdentifications} 
\end{figure}

As an example, consider an instanon on a square face with opposite edges separating it from a given {\em same} face and let us apply the instanton backreaction using the local recipe. The global identification leads to a non-contractible cycle along the redundant handle, as shown in \fref{Fig:GlobalIdentifications}.a. This cycle precisely specifies where the bridge must be, instead of through a spurious handle. Instead, if the correct local prescription is implemented, there is no non-contractible cycle, as shown in \fref{Fig:GlobalIdentifications}.b.

This non-contractible cycle indicates that two faces are identified by the additional handle. For instance, in the $dP_1$ example presented in \fref{DimerdP1UnitBack}, the local recipe produces an additional handle as shown in \fref{Fig:GlobalIdentifications}.c. This handle connects face 1 and 2, supports a non-contractible cycle and spoils the tiling.

\begin{figure}
    \centering
	\begin{subfigure}[t]{0.24\textwidth }
		\begin{center} 
		\includegraphics[width=\textwidth]{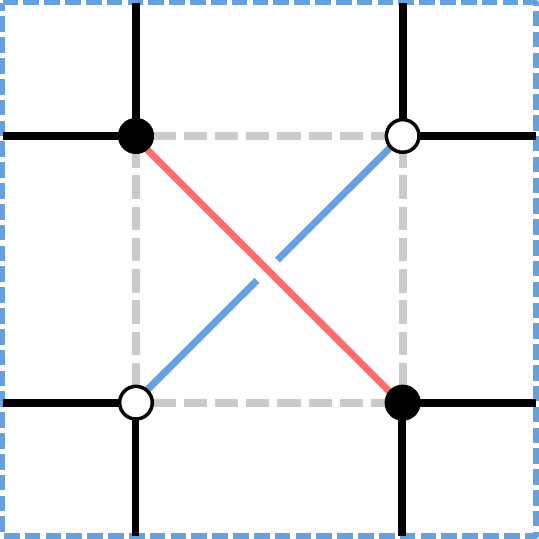}
		\caption{}
		\label{Fig:F0Local}
		\end{center}
    \end{subfigure}   \hspace{10mm}
    \begin{subfigure}[t]{0.24\textwidth }
        \begin{center} 
		\includegraphics[width=\textwidth]{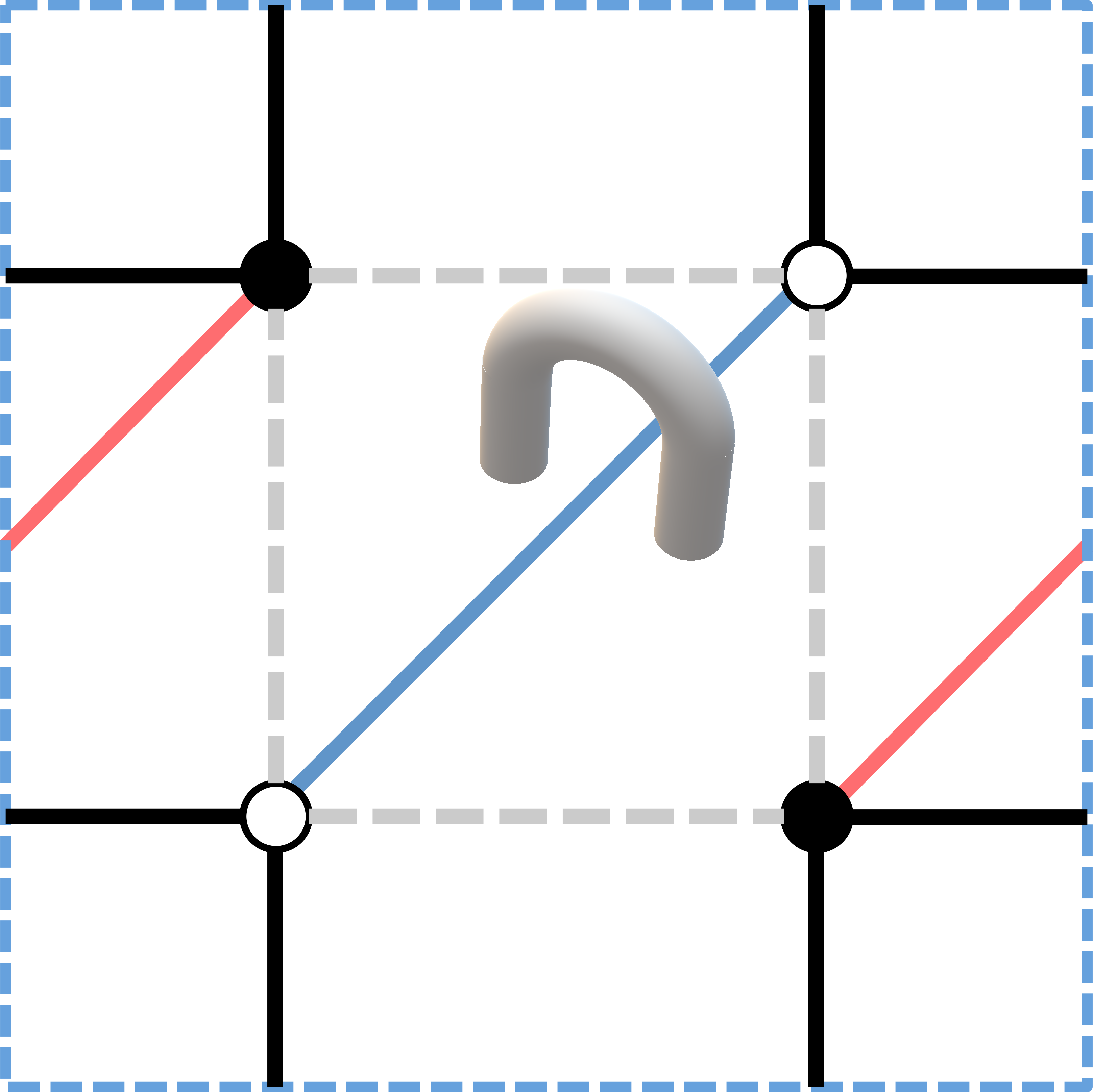}
		\caption{}
		\label{Fig:Bridge2}
		\end{center}
    \end{subfigure} \hspace{10mm}
	\begin{subfigure}[t]{0.24\textwidth }
		\begin{center} 
		\includegraphics[width=\textwidth]{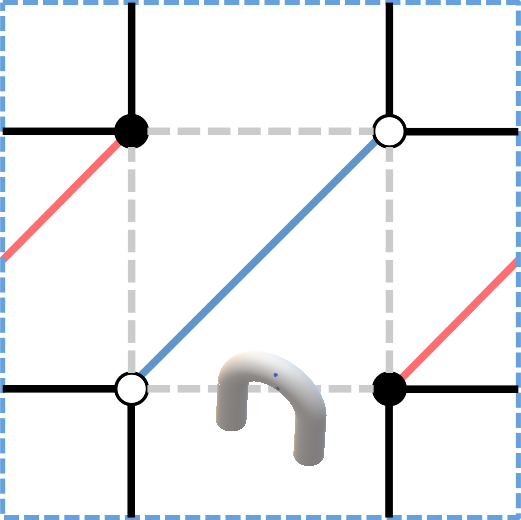}
		\caption{}
		\label{Fig:Bridge3}
		\end{center}
    \end{subfigure}
    \caption{a) Application of the local recipe of the backreaction for an instanton on face 1 of $F_0$. b) It produces a theory identical to the correct (global) backreaction, with an additional spurious handle. c) By a continuous deformation, the spurious handle manifestly connects the face to itself, showing it is not a consistent tiling.}\label{Fig:BridgesF0} 
\end{figure}

Similar comments apply to the case of the $F_0$ theory, for which the correct global recipe for backreaction was implemented in \fref{F0_to_conifold}. The only subtlety is that the local recipe, shown in \fref{Fig:BridgesF0}.a  would seem to produce the correct theory, but there is still a spurious handle, so it does not define a proper BFT, see \fref{Fig:BridgesF0}.b. By moving one of the legs off the bridge, the spurious handle manifestly connects the face to itself, as shown in \fref{Fig:BridgesF0}.c. 

We conclude by emphasizing that there is an unambiguous recipe for backreaction on the dimer, described in \sref{sec:dimer-backreaction-general}, namely removal of edges and recombination of nodes in the most economic way. In the generic case where the instanton has no repeated neighboring faces, this agrees with the local recipe, which therefore provides a simple surgery prescription for the generic theory. In most of the remainder of this paper, we focus on this generic situation.

\subsection{Extension to General BFTs}
\label{section_extension}

In the previous section we introduced a graphic implementation of the backreaction of a D-brane instanton on a face of a dimer. It is natural to extend this operation to the case in which the starting point is a general BFT, i.e. with arbitrary genus and number of boundaries. Physically, the initial BFT might be the result of backreacting additional instantons, which would change the genus, with boundaries, if present, generated by flavor D7-branes along the lines of \cite{Franco:2013ana}. At this point, it is unknown whether all BFTs can be obtained by this procedure. This is an interesting question that we postpone for future work.

Regardless of whether this operation can always be associated to a D-brane instanton, it is interesting to add it to the list of basic transformations that act on general BFTs, together with the condensation of 2-valent nodes, the square move and bubble reduction (see \cite{Franco:2012mm} and references therein for detailed discussions of these operations and their physical interpretation). In particular, it would be interesting to study its effect in the diverse applications of BFTs, e.g. in the context of scattering amplitudes, where the bipartite graphs are interpreted as on-shell diagrams \cite{ArkaniHamed:2012nw,Franco:2012mm,Franco:2013nwa,Franco:2014nca,Arkani-Hamed:2014bca,Franco:2015rma,Bourjaily:2016mnp}.

\begin{figure}[!ht]
\begin{center}
\includegraphics[width=\textwidth]{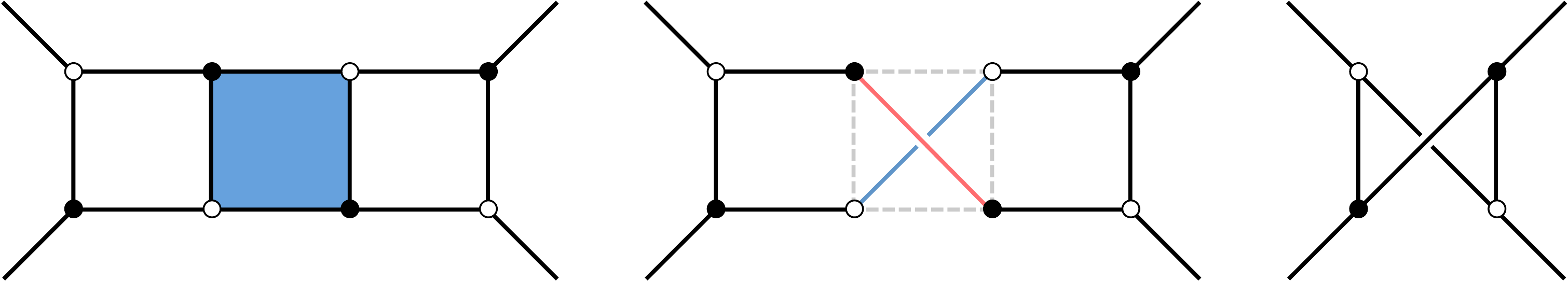}
\caption{Instanton backreaction for a general BFT on a disk, which is left implicit in the figure.}
\label{example_disk_BFT}
\end{center}
\end{figure}

\fref{example_disk_BFT} illustrates this operation for a BFT on a disk. The initial graph is {\it reducible}, namely it is possible to decrease the number of internal faces by a combination of square moves and bubble reductions \cite{ArkaniHamed:2012nw,Franco:2012mm}. The resulting non-planarity is reminiscent to the one that is necessary to capture the full matroid stratification of the Grassmannian in terms of on-shell diagrams, as discussed in  \cite{Franco:2013nwa}. 

\subsection{BFT Genus and Instanton Backreaction}

\label{section_BFT_genus_backreaction}

Let us try to understand the change in genus on the BFT side in more detail, in the generic case with no global identifications. The discussion below builds on and extends \sref{sec:dimer-backreaction-general}. It is phrased in a way that it easily applies to instantons on a $2k$-sided face of the tiling, for general $k$. For concreteness, we will focus on a $k=4$ example.

\begin{figure}[ht]
	\centering
	\includegraphics[width=11cm]{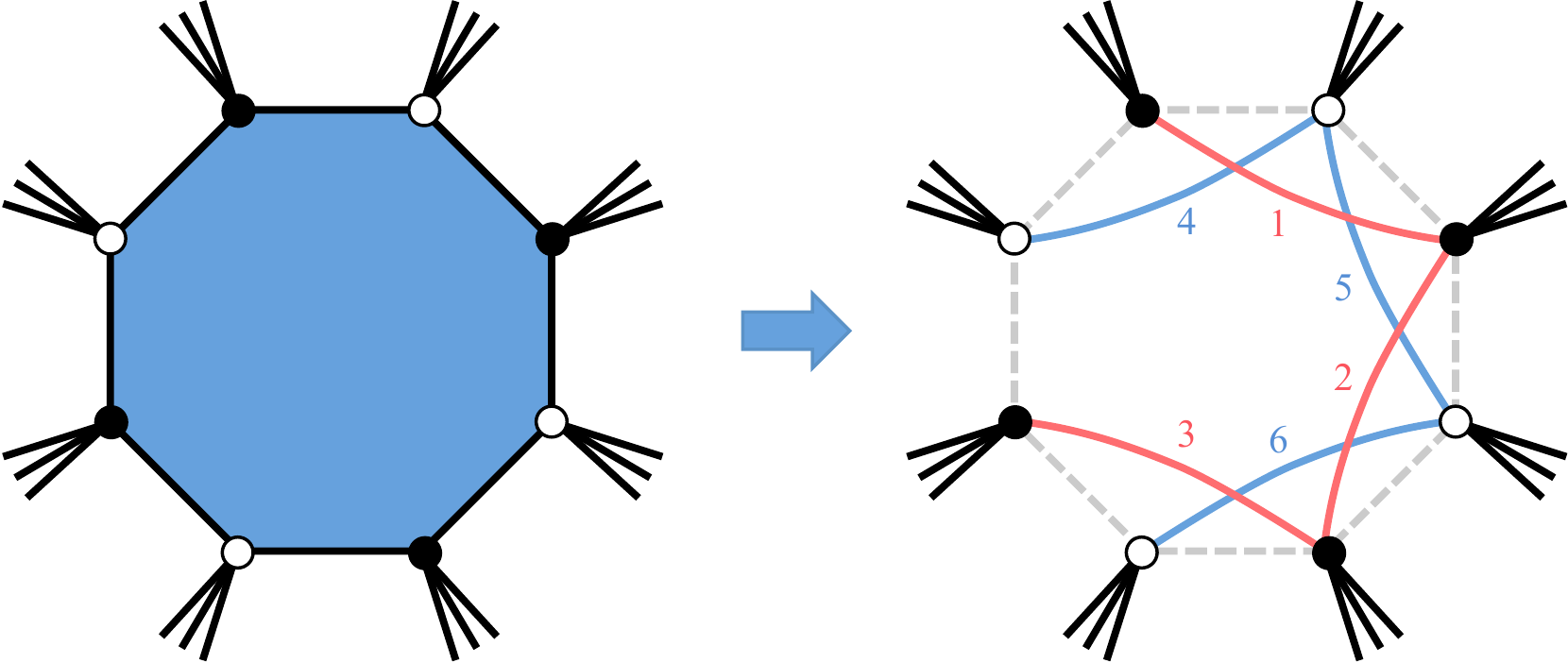}
\caption{Backreaction of an instanton on an octagonal face of the dimer, i.e. $k=4$.}
	\label{backreaction_k=4}
\end{figure}

\fref{backreaction_k=4} shows the backreaction of the instanton on the dimer. This is the $k=4$ analogue of Figures \ref{dimer-generic-4side} and \ref{dimer-generic-6side}. The face disappears and the nodes at its corners are recombined. This can be achieved by introducing $k-1$ bridges between white nodes (shown in blue) and $k-1$ bridges between black nodes (shown in red). We label the bridges to facilitate their identification.

According to our earlier discussion, in the generic case the genus of the BFT Riemann surface changes by $\Delta g = k-1$. We now device a simple graphical representation that makes the topology of the extra handles manifest.

We can think about the change in the Riemann surface as the result of cutting a hole on the original surface and gluing to it a genus $k-1$ ``handle" with an identical hole.\footnote{For brevity, we will use the term handle even for genus greater than 1.} The new edges associated with the bridges responsible for recombining the corners of the instanton face are the only ones living on the handle. The rest of the bipartite graph remains on the original Riemann surface.

\begin{figure}[ht]
	\centering
	\includegraphics[height=7cm]{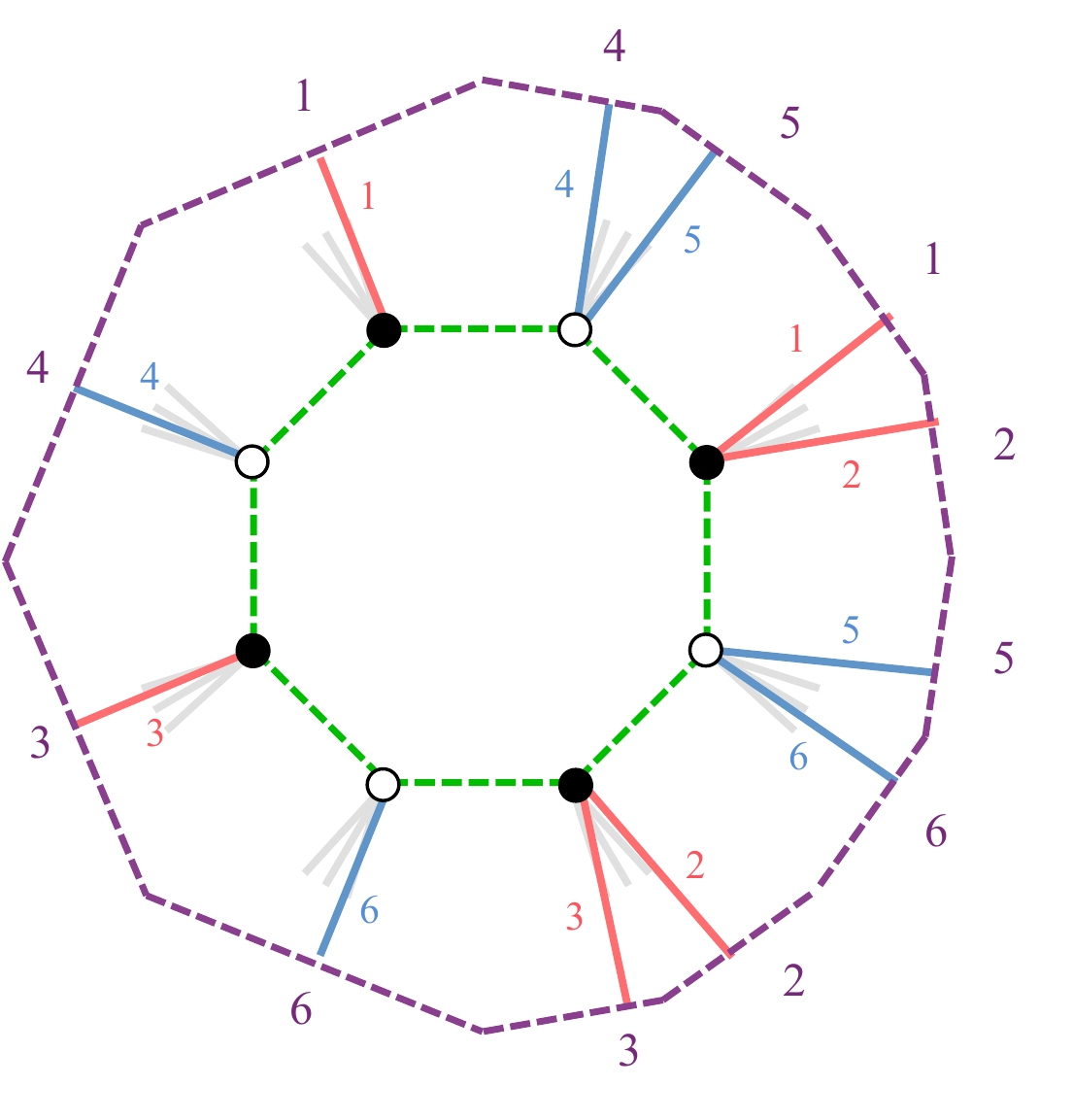}
\caption{Change in the Riemann surface by gluing a $\Delta g = k-1=3$ handle.}
	\label{new_handle}
\end{figure}

\fref{new_handle} shows the change in the Riemann surface for $k=4$.

\fref{new_handle} should be interpreted as follows.

\begin{itemize}

\item The green dashed loop indicates the cut at which the handle is glued to the Riemann surface. It is very natural to place this cut at the boundary of the original face.

\item A genus $g$ Riemann surface can be represented by a $4g$-gon with pairwise identification of edges. Each of this pairs corresponds to one of the $2g$ fundamental cycles. In \fref{new_handle}, the handle has genus 3, so it is presented by the 12-sided dashed purple polygon. This handle has a hole, whose boundary is the green loop, along which it is glued to the original Riemann surface.

\item As shown in \fref{new_handle}, each of the fundamental cycles is used exclusively by one of the new bridges. Thus it is natural to label the corresponding pair of sides in the $4g$-gon with the same name of the corresponding bridge. These labels are shown in purple in the figure. It becomes clear how this configuration avoids crossings between bridges and how it is generalized to arbitrary $k$.

\item The fact that each fundamental cycle is used by a single bridge makes the computation of the toric diagram for the new BFT straightforward, as we will explain in the coming section.

\end{itemize}

\begin{figure}[ht]
	\centering
	\includegraphics[height=7cm]{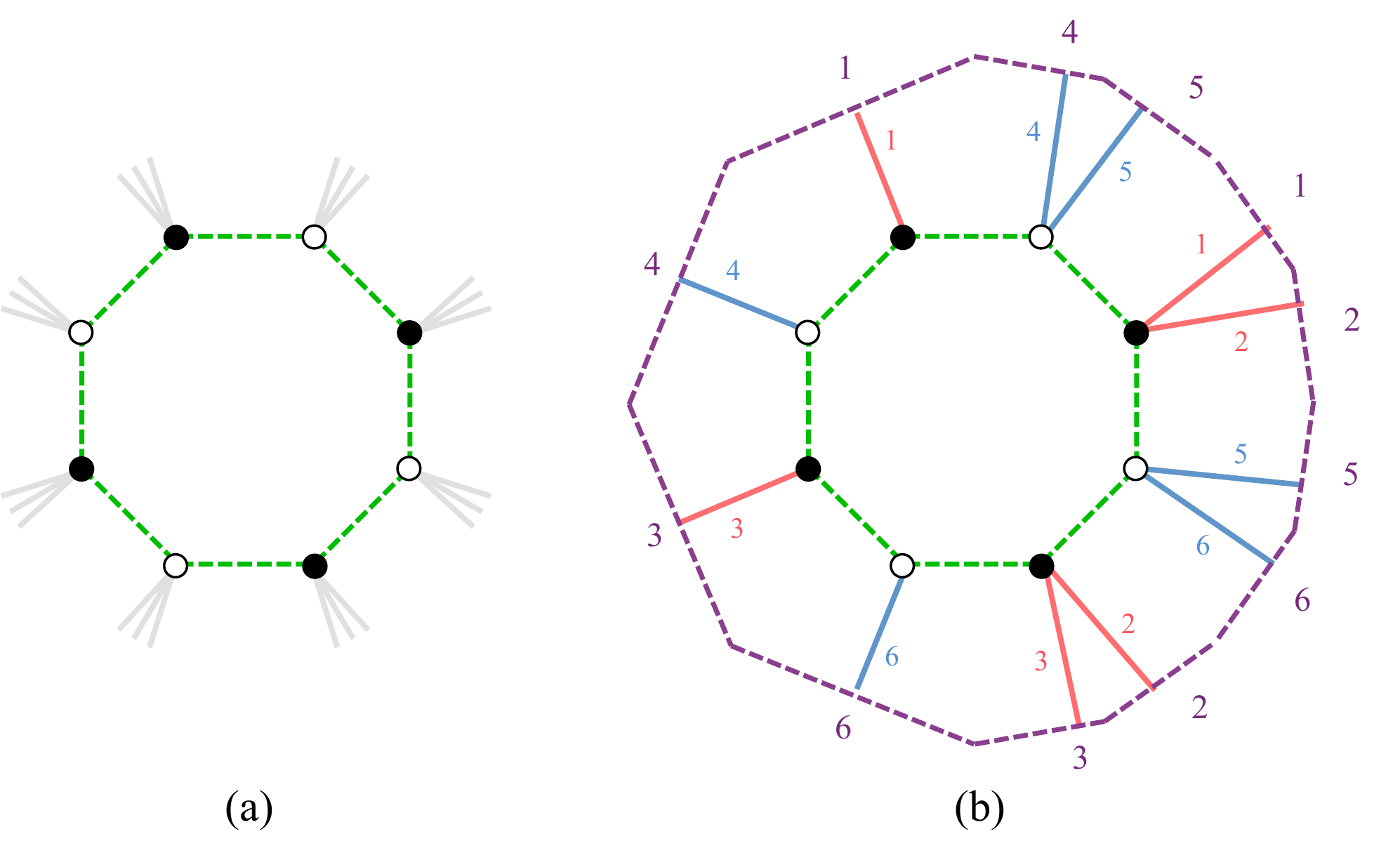}
\caption{a) Part of the dimer that remains on the original Riemann surface. b) The new edges live on the genus $k-1=3$ handle. Both surfaces have matching holes. They are glued along the boundaries, which are shown in green.}
	\label{new_handle_separated}
\end{figure}

\fref{new_handle_separated} is identical to \fref{new_handle}, but shows the original Riemann surface and the handle, both with matching holes, separately.

\section{The Toric Geometry of Backreacted Dimers}

\label{section_toric_geometry_backreacted_dimers}

Since the backreacted theory is a BFT, its moduli space of vacua is a toric CY. This geometry encodes important information about the gauge theory. Following the general discussion in \cite{Franco:2012mm}, the moduli space of a genus $g$ BFT is a CY $(2g+1)$-fold, which has a $2g$-dimensional toric diagram.\footnote{Here we assume the BFT has no boundaries. It is straightforward to incorporate them to our discussion.} Points in the toric diagram correspond to (collections of) perfect matchings of the bipartite graph.

It is certainly straightforward to directly determine the toric diagram for the resulting BFT (see e.g. \cite{Franco:2012mm}). However, it is instructive to understand how the new toric CY relates to the original one. Below we do this in two steps: we first find the perfect matchings of the final theory and then we determine their positions in the toric diagram.

\subsection{Perfect Matchings}

\label{section_removed_perfect_matchings}

In order to identify the new perfect matchings, it is convenient to condense all the bridges, i.e. the corresponding 2-valent nodes. We will later reintroduce them to determine 
the final toric diagram. 

Let us decompose every perfect matching as $p_\mu=p_\mu^{int}+p_\mu^{ext}$, where $p_\mu^{int}$ contains the edges in $p_\mu$ that belong to the instanton face, while $p_\mu^{ext}$ contains all the other edges. After backreacting the instanton and integrating out bridges, all internal edges disappear and  $p_\mu \to p_\mu^{ext}$. Below we will study the conditions under which $p_\mu^{ext}$ is a perfect matching of the backreacted dimer. It is important to remark that since only external edges survive backreaction, the $p_\mu^{ext}$'s contain all possible perfect matchings of the final theory.

\paragraph{Removed perfect matchings.} 
We refer to the perfect matchings that do not survive this process as {\it removed perfect matchings}. It is possible to identify them as follows. 

Condensing all bridges, the number of superpotential terms is reduced by $(2k-2)$. Consequently, the number of edges in a perfect matching is reduced by $(k-1)$. Since all surviving edges are external to the instanton face and the exterior content of the perfect matching remains unchanged, we conclude that this change must correspond to internal edges in the perfect matching. Denoting the number of internal edges in $p_\mu$ as $E(p_\mu^{int})$, we conclude that iff
\beq
E(p_\mu^{int}) \neq k-1
\eeq
the perfect matching is removed, i.e. $p_\mu^{ext}$ is not a perfect matching after backreaction.

It is straightforward to show that an equivalent condition is that perfect matchings are removed iff $p_\mu^{ext}$ contains more than one corner of the instanton face of a given color. In such a case, $p_\mu^{ext}$ is not a perfect matching after corner identification, since it contains more than one edge terminating on some of the nodes.

\subsection{The New Toric Diagram}
\label{section_new-toric-diagram}

In order to assign coordinates in the toric diagram to the surviving perfect matchings, it is convenient to reintroduce the bridges, i.e. to integrate in the corresponding massive pairs of edges.\footnote{Of course, as already mentioned, it is also possible to directly find these coordinates, without introducing the bridges \cite{Franco:2012mm}.} Our prescription will generate coordinates in $\mathbb{Z}^{2g}$. 

For concreteness, let us assume we start from a dimer, i.e. from a BFT on $\mathbb{T}^2$.\footnote{In the absence of global identifications, the new BFT has genus $g=1+\Delta g=k$, and the toric diagram lives in $\mathbb{Z}^{2k}$.} By convention, we will identify the first two coordinates with those in the original toric diagram. They remain unchanged, provided that the instanton face does not intersect the boundaries of the original unit cell. If the parent dimer is sufficiently large, it is always possible to define the unit cell to avoid such crossings. This condition is satisfied in all the explicit examples that we consider below.  The remaining $2\Delta g$ coordinates are related to the new cycles introduced with the handle, as discussed in \sref{section_BFT_genus_backreaction}.

After reintroducing the bridges, we complete every surviving $p_\mu^{ext}$ into a perfect matching. This completion is unique. Given such a completion, there are two standard approaches for establishing its position in the toric diagram:
\begin{itemize}
\item \underline{Method 1}: each coordinate is given by the net intersection number between the edges in the perfect matching, counted with orientation, and the corresponding cycle.\footnote{By convention, we orient edges in the graph from white to black nodes.} 
\item \underline{Method 2}: perfect matchings are mapped to oriented cycles by subtracting an arbitrary reference perfect matching. Coordinates correspond to winding numbers of the resulting cycles or, equivalently, to the monodromies of the height function.
\end{itemize}
In the coming section, we will illustrate both of them in an explicit example. In practice, the first approach is typically simpler to implement.

\begin{figure}[ht]
	\centering
	\includegraphics[width=\textwidth]{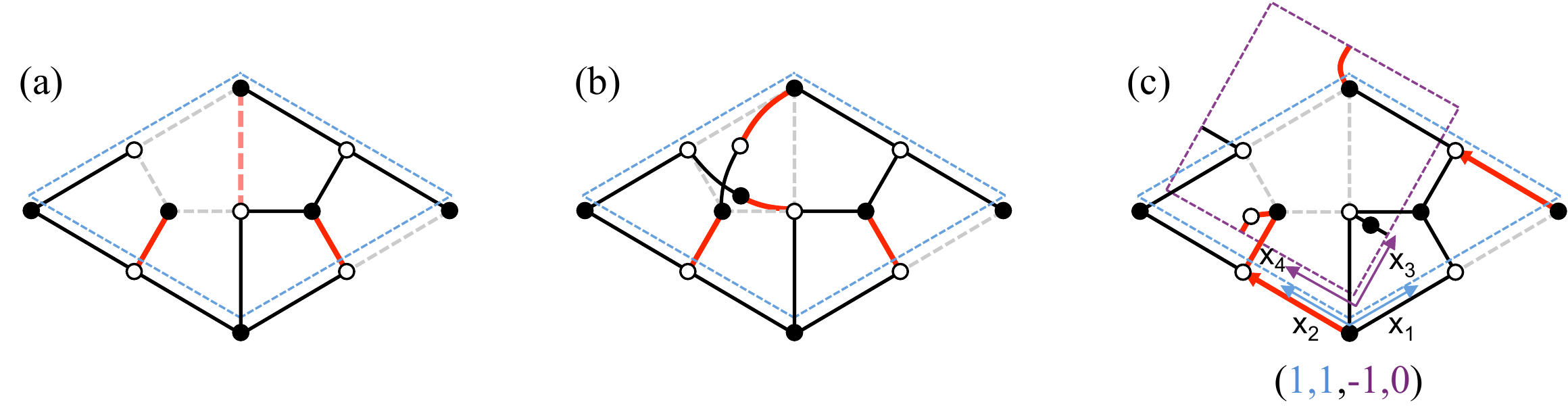}
\caption{An example in phase 1 of $dP_3$. a) The original perfect matching. b) Backreaction including bridges. c) Explicit introduction of the corresponding genus 1 handle.}
	\label{example_coordinates_BFT_dP3_1}
\end{figure}

\fref{example_coordinates_BFT_dP3_1} shows an example for phase 1 of $dP_3$. The instanton is located at the top-left  face, which is a square, so it has $k=2$. In \fref{example_coordinates_BFT_dP3_1}.a, we show the original perfect matching under consideration. It survives in the final BFT because it satisfies the condition that the number of edges in $p_\mu^{int}$ is equal to $k-1=1$. In \fref{example_coordinates_BFT_dP3_1}.b we add the bridges and the corresponding edges to form a perfect matching. In \fref{example_coordinates_BFT_dP3_1}.c, we implement the backreaction of the Riemann surface using the approach outlined in \sref{section_BFT_genus_backreaction}, introducing the genus $k-1=1$ handle.\footnote{For genus 1 handles, we label the boundaries of the handle's fundamental domain according to the transverse axes. For $k>2$ it is convenient to label them according to the corresponding bridges, as in \fref{new_handle_separated}.} We determine the corresponding coordinates in the toric diagram from the intersections between the perfect matching and the various fundamental cycles. 

\subsection{Coordinates from Bridges}
\label{section_coordinates-bridges}

\begin{figure}[ht]
	\centering
	\includegraphics[width=11cm]{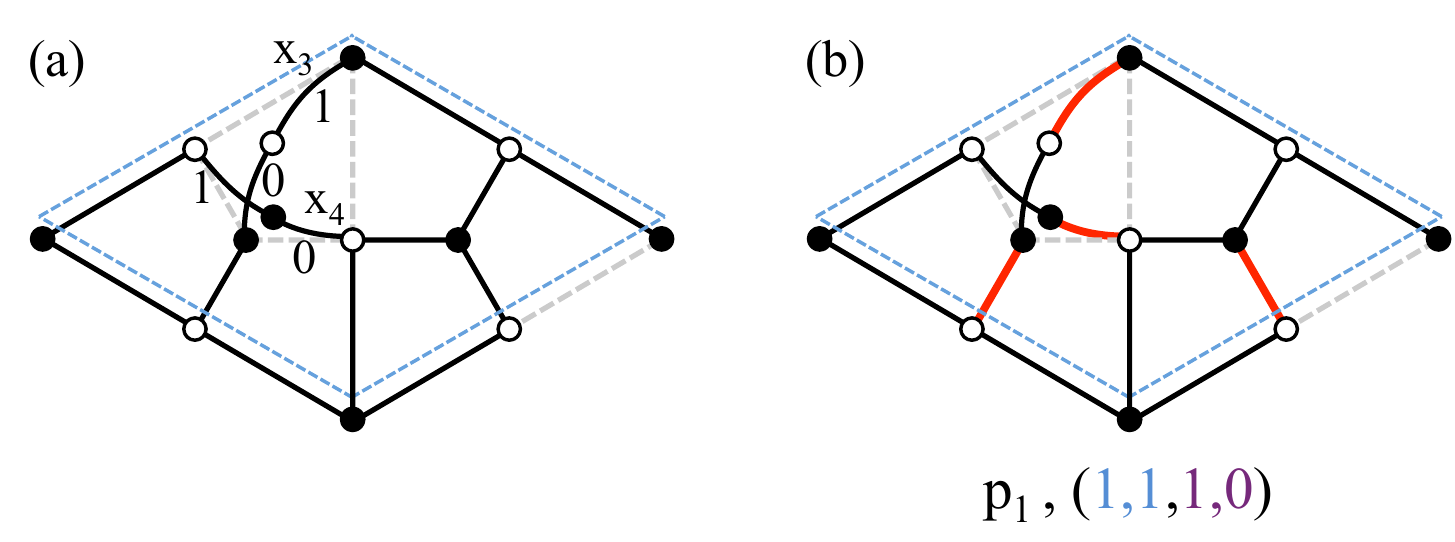}
\caption{a) Prescription for assigning new coordinates to the edges on the bridges. b) The perfect matching of \fref{example_coordinates_BFT_dP3_1} and the resulting coordinates.}
	\label{coordinates_bridges_dP3}
\end{figure}

We discussed a detailed visualization of the new genus $(k-1)$ handle in terms of a $4(k-1)$-gon and explained how to use this construction for determining the new coordinates of perfect matchings. It is however desirable to introduce a simpler prescription in which the coordinates can be directly read from the bridges. This is straightforward, since bridges are in one-to-one correspondence with the new cycles/coordinates. To do so, we draw bridges with the intermediate 2-valent nodes. By convention, we associate $0$ and $1$ contributions to the corresponding coordinate to the two edges on each bridge. It is always possible to avoid $(-1)$ contributions, which can certainly be generated by the prescription introduced in the previous section, by an appropriate choice of the relative position of the middle point of bridges with respect to the corresponding boundaries of the fundamental domain of the handle. Equivalently, this simply translates into a choice of the positive direction for each of the cycles. In \fref{coordinates_bridges_dP3} we illustrate this rule for the example in \fref{example_coordinates_BFT_dP3_1}.

The fact that new coordinates can only take values 0 and 1 constraints the BFTs that can be generated by instantons. In particular, we cannot obtain BFTs with toric diagrams that are ``too wide" in more than two directions (the ones for the original dimer). This argument applies even for multiple instantons.

\subsection{Example: $dP_3$}
\label{section_dp3}

\begin{figure}[H]
	\centering
	\includegraphics[width=\linewidth]{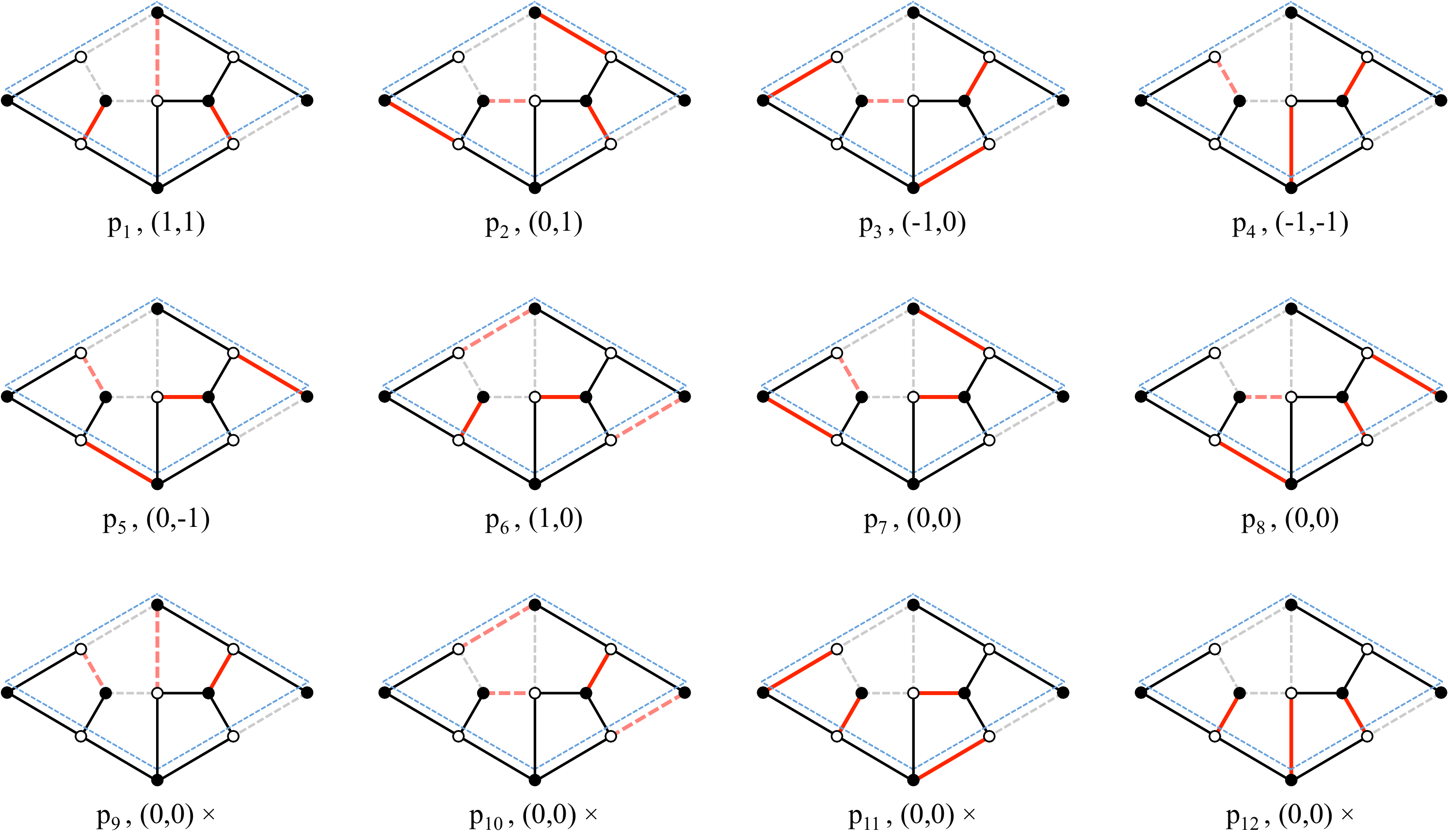}
\caption{The 12 perfect matchings for phase 1 of $dP_3$.}
	\label{original_pms_dP3_1}
\end{figure}

We now illustrate the ideas introduced in the previous section in an explicit example. Let us consider phase 1 of $dP_3$. \fref{original_pms_dP3_1}  presents the perfect matchings for this theory and their positions in the toric diagram, computed from their intersections with the boundaries of the unit cell. We also identify the removed perfect matchings with a cross. In this case, perfect matchings for all points in the original toric diagram survive.\footnote{Generically, however, there can be cases in which all the perfect matchings for a given point in the original toric diagram disappear.} However, different perfect matchings for a given point, in this case $p_7$ and $p_8$, have different lifts.

\begin{figure}[H]
	\centering
	\includegraphics[width=\linewidth]{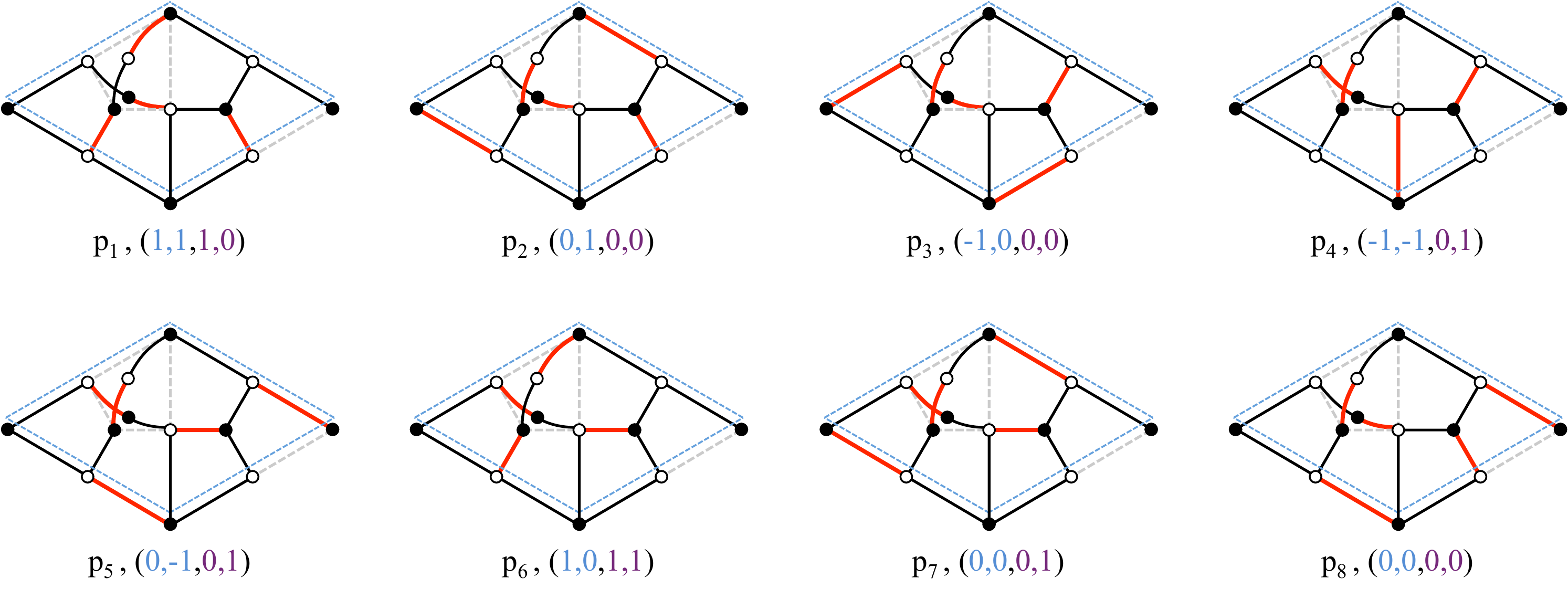}
\caption{Surviving perfect matchings. The new coordinates are determined by the edge content on the bridges, using the convention in \fref{coordinates_bridges_dP3}.}
	\label{final_pms_dP3_1_simple}
\end{figure}

\fref{final_pms_dP3_1_simple} presents the surviving perfect matchings, $p_1,\ldots,p_8$, in the backreacted dimer. This example illustrates how to proceed in general: in order to complete perfect matchings we must include edges on the bridges, which in turn determine the new coordinates in the final toric diagram.

\begin{figure}[H]
	\centering
	\includegraphics[width=\linewidth]{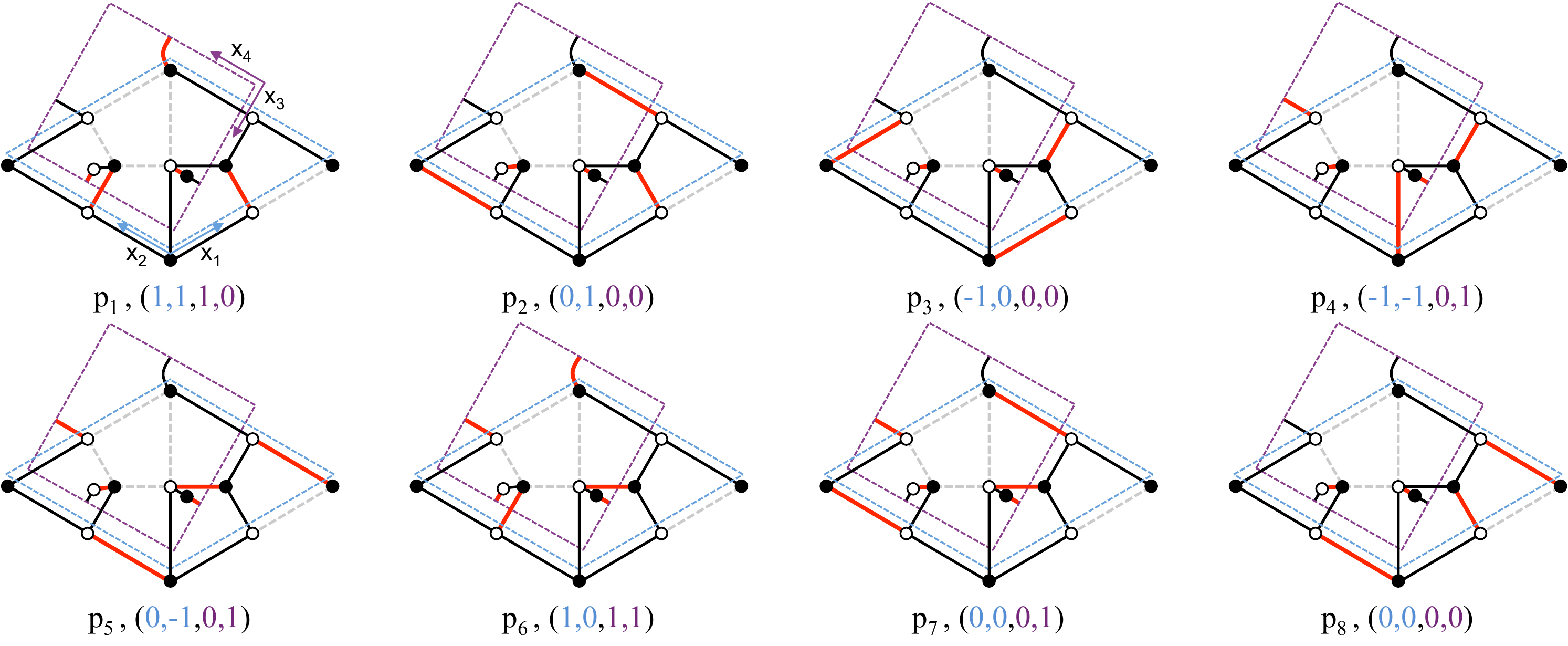}
\caption{Surviving perfect matchings with the bridges and the genus 1 handle.}
	\label{final_pms_dP3_1}
\end{figure}

The same perfect matchings are presented in \fref{final_pms_dP3_1}, this time explicitly showing the $k-1=1$ handle in purple. 

\begin{figure}[H]
	\centering
	\includegraphics[width=\linewidth]{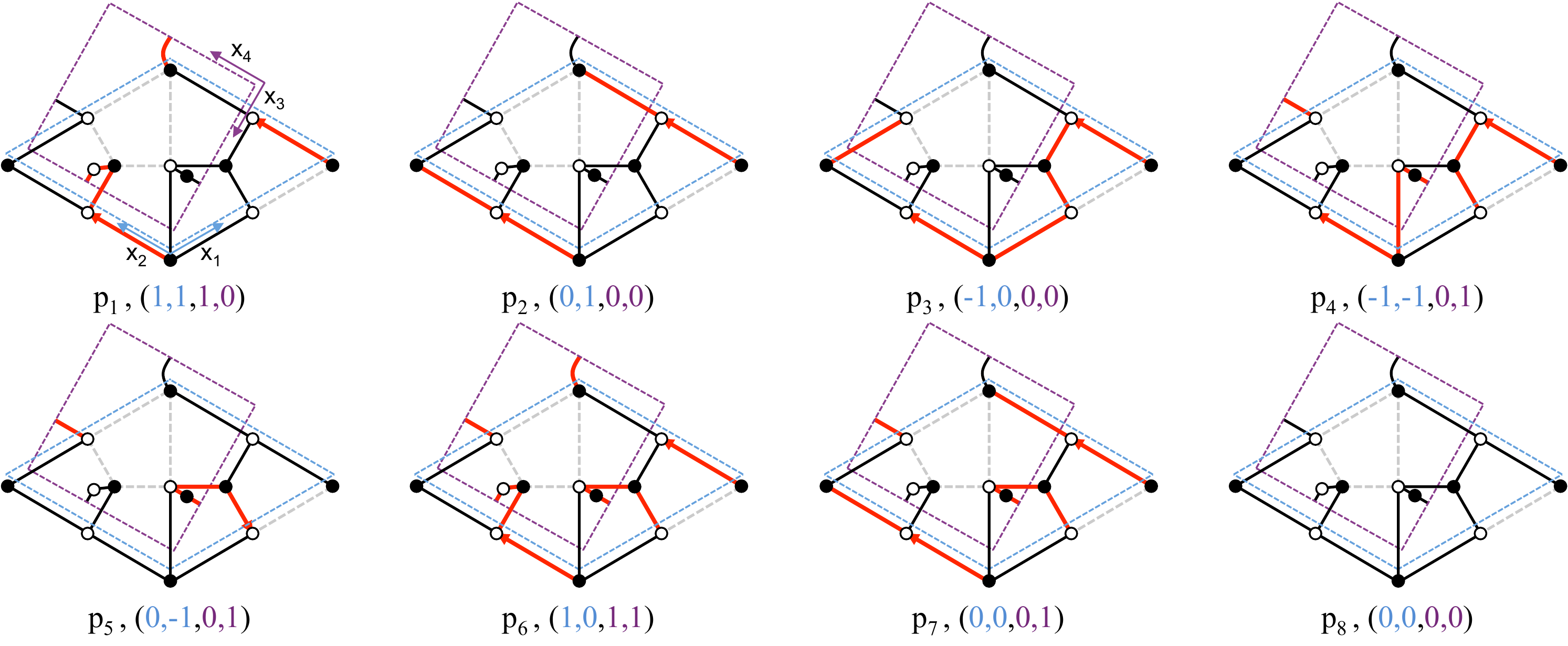}
\caption{Cycles for the surviving perfect matchings using $p_8$ as reference.}
	\label{final_pms_dP3_1_cycles}
\end{figure}

In \fref{final_pms_dP3_1_cycles} we map the perfect matchings to cycles, using $p_8$ as reference. Since the coordinates for $p_8$ are $(0,0,0,0)$, the winding numbers agree with the coordinates previously computed from the intersection numbers. Otherwise, they would simply differ by a constant shift, given by the coordinates of the reference perfect matching.

The resulting toric diagram is given by the following matrix
\beq
\left(
\begin{array}{cccccccc}
\ p_1 \ & \ p_2 \ & \ p_3 \ & \ p_4 & \ p_5 \ & \ p_6 \ & \ p_7 \ & \ p_8 \ \\ \hline
 1 & 0 & -1 & -1 & 0 & 1 & 0 & 0 \\
 1 & 1 & 0 & -1 & -1 & 0 & 0 & 0 \\
 1 & 0 & 0 & 0 & 0 & 1 & 0 & 0 \\
 0 & 0 & 0 & 1 & 1 & 1 & 1 & 0 \\
\end{array}
 \right) \quad \to \quad 
 \left(
 \begin{array}{cccccccc}
\ p_1 \ & \ p_2 \ & \ p_3 \ & \ p_4 & \ p_5 \ & \ p_6 \ & \ p_7 \ & \ p_8 \ \\ \hline
 1 & 0 & 0 & 0 & 0 & 1 & 0 & 0 \\
 0 & 1 & 0 & 0 & 0 & 0 & 1 & 0 \\
 0 & 0 & 1 & 0 & -1 & -1 & -1 & 0 \\
 0 & 0 & 0 & 1 & 1 & 1 & 1 & 0 \\
\end{array}
\right) ,
\eeq
where on the right hand side we have row-reduced it to give it a simpler form and verify that all coordinates are indeed independent. We conclude the toric diagram is $4d$, i.e. it corresponds to a CY 5-fold.

\section{Seiberg Duality}
\label{section_seiberg-duality}

\begin{figure}[ht]
	\centering
	\includegraphics[width=8cm]{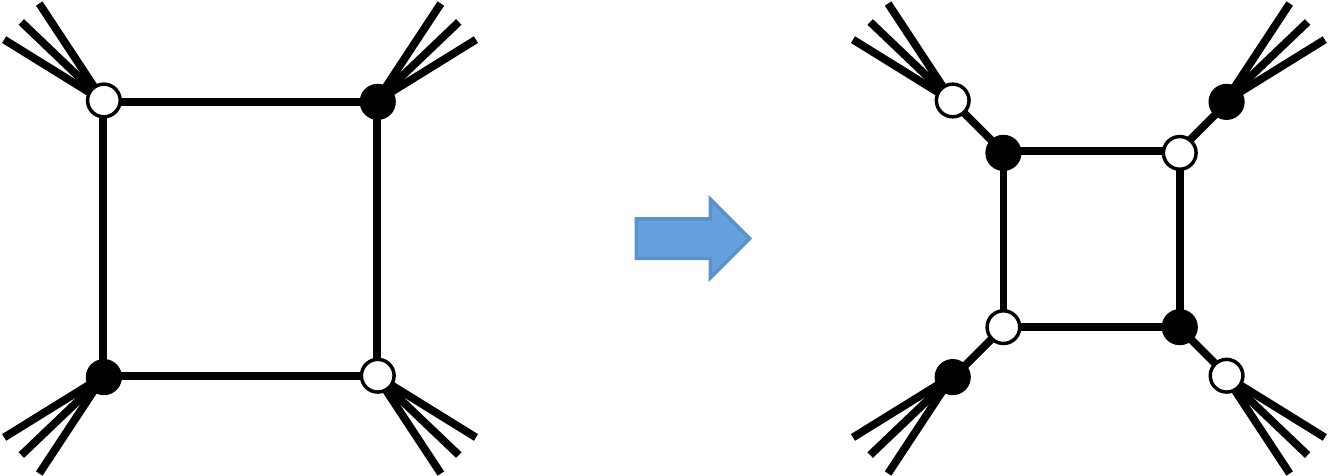}
\caption{Square move implementing Seiberg duality on a 4-sided face of a BFT.}
	\label{square_move}
\end{figure}

Seiberg duality admits a simple graphical implementation for BFTs (see \cite{Franco:2005rj} for the original discussion for dimers and \cite{Franco:2012mm,Xie:2012mr} for general BFTs). More precisely, Seiberg duality acting on a gauge group associated to a 4-sided face of a BFT corresponds to the so-called {\it square move}, which is shown in \fref{square_move}, and generates a new theory which is also of BFT type.\footnote{Acting with Seiberg duality on a face with more than 4 sides leads to a dual theory that is not a BFT, namely that is not described by a bipartite graph. While we will not consider this possibility in this paper, it is perfectly fine and interesting from a physical standpoint.} It is natural to investigate the interplay between Seiberg duality and instanton backreaction. There are three distinct possibilities, depending on whether Seiberg duality acts on:
\begin{itemize}
\item[{\bf a)}] The instanton face. Since we do not wrap regular D-branes on the face occupied by an instanton, there is no corresponding gauge group. By Seiberg dualizing the instanton face, we mean performing a square move on it. 
\item[{\bf b)}] A face that is adjacent to the instanton one, i.e. which have some common edge(s) with it.
\item[{\bf c)}] A non-adjacent face.
\end{itemize}
Below we discuss the first two possibilities, comparing the results of backreaction in the original and in the Seiberg dual theories. Case (c) is straightforward: since both instanton backreaction and Seiberg duality are local operations in the BFT that at most affect neighboring faces, it is clear that the two operations commute in this case.

\subsection{Seiberg Duality on the Instanton Face}
\label{section_seiberg_same}

\begin{figure}[!ht]
\begin{center}
\includegraphics[height=8.5cm]{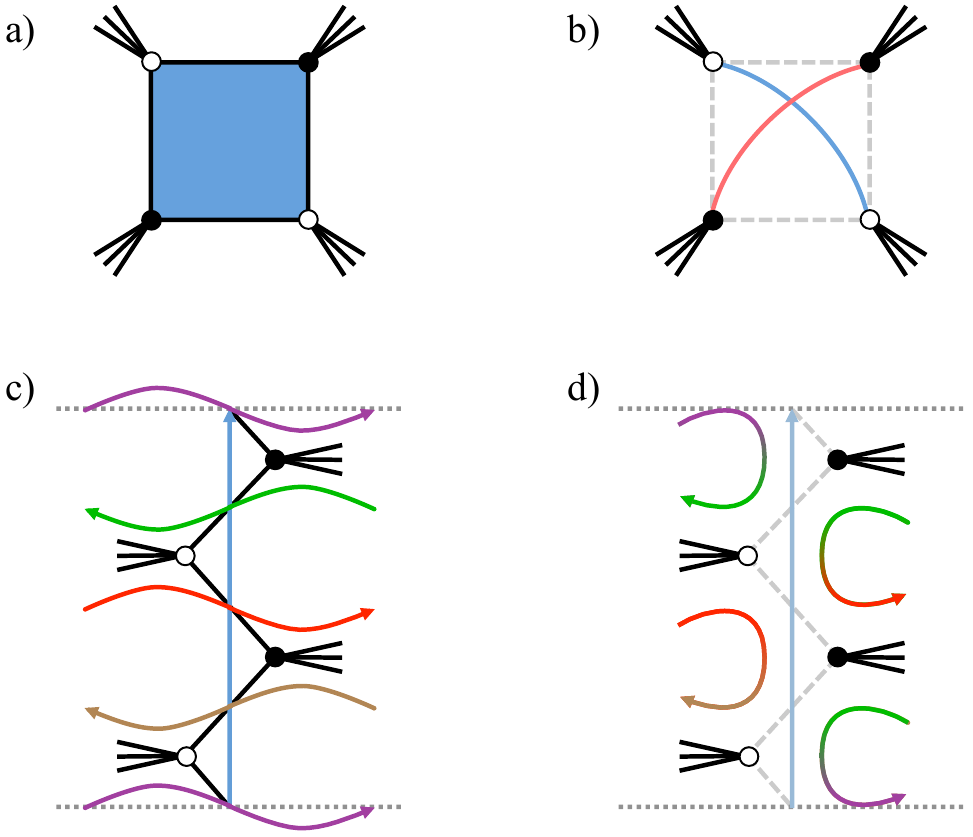}
\caption{a) Local piece of a BFT with an instanton on a 4-sided face. b) Backreacted BFT, with the identification of corner nodes indicated by bridges. c) The initial configuration in the mirror. The instanton wraps the length 4 blue 1-cycle. d) Effect of backreaction in the mirror.}
\label{seiberg-instanton1}
\end{center}
\end{figure}

Consider a theory with an instanton on a 4-sided face producing a (possibly higher genus) BFT via its backreaction, as illustrated in \fref{seiberg-instanton1}. (a) and (b) show the backreaction from the BFT perspective. (c) and (d) show the same process from the mirror viewpoint.

\begin{figure}[!ht]
\begin{center}
\includegraphics[height=8.5cm]{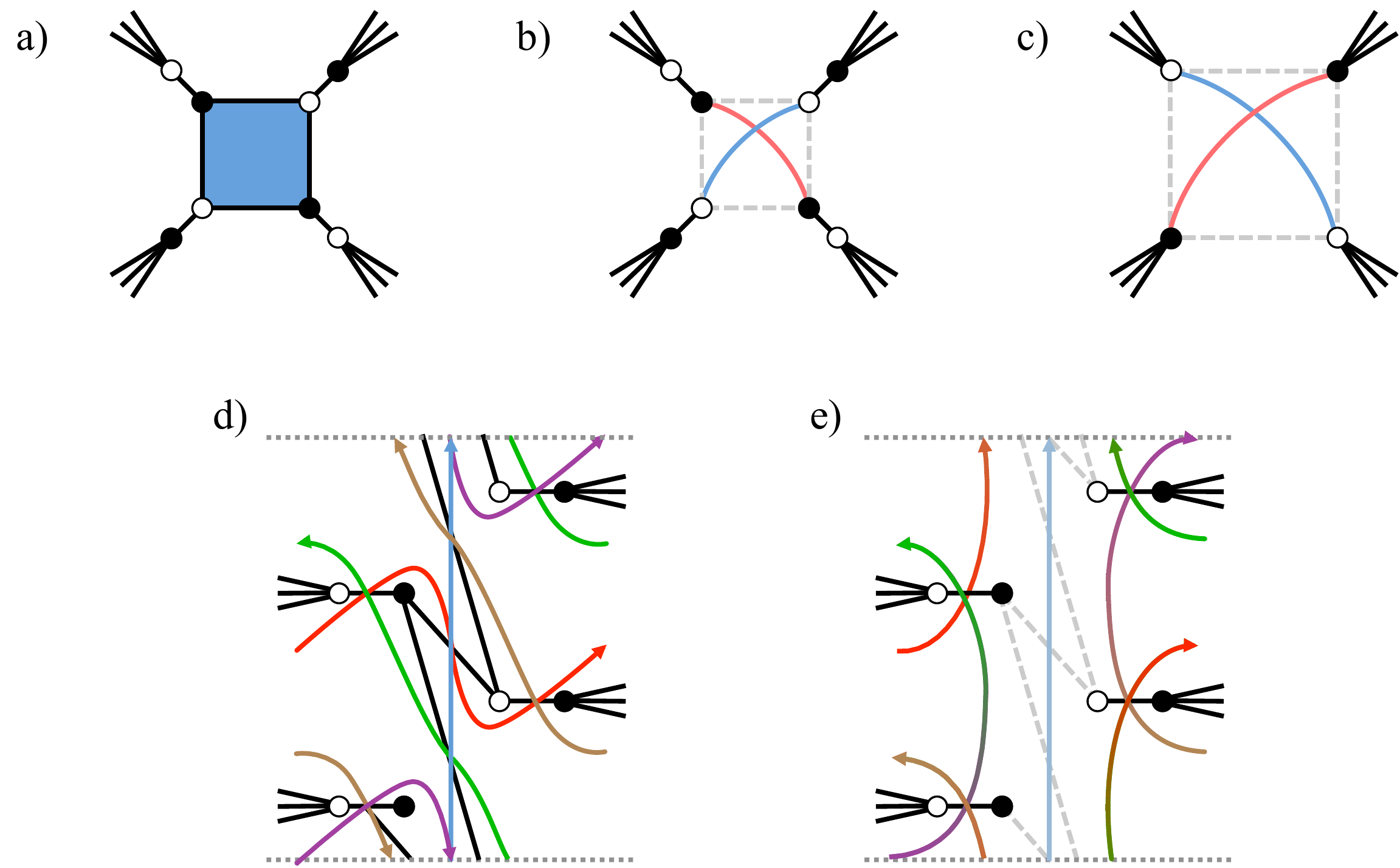}
\caption{a) Seiberg dual of the local configuration in \fref{seiberg-instanton1}.a. b) Backreacted BFT, with the identification of corner nodes indicated by bridges. c) After integrating out massive fields we obtain \fref{seiberg-instanton1}.b. d) The Seiberg dual configuration in the mirror. The instanton wraps the length 4 blue 1-cycle. e) Effect of backreaction in the mirror.}
\label{seiberg-instanton2}
\end{center}
\end{figure}

Let us now compare it with the theory obtained by first Seiberg dualizing the node on which the instanton sits and then backreacting the instanton. This process is shown in \fref{seiberg-instanton2}. The result is the same as the one obtained by backreacting the instanton on the original theory. From the BFT point of view, we see that \fref{seiberg-instanton2}.c is identical to \fref{seiberg-instanton1}.b. The field theory analysis is straightforward and can be directly inferred from the bipartite graph, so we skip it.

\subsection{Seiberg Duality on an Adjacent Face}

\label{section_seiberg_neighbor}

\begin{figure}[!ht]
\begin{center}
\includegraphics[width=\linewidth]{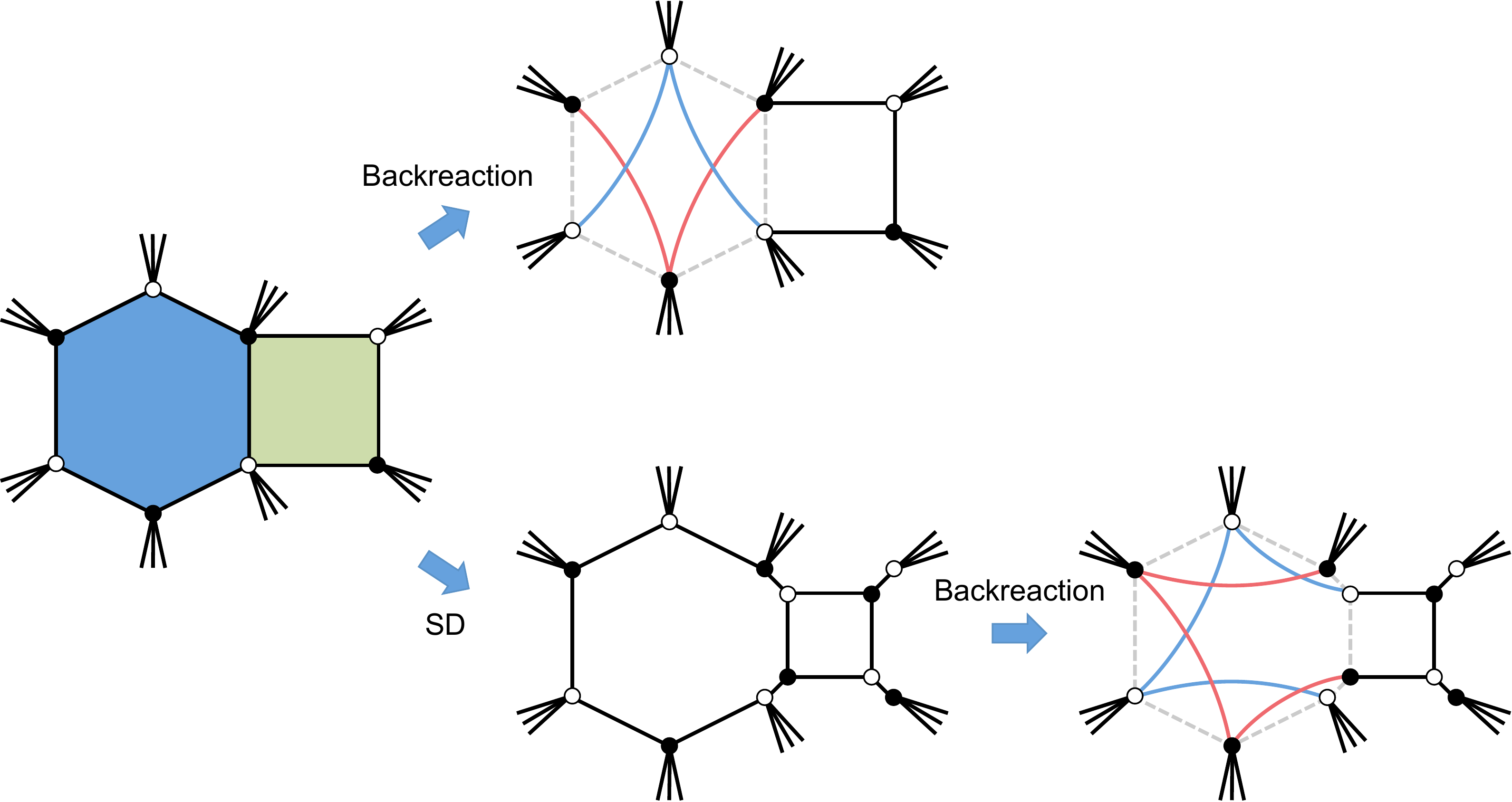}
\caption{Seiberg duality on a face (green) adjacent to a D-brane instanton (blue). On the first row we show the instanton backreaction on the original theory. On the second row, we first apply Seiberg duality and then backreact the face originally occupied by the instanton.}
\label{SD_adjacent_faces}
\end{center}
\end{figure}

Let us now consider Seiberg dualizing a face that is adjacent to the one with an instanton, as illustrated in \fref{SD_adjacent_faces}. We indicate the dualized and instanton faces in green and blue, respectively. We restrict the green face to be a square, so that we remain within the BFT class of theories. The instanton face can have an arbitrary number of edges. Without loss of generality, we take it to be an hexagon in this example.

On the first row of \fref{SD_adjacent_faces}, we backreact the instanton on the original BFT. On the second row, instead, we Seiberg dualize the green face before backreacting the instanton. More specifically, what we mean by this is that in the Seiberg dual we backreact an instanton that occupies the same face as the original one. In \sref{section_SD_brane charges} we will elaborate on the relation between the cycles wrapped by the instantons in both theories. While the details of the final result are example dependent and not so important, a lesson from \fref{SD_adjacent_faces} is that these two procedures generically lead to different BFTs. As illustrated below in an example, such BFTs are in general not even Seiberg dual.

\subsubsection{Different Results: $dP_2$}

\begin{figure}[!ht]
\begin{center}
\includegraphics[width=13cm]{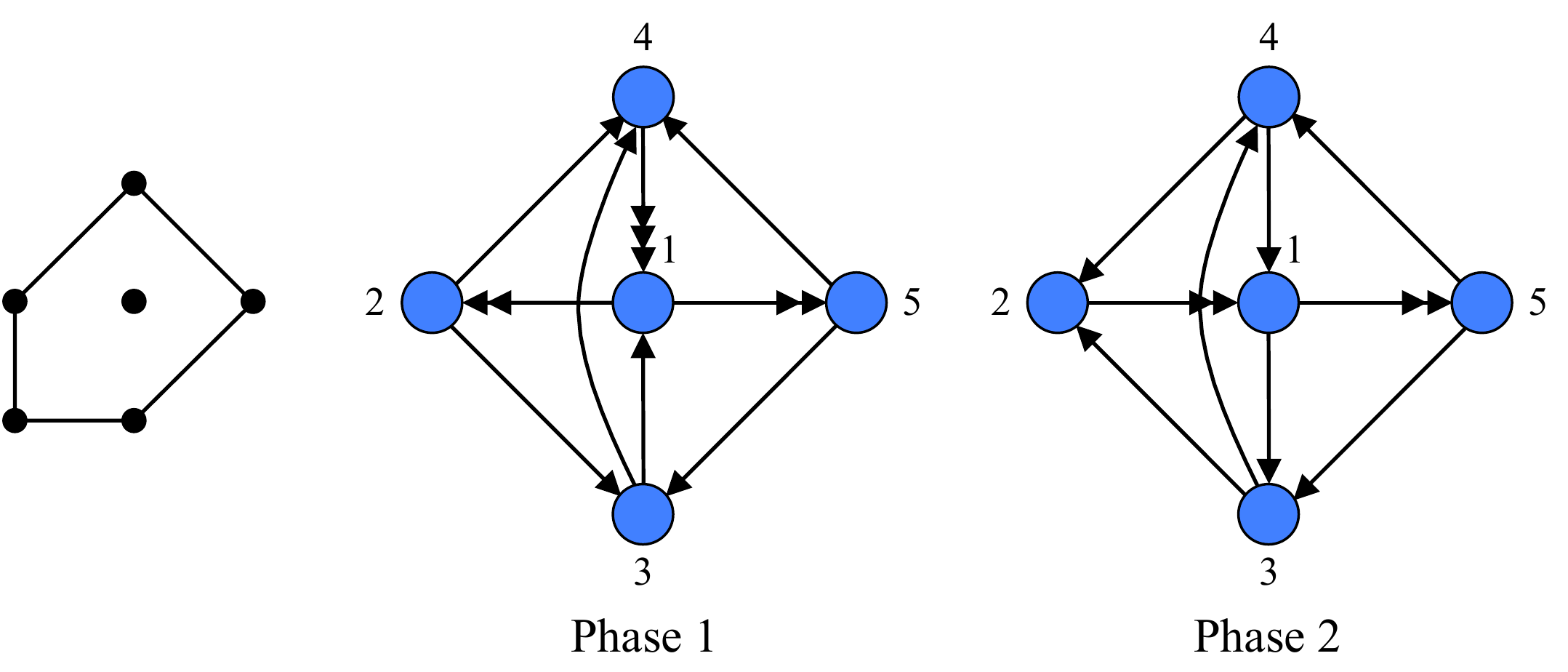}
\caption{Toric diagram for $dP_2$ and quivers for its two toric phases. The two phases are connected by Seiberg duality on node 2.}
\label{dP2_toric_quivers}
\end{center}
\end{figure}

We first consider an example in which, as generically expected, the two operations produce different BFTs. \fref{dP2_toric_quivers} shows the toric diagram for $dP_2$ and the quivers for its two toric phases.

\begin{figure}[!ht]
\begin{center}
\includegraphics[width=6cm]{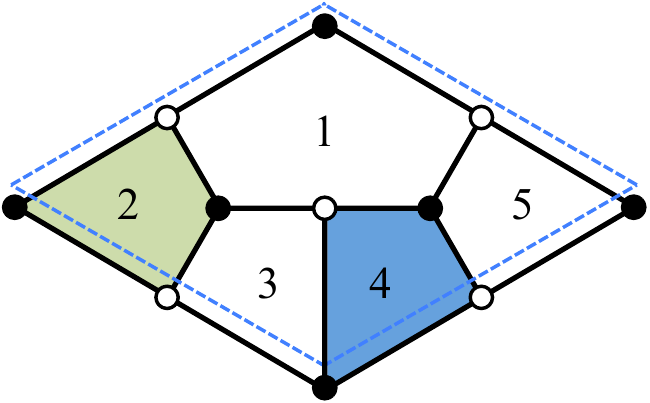}
\caption{Phase 2 of $dP_2$. We will consider an instanton on face 4 (blue) and Seiberg duality on face 2 (green).}
\label{SD_instanton_dP2}
\end{center}
\end{figure}

Let us start from phase 2, whose dimer is shown in \fref{SD_instanton_dP2}. We will consider an instanton on node 4 (blue) and Seiberg duality on node 2 (green). Notice that these two faces are adjacent once the periodicity of $\mathbb{T}^2$ is taken into account. Below we study carefully what happens when these two operations are implemented in different orders. 

\paragraph{Backreaction First.}

\begin{figure}[!ht]
\begin{center}
\includegraphics[width=10cm]{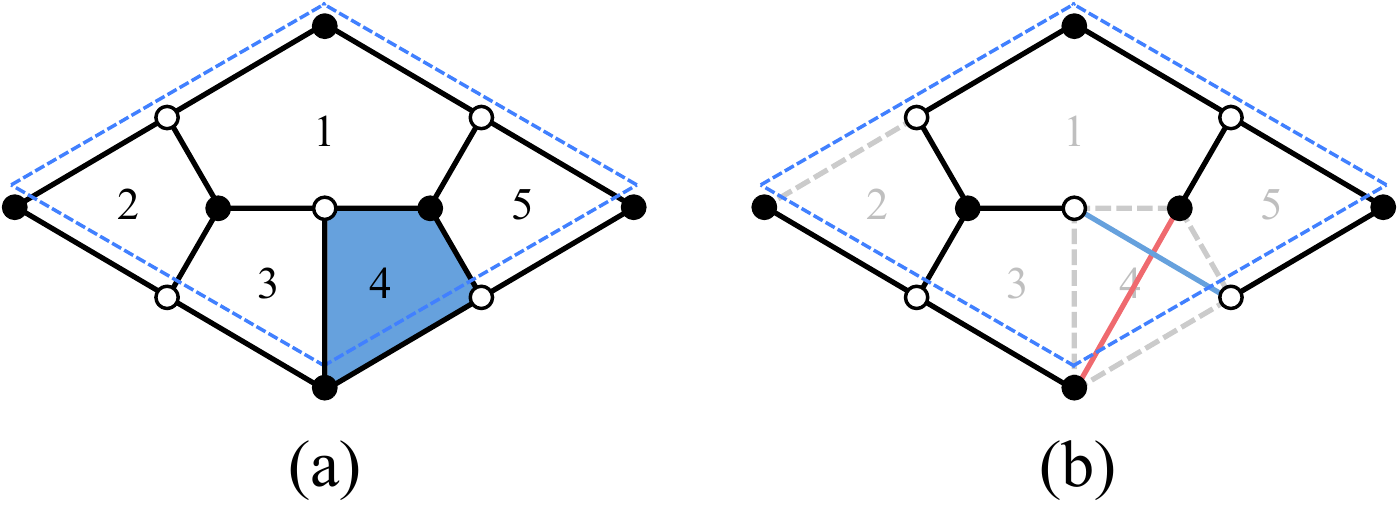}
\caption{Backreaction of an instanton on face 4 of phase 2 of $dP_2$.}
\label{dP2_2_dimer_backreaction}
\end{center}
\end{figure}

Let us first backreact the instanton on face 4. The result is shown in \fref{dP2_2_dimer_backreaction}.b

\begin{figure}[!ht]
\begin{center}
\includegraphics[width=3.5cm]{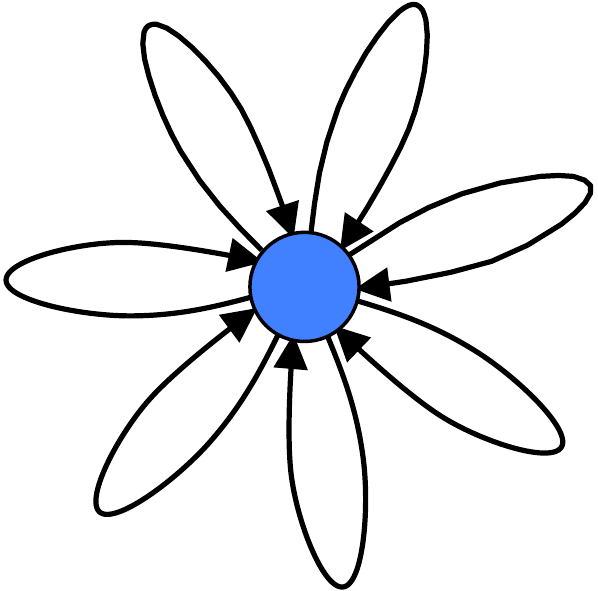}
\caption{Quiver for the genus 2 BFT obtained by backreacting an instanton on face 4 of phase 2 of $dP_2$.}
\label{7_chiral_quiver}
\end{center}
\end{figure}

It is impossible to avoid the crossing of bridges and we have $\Delta g=1$. It is a rather straightforward, albeit tedious, exercise to explicitly embed the final BFT into a genus 2 Riemann surface. This is not very illuminating, so let us exploit the information at hand. After integrating out massive fields in \fref{dP2_2_dimer_backreaction}.b we are left with 4 nodes and 7 edges. 
Combined with the knowledge that this is a genus 2 BFT, we conclude it has a single face. The 5 faces of the original theory get combined into a single one wrapped over the genus 2 Riemann surface. The corresponding quiver consists of a single gauge group and 7 chirals transforming in the adjoint representation, as shown in \fref{7_chiral_quiver}. Its superpotential can be read from the bipartite graph and contains two cubic and two quartic terms. This theory is Model 7.4 in the classification of \cite{Cremonesi:2013aba}.

\paragraph{Seiberg Duality First.}

Starting from phase 2 of $dP_2$ and acting with Seiberg duality on node 2 first, we obtain phase 1, whose quiver is shown in \fref{dP2_toric_quivers}. 

\begin{figure}[!ht]
\begin{center}
\includegraphics[width=\textwidth]{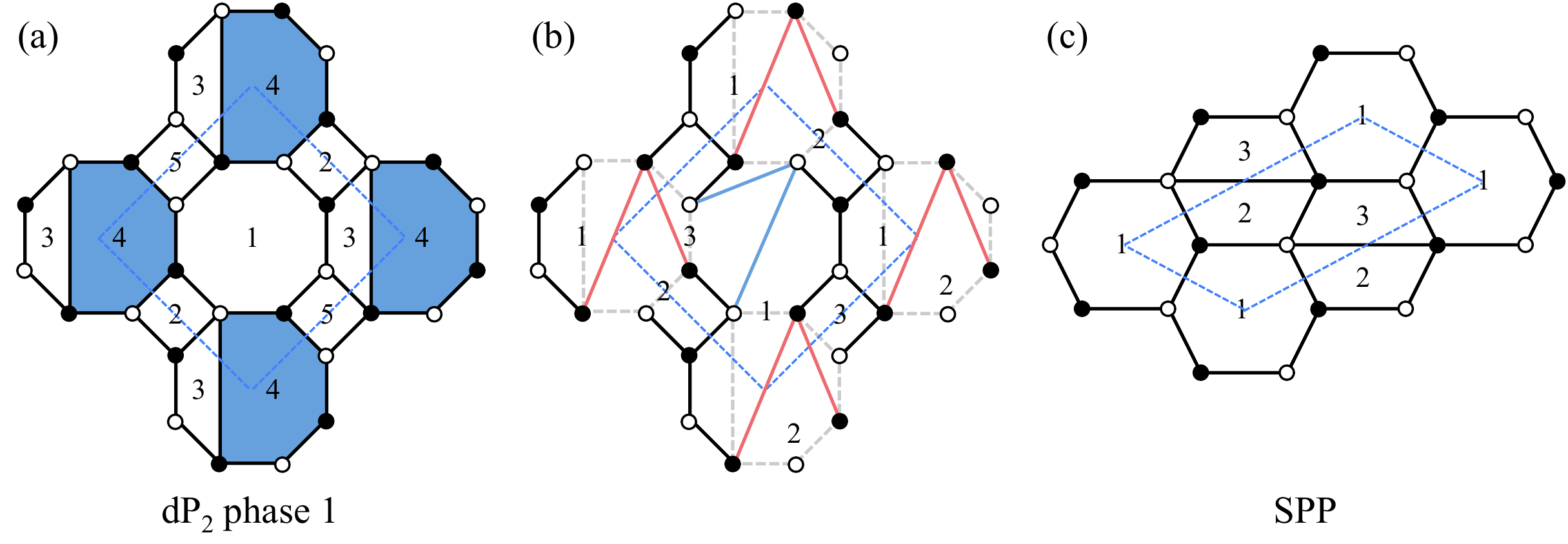}
\caption{Instanton backreaction from phase 1 of $dP_2$ to SPP.}
\label{dP2_1_dimer_backreaction}
\end{center}
\end{figure}

\begin{figure}[!ht]
\begin{center}
\includegraphics[width=8.7cm]{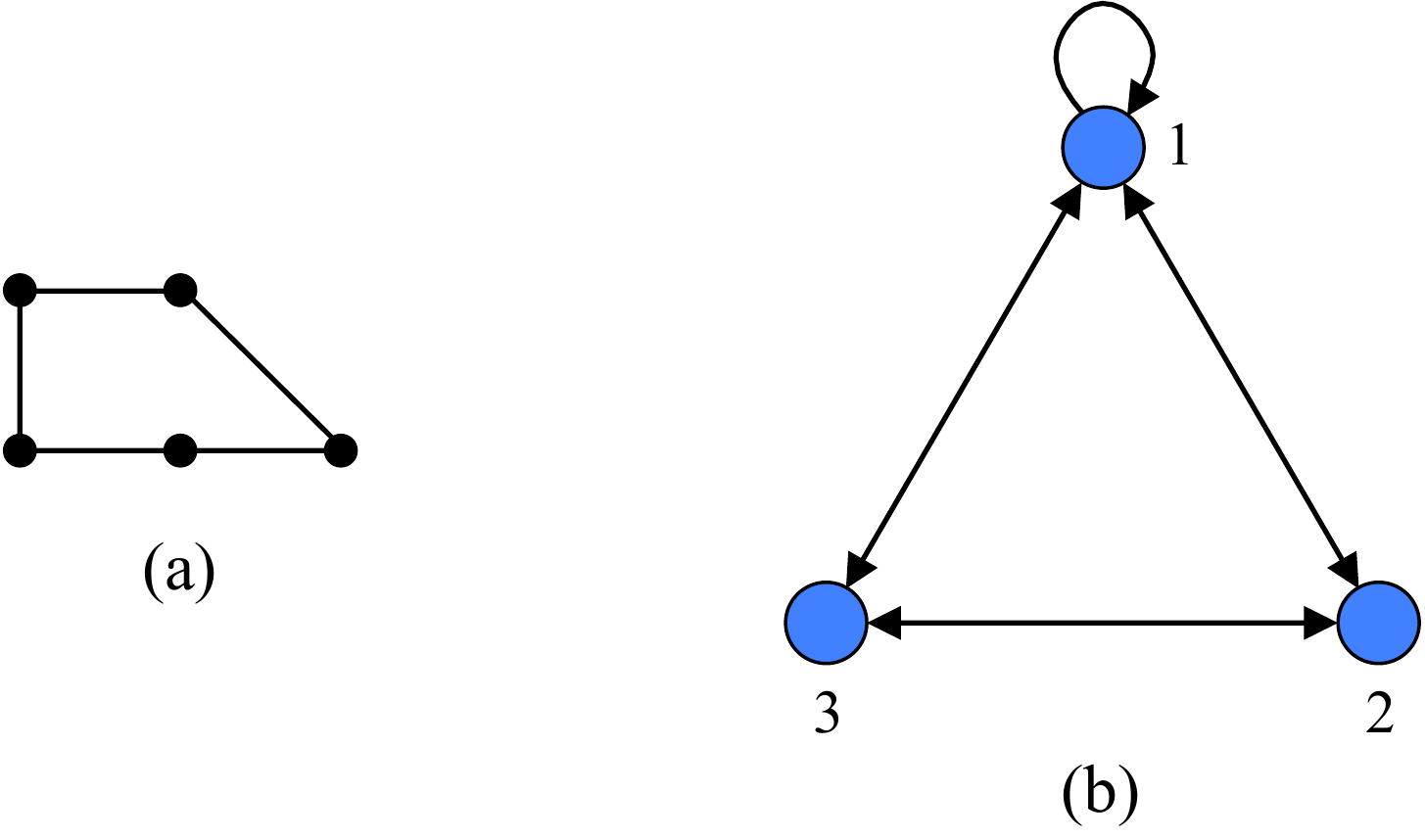}
\caption{a) Toric diagram and b) quiver for the SPP.}
\label{toric_quiver_SPP}
\end{center}
\end{figure}

The corresponding dimer, with an instanton on face 4, is presented in \fref{dP2_1_dimer_backreaction} (a). The instanton backreaction is shown in (b). Face 4 has six sides, i.e. $k=3$, so we would naively expect the genus of the BFT to change by $\Delta g=2$. Interestingly, as shown in the figure, global identifications make it possible to pick bridges such that there are no crossings. As a result, we obtain a new genus 1 BFT. Integrating out massive chiral fields and rearranging the graph, we obtain (c), which is the dimer model for the suspended pinch point (SPP) \cite{Franco:2005rj}. The quiver and toric diagram for the SPP are shown in \fref{toric_quiver_SPP}.

It is clear that this theory is different from the one obtained by backreacting the instanton first. In fact it is not even Seiberg dual to it, since the number of gauge groups, BFT genus and moduli space (even its dimension) are different.

\subsubsection{Same Result: $F_0$}

\begin{figure}[!ht]
\begin{center}
\includegraphics[width=\linewidth]{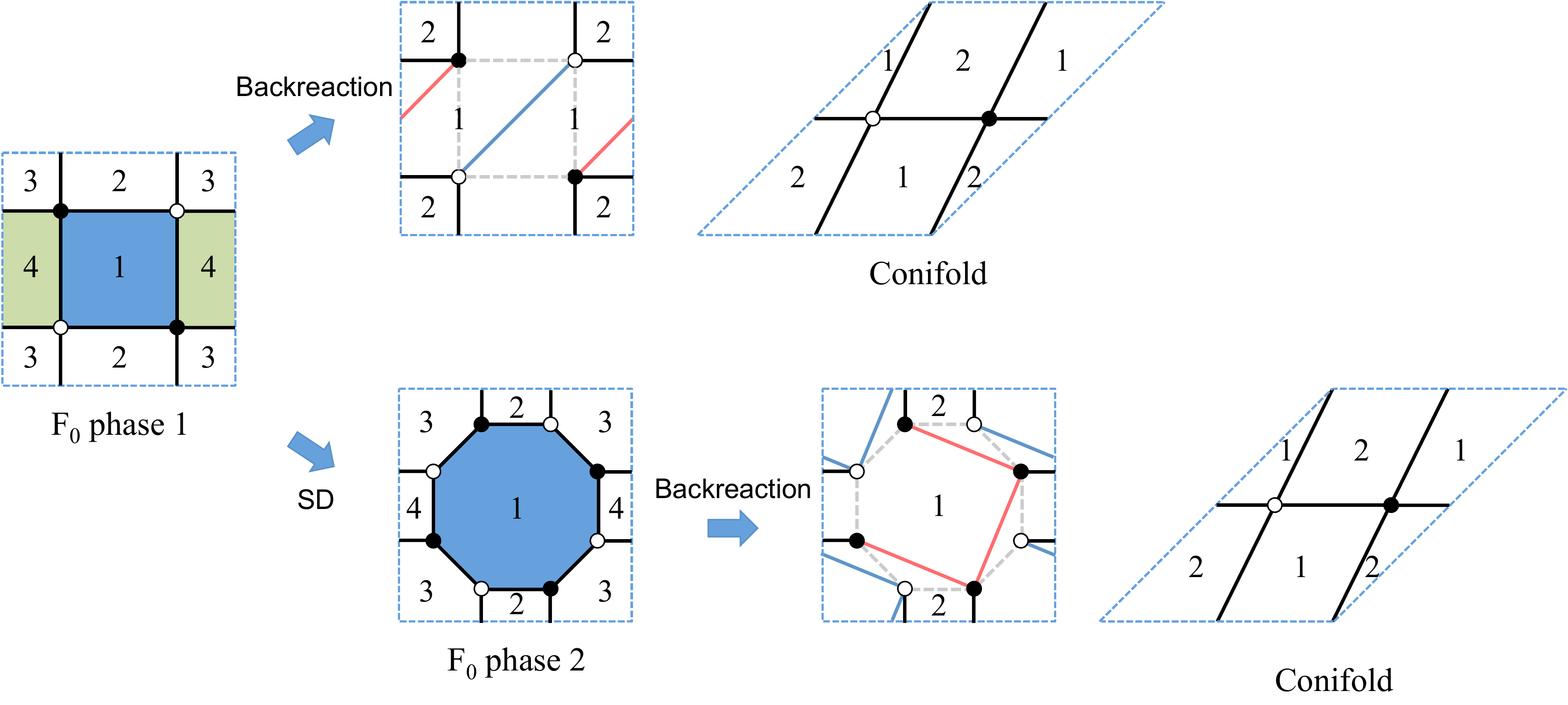}
\caption{Starting from phase 1 of $F_0$, we compare the backreaction of an instanton on face 1 before and after Seiberg duality on the adjacent face 4.}
\label{F0_SD_backreaction_to_conifold}
\end{center}
\end{figure}

While generically backreaction in the original and the Seiberg dual theories do not lead to the same BFT, this can occur in simple models. This is the case for $F_0$, as we now explain. As shown in \fref{F0_SD_backreaction_to_conifold}, we will start from phase 1 and consider an instanton on face 1 and Seiberg duality on face 4. Backreaction of the instanton leads to the conifold, as discussed in \sref{section_examples_global_identifications} and summarized in the top row of the figure. Dualizing face 4 first, we obtain phase 2 of $F_0$. In this case, backreaction of the instanton on face 1 also leads to the conifold.  

\subsubsection{Seiberg Duality and D-brane Charges}

\label{section_SD_brane charges}

We can understand in further detail why instanton backreaction and Seiberg duality on an adjacent face generically do not commute by considering how Seiberg duality transforms the cycles wrapped by different stack of D-branes or, equivalently, their D-brane charges. 

\begin{figure}[!ht]
\begin{center}
\includegraphics[width=6cm]{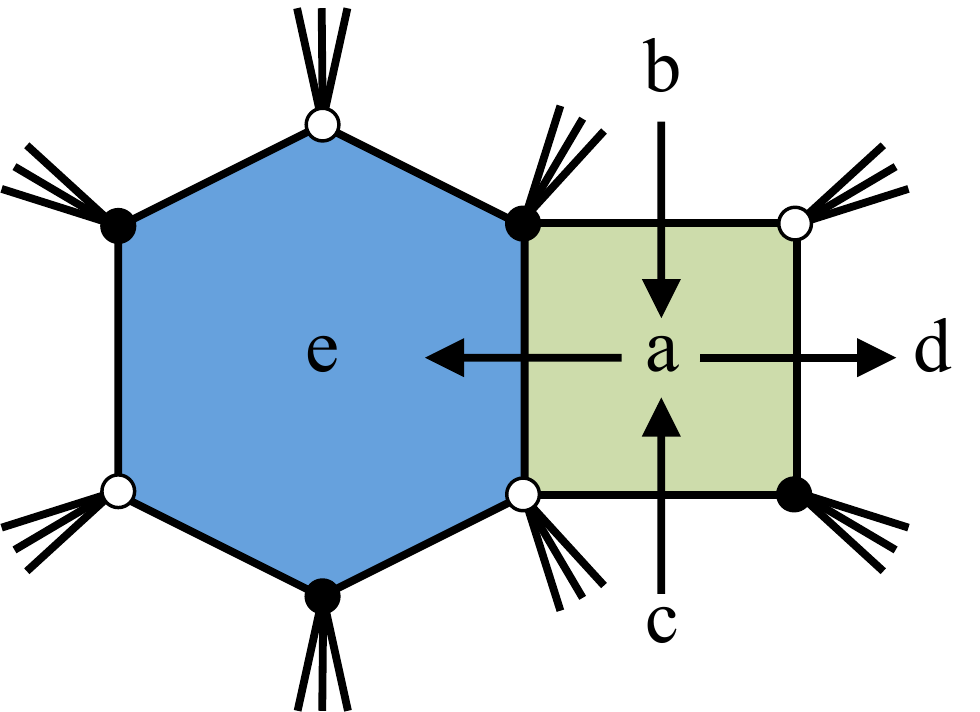}
\caption{Local configuration showing a face to be Seiberg dualized (green) which is adjacent to a D-brane instanton (blue).}
\label{quiver_Seiberg_instanton}
\end{center}
\end{figure}

\fref{quiver_Seiberg_instanton} shows the local configuration we are interested in. Seiberg duality will act on face $a$. The four adjacent faces are labeled $b$, $c$ $d$ and $e$, with the D-brane instanton located on face $e$. We also explicitly show the arrows representing the bifundamental chiral fields connecting $a$ to the four adjacent faces. Without loss of generality, we assume that the chiral field connecting $a$ and $e$ goes from $a$ to $e$.\footnote{As usual, the orientation of all the arrows can be inverted by flipping the convention for fundamental and antifundamental representations.}

The intersection numbers between branes indicate the number of arrows connecting them, with their orientation determined by the sign. In this case, we have
\beq
\begin{array}{ccccccc}
(\left[b\right] \cdot \left[a\right]) & = & 1 & \ \ \ \ \ \ (\left[d\right] \cdot \left[a\right]) & = & -1 \\
(\left[c\right] \cdot \left[a\right]) & = & 1 & \ \ \ \ \ \ (\left[e\right] \cdot \left[a\right]) & = & -1 
\end{array}
\eeq

Acting with Seiberg duality on $[a]$, the different branes transform as follows (see e.g. \cite{Cachazo:2001sg,Feng:2002kk}): 
\beq
\begin{array}{ccl}
\left[a'\right] & = & -\left[a\right] \\[.2cm]
\left[b'\right] & = & \left[b\right] + (\left[b\right] \cdot \left[a\right]) \left[a\right]  = \left[b\right] + \left[a\right] \\
\left[c'\right] & = & \left[c\right] + (\left[c\right] \cdot \left[a\right]) \left[a\right]  = \left[c\right] + \left[a\right] \\[.2cm]
\left[d'\right] & = & \left[d\right] \\
\left[e'\right] & = & \left[e\right] 
\end{array}
\label{transformation_D-brane_charges}
\eeq
The orientation of the brane for the dualized gauge group is reversed. The branes connected to incoming flavors pick a contribution proportional to $[a]$ and the relative intersection numbers. Finally, the branes connected to outgoing flavors remain unchanged. In particular, the instanton $[e]$ is invariant.\footnote{In the convention that inverts all the arrows in the quiver, the roles of $(b,c)$ and $(d,e)$ are exchanged. In particular, $d$ and $e$ are transformed while $b$ and $c$ stay the same. The final results are independent of this choice.}

Let us now consider the new intersection numbers between the instanton and other D-branes. From \eref{transformation_D-brane_charges}, 
\beq
(\left[e' \right] \cdot \left[ a' \right]) = - (\left[e\right] \cdot \left[ a \right])
\eeq
and
\beq
\begin{array}{ccccc}
(\left[e' \right] \cdot \left[ b' \right]) & = & (\left[e \right] \cdot \left[ b \right]) +  (\left[e \right] \cdot \left[ a \right]) & = & (\left[e \right] \cdot \left[ b \right]) - 1 \\
(\left[e' \right] \cdot \left[ c' \right]) & = & (\left[e \right] \cdot \left[ c \right]) +  (\left[e \right] \cdot \left[ a \right]) & = & (\left[e \right] \cdot \left[ c \right]) - 1 
\end{array}
\eeq
As a consequence of these new intersections, the instanton on the dual theory breaks a different $U(1)$ subgroup of the global symmetry. Hence, as expected, after introducing the instanton generated field theory operator we do not obtain the Seiberg dual of the original theory plus instanton.

\section{Multi-Instantons and Complex Deformations}
\label{section_multi-complex}

Our previous discussion has focused on the case of single instantons, namely those associated to a single face in the original theory. Although the discussion of the richer class of general multiple instantons is left for future work, we would now like to delve into a particularly interesting class, corresponding to (generically) multiple instantons triggering complex deformations of the original geometry.

The effect of backreaction is to pinch off the cycle in the mirror Riemann surface $\Sigma$ wrapped by the D-brane instanton, which in turn triggers the recombination of the D-branes that intersect it. Interestingly, it is sometimes possible to wrap the instanton on a cycle such that shrinking it to zero size splits $\Sigma$ into two disconnected components $\Sigma_1$ and $\Sigma_2$.\footnote{This is the simplest possibility. Splitting $\Sigma$ into more components is also possible.}

Since backreaction preserves the original punctures, whenever such decomposition occurs, the punctures get distributed between $\Sigma_1$ and $\Sigma_2$. The mirror Riemann surface corresponds to thickening the $(p,q)$-web dual to the toric diagram. Hence, the $\Sigma\to \Sigma_1 + \Sigma_2$ splitting corresponds to decomposing the web into two subwebs in equilibrium, i.e. webs for which the $(p,q)$ charges of external legs sum up to zero. Such decomposition of the web represents a complex deformation of the underlying CY$_3$ \cite{Franco:2005fd,GarciaEtxebarria:2006aq}, which generalizes the well-known deformation of the conifold \cite{Klebanov:2000hb}. Generalizing the conifold case, such deformations can be triggered by the so-called {\it deformation fractional branes} in the general classification of \cite{Franco:2005zu}. From a quiver perspective, fractional branes correspond to anomaly free modifications of the ranks of gauge groups, i.e. of faces in the dimer. Deformation fractional branes are such that, at low energies, the dynamics of the gauge theory on them is (possibly partial) confinement, which translates into the complex deformation of the associated CY$_3$. 

Recently, it was noted that precisely the same deformation is achieved if the fractional branes are replaced by D-brane instantons, namely if we wrap D-brane instantons over the corresponding cycle \cite{Tenreiro:2017fon}. Equivalently, this corresponds to locating the instanton on the faces of the dimer associated to the fractional branes. This is perhaps not surprising, since D-brane instanton effects in various CY 3-folds can be understood as the IR dynamics of theories with duality cascades generated by fractional branes \cite{Aharony:2007pr,Argurio:2012iw,Franco:2015kfa}. 

\begin{figure}[!ht]
\begin{center}
\includegraphics[width=10cm]{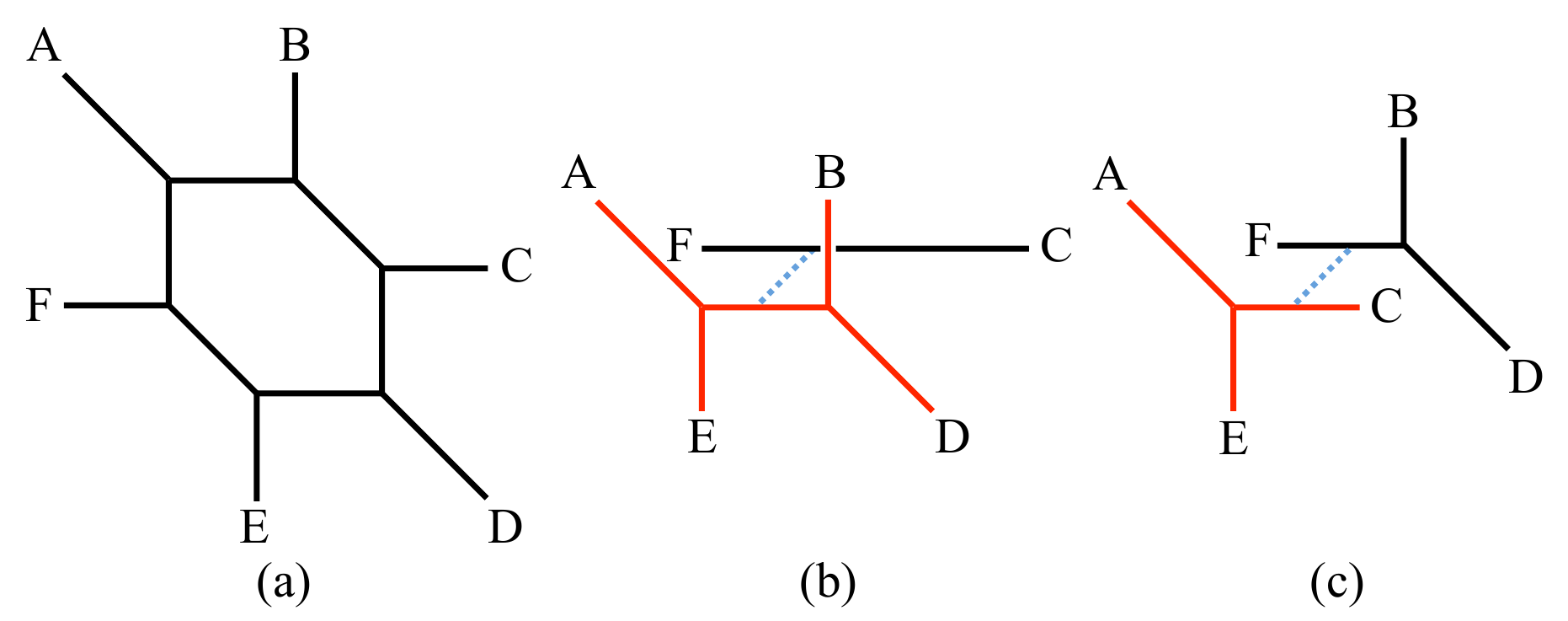}
\caption{Web diagrams of the two possible complex deformations of $dP_3$. The dashed segments indicate $S^3$'s.}
\label{web_deformations_dP3}
\end{center}
\end{figure}

In order to illustrate these ideas, let us consider the $dP_3$ theory. 
This geometry admits two independent complex deformations, which are shown in \fref{web_deformations_dP3} in terms of $(p,q)$-webs. In the context of deformation fractional branes, these two deformations were considered in \cite{Franco:2005fd}. When generated by D-brane instanton backreaction, they were studied in \cite{Tenreiro:2017fon}, focusing on the mirror perspective. Below we discuss how they are captured by implementing instanton backreaction directly at the level of the dimer. 

It would be interesting to investigate whether D-brane instantons that start from BFTs of genus different from 1 can lead to similar deformations of the corresponding toric CYs.

\subsection*{From $dP_3$ to the Conifold}

\begin{figure}[!ht]
\begin{center}
\includegraphics[width=12cm]{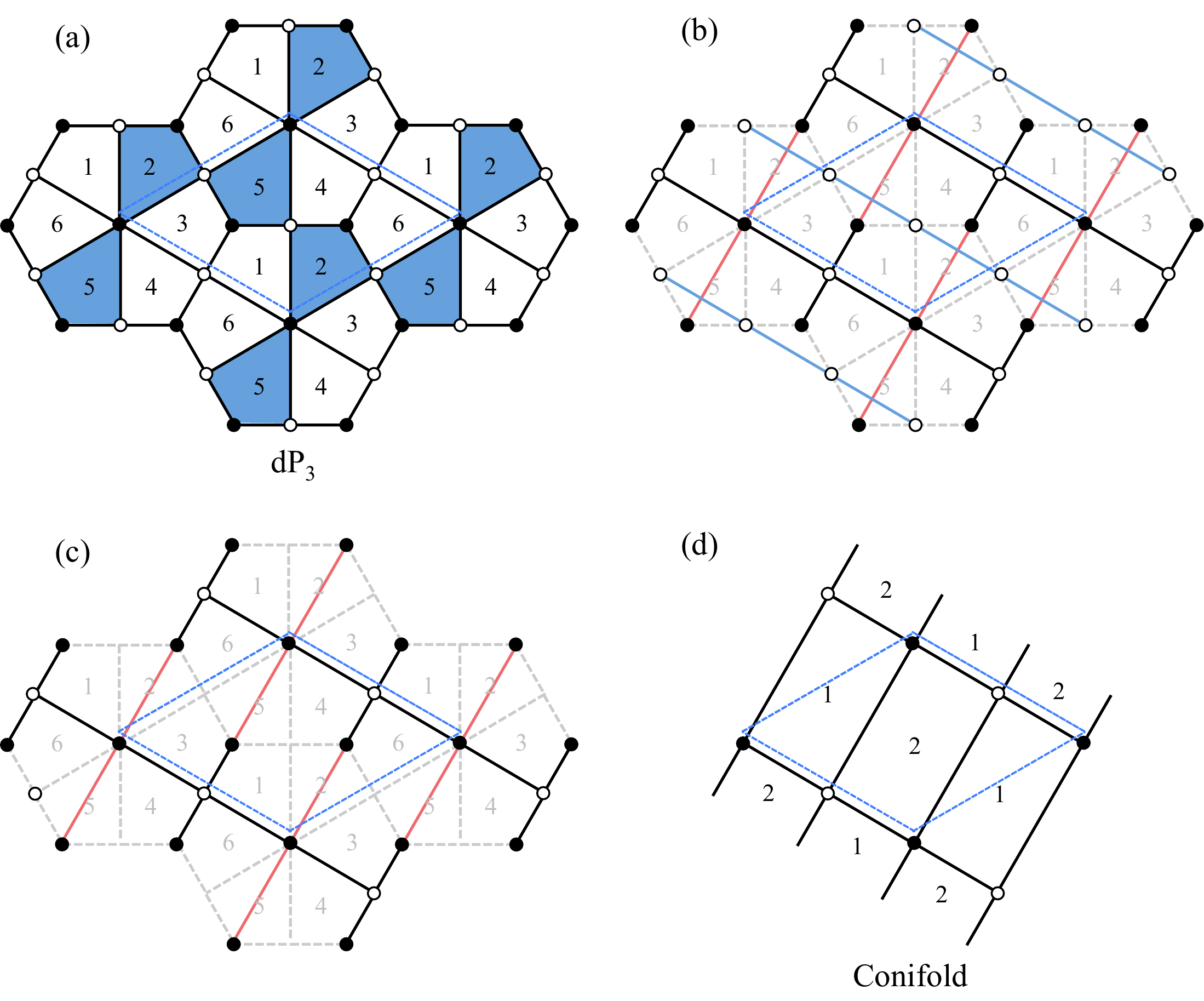}
\caption{Instanton backreaction from phase 1 of $dP_3$ to the conifold.}
\label{deformation_dP3_conifold_dimer}
\end{center}
\end{figure}

Let us first consider the deformation in \fref{web_deformations_dP3}.b, that goes from $dP_3$ to the conifold. \fref{deformation_dP3_conifold_dimer} goes through the process step by step. In order to achieve this deformation, the D-brane instanton must wrap a cycle that covers faces 2 and 5 of the dimer, as shown in (a). The instanton backreaction is shown in (b). Interestingly, the blue bridges form a ``necklace" that is disconnected from the rest of the graph and that disappears at low energies since it consists entirely of massive fields. Removing the blue bridges we obtain (c). After integrating out the massive fields in the red bridges, we obtain (d), which is the dimer for the conifold.

\subsection*{From $dP_3$ to $\mathbb{C}^3$}

We now consider the deformation from $dP_3$ to $\mathbb{C}^3$ of \fref{web_deformations_dP3}.c. In this case, the instanton must be placed on faces 2, 4 and 6 as shown in (a). The backreaction is presented in (b). The red bridges form a hexagonal lattice of massive fields, which is decoupled from the rest of the graph and disappears at low energies, leaving the configuration in (c). Blue bridges form triangles that collapse into single white nodes when massive fields are integrated out. The final result (d) is the dimer for $\mathbb{C}^3$.

\begin{figure}[!ht]
\begin{center}
\includegraphics[width=12cm]{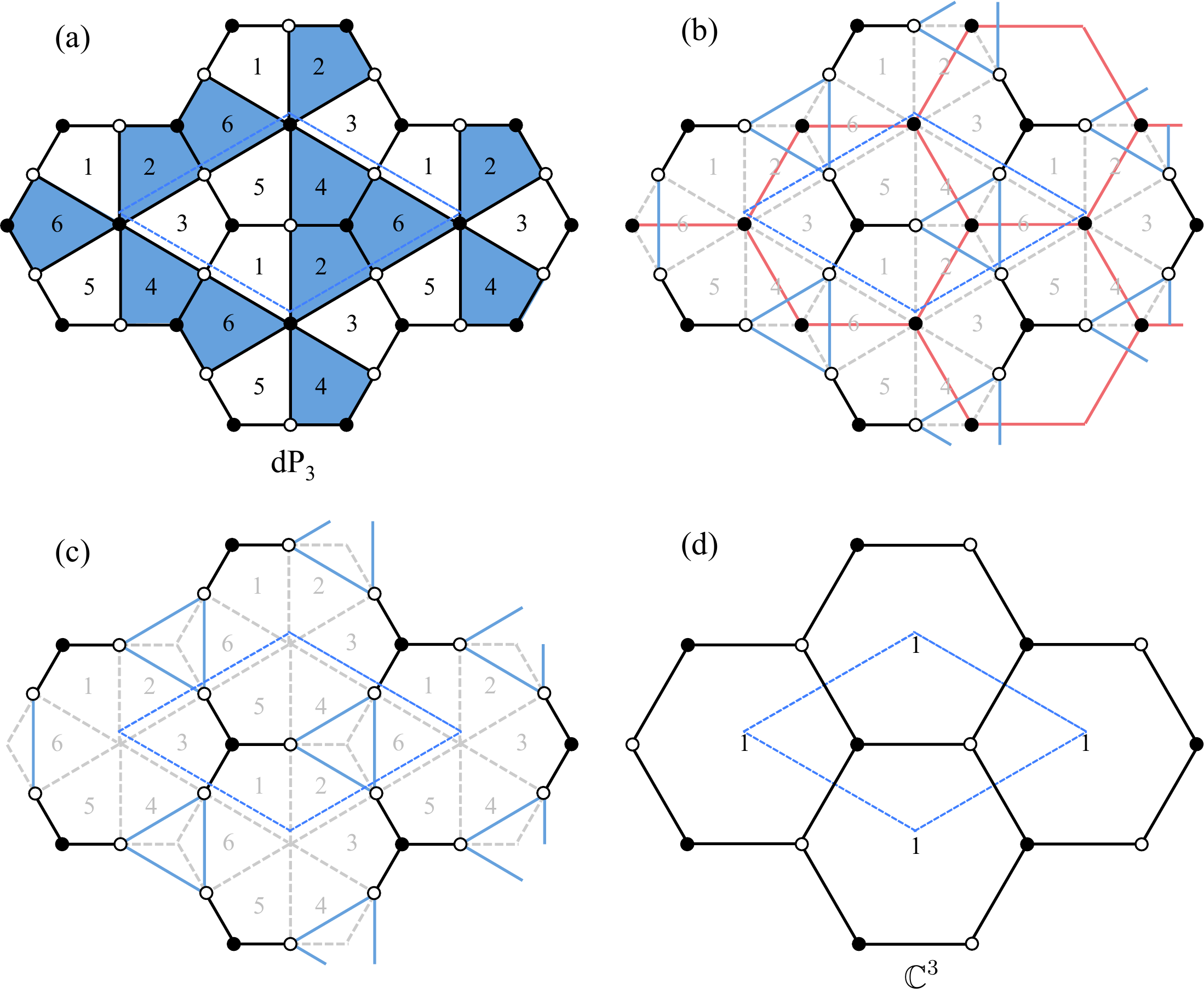}
\caption{Instanton backreaction from phase 1 of $dP_3$ to $\mathbb{C}^3$.}
\label{deformation_dP3_C3_dimer}
\end{center}
\end{figure}

\section{The Inverse Problem}
\label{section_inverse}

We have seen that starting from a BFT and a choice of instanton we generate a second BFT. It is interesting to consider the {\it inverse problem}, namely: given a BFT and assuming it arises from an instanton backreaction on a parent BFT, can we reconstruct the latter? In general, two different BFTs with two different instantons can produce the same BFT upon backreaction.\footnote{For simplicity, we restrict to single instantons. Our discussion extends straightforwardly to multi-instantons and even theories with different numbers of instantons.} The underlying toric geometries can be used as a guide, following the transformation of the toric diagrams discussed in \sref{section_toric_geometry_backreacted_dimers}. Below we present an explicit example, in which the same BFT is obtained in two different ways.

\subsection*{$dP_2$ to SPP}
\begin{figure}[!ht]
\begin{center}
\includegraphics[width=\textwidth]{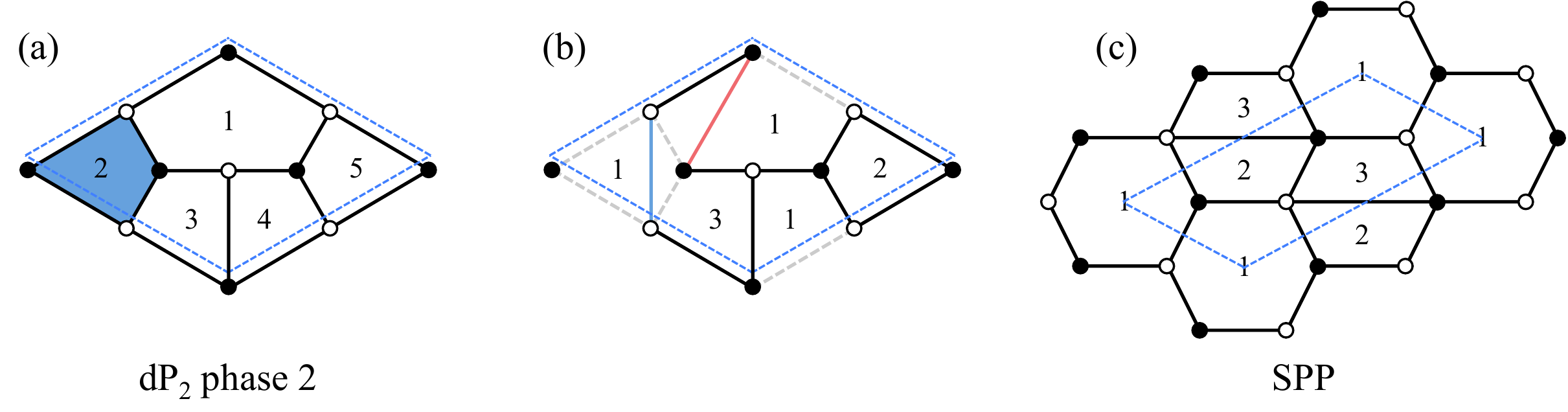}
\caption{Instanton backreaction from phase 2 of $dP_2$ to SPP.}
\label{dP2_to_SPP_dimer}
\end{center}
\end{figure}

Let us first consider phase 2 of $dP_2$, which we already discussed in the previous section. Its toric diagram and quiver were given in \fref{dP2_toric_quivers}. Let us now consider an instanton on face 2, as shown in \fref{dP2_to_SPP_dimer}.a. The backreaction of this instanton is presented in \fref{dP2_to_SPP_dimer}.b, where we have eliminated the instanton face, added the bridges and given new labels to the resulting faces. Interestingly, as shown in the figure, global identifications in this model make it possible to avoid bridge crossings and remain on genus 1. Integrating out massive chiral fields and rearranging the graph, we obtain \fref{dP2_to_SPP_dimer}.c, the dimer for the SPP.

\subsection*{$PdP_2$ to SPP}

Let us now consider $PdP_2$, whose toric diagram and quiver were presented in \fref{diagrams_PdP2}. Comparing Figures \ref{diagrams_PdP2} and \ref{dP2_toric_quivers}, we see that the quivers for this theory and for the phase 2 of $dP_2$ we have just considered are very similar, differing only by a pair of chiral fields associated to a bidirectional arrow connecting nodes 2 and 5. However, the superpotentials are rather different, as encoded in the corresponding dimer models.

\begin{figure}[!ht]
\begin{center}
\includegraphics[width=\textwidth]{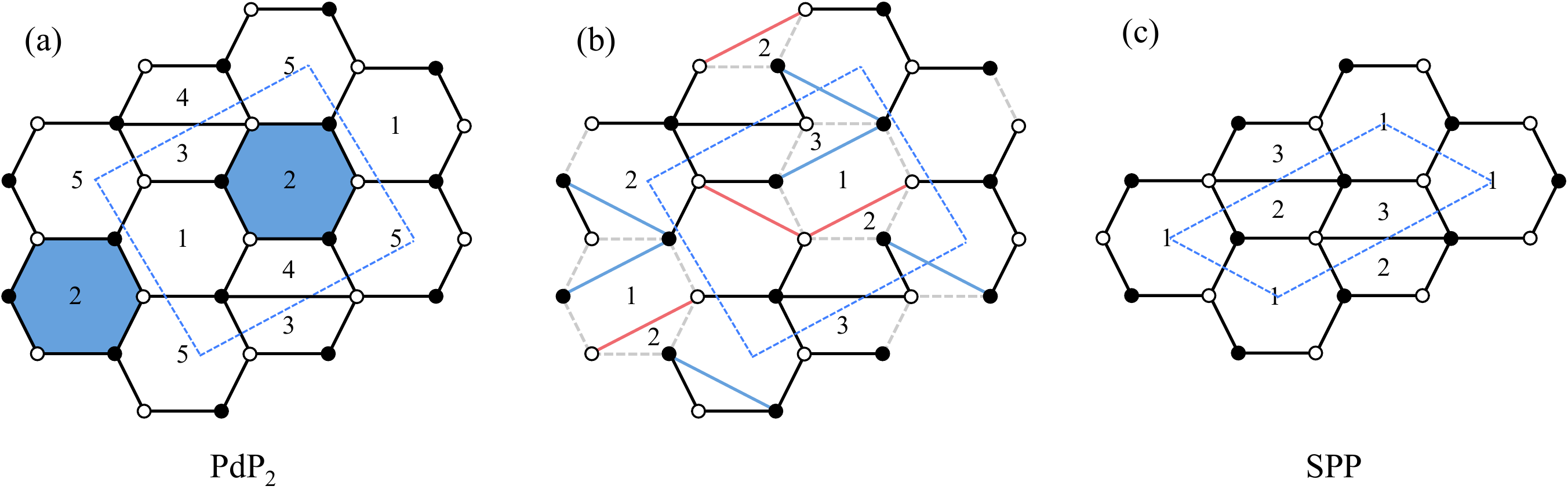}
\caption{Instanton backreaction from $PdP_2$ to SPP.}
\label{PdP2_to_SPP_dimer}
\end{center}
\end{figure}

In \sref{sec:example-pdp2} we considered the effect of an instanton on face 4. Let us now study, instead, an instanton on face 2, as shown in \fref{PdP2_to_SPP_dimer}. (a) shows the dimer with the instanton. In (b), we show the instanton backreaction and have relabeled the surviving faces of the dimer. Unlike what happens for an instanton on face 4, when the instanton is on face 2 global identifications make it possible to avoid bridge crossing without increasing the genus of the BFT. After integrating out massive fields we obtain (c), which is the dimer for the SPP.

\begin{figure}[!ht]
\begin{center}
\includegraphics[width=8.5cm]{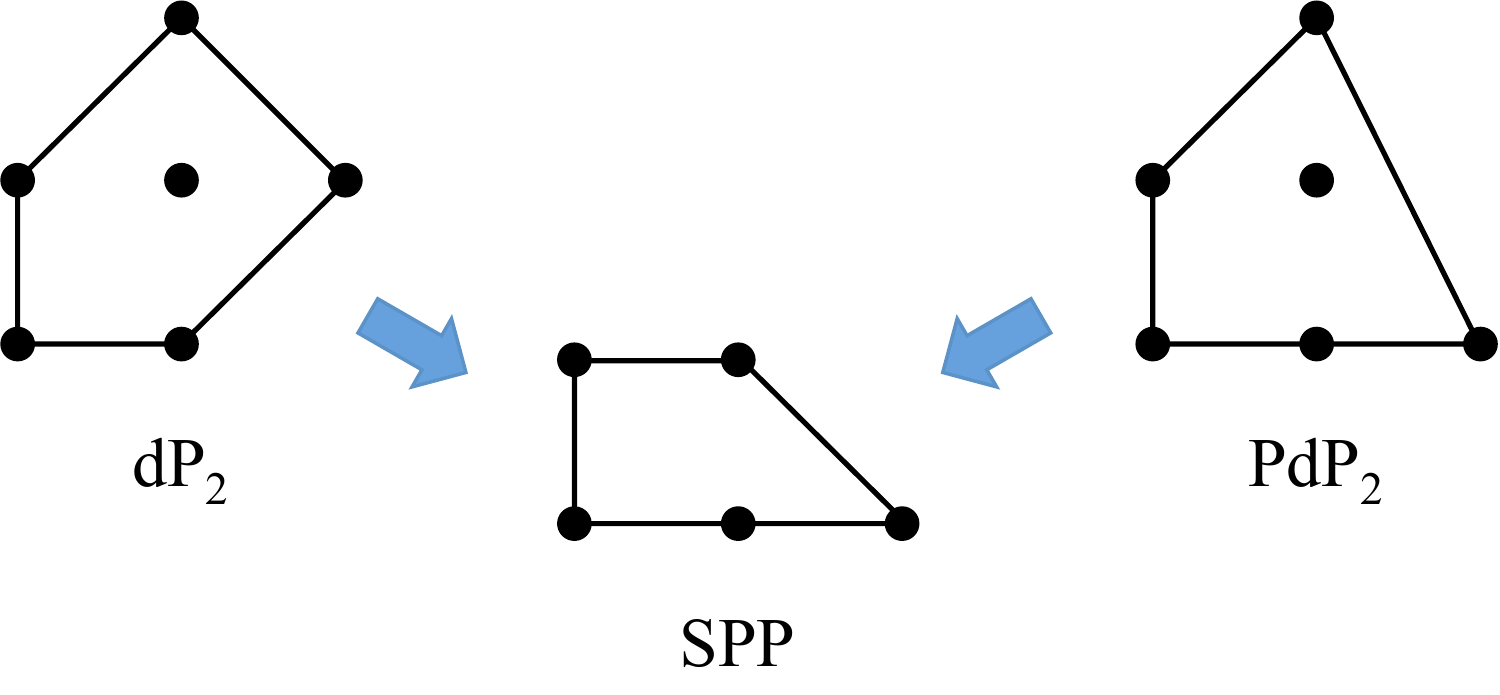}
\caption{The SPP can be obtained from backreaction of instantons on $dP_2$ and $PdP_2$.}
\label{dP2_PdP2_to_SPP_toric}
\end{center}
\end{figure}

We conclude that, as summarized in \fref{dP2_PdP2_to_SPP_toric}, we can reach the BFT for the SPP by backreacting D-brane instantons on either $dP_2$ or $PdP_2$.

We have deliberately made the similarities between phase 2 of $dP_2$ and $PdP_2$, and between the instantons we considered, as manifest as possible. However, it is important to emphasize that, in general, instantons on significantly different theories can produce the same BFT.

\section{Conclusions}
\label{section_conclusions}

In this paper we have provided a simple description of the effect of D-brane instantons in systems of D3-branes at toric CY$_3$ singularities, in terms of a combinatorial recipe in the corresponding bipartite dimer diagram. Interestingly, the prescription brings generically higher-genus BFTs into the game. In this sense, it provides a new physical interpretation for the latter, which adds to those already in the literature, and in fact the first directly relating BFTs to realizations in string theory.

The combinatorial recipe can be generalized to arbitrary BFTs, and provides a new operation relating BFTs on Riemann surfaces generically of different genus. It would be interesting to explore the implications of this operation in the interpretation of bipartite graphs as describing on-shell scattering amplitudes.

Further interesting directions for future work include:

\begin{itemize}

\item Backreaction effects of non-compact instantons, corresponding to Euclidean D3-branes on non-compact 4-cycles. Their proper understanding should connect with the general considerations in \cite{Forcella:2008au}.

\item Systematic study of instanton effects on the D3-brane systems with flavor D7-branes. Since the latter provide the natural arena for the string theory embedding of (low-genus) BFTs with boundaries \cite{Franco:2013ana}, the introduction of handles via D-brane instanton backreaction presumably allows the embedding of the general class of BFTs in string theory.

\item We have taken first steps towards the discussion of multi-instanton backreaction, recovering and explaining earlier results in the case of complex deformations. We expect a systematic discussion of general multi-instanton backreaction to reveal other interesting geometric operations.

\item The higher-dimensional toric data obtained for the higher-genus BFT resulting from instanton backreaction on a CY 3-fold D-brane gauge theory corresponds to a higher-dimensional geometry, whose physical realization is still lacking. It would be interesting to identify physical objects potentially related to this higher-dimensional variety, and its role in the non-perturbative dynamics of the D-brane system.

\end{itemize}

We hope to come back to these and other related questions in future work.

\section*{Acknowledgments}

We would like to thank L. Martucci for useful discussions. The work of S. F. is supported by the U.S. National Science Foundation grant PHY-1518967. He also gratefully acknowledges the Tsinghua Sanya International Mathematics Forum (TSIMF) for hosting the Workshop on SCFTs in Dimension 6 and Lower, where part of this project was carried out. E. G. and A. U. are partially supported by the grants FPA2015-65480-P from the MEIC/FEDER, the ERC Advanced Grant SPLE under contract ERC-2012-ADG-20120216-320421 and the grant SEV-2016-0597 of the ``Centro de Excelencia Severo Ochoa" Programme.

\bibliographystyle{JHEP}
\bibliography{mybib}


\end{document}